\documentclass[a4paper,11pt]{article}
\pdfoutput=1 
\usepackage{jcappub} 
\usepackage{amsmath}
\usepackage{amssymb}
\usepackage{float}
\usepackage{dcolumn}
\usepackage[usestackEOL]{stackengine}
\usepackage{multirow}
\usepackage{color}
\usepackage{graphicx}
\usepackage{dcolumn}
\usepackage{bm}
\usepackage{mathrsfs}
\usepackage{hyperref}

\graphicspath{{./}{figures/}}

\begin{document}

\author{Qixin Yu}
\author{and Dieter Horns}
\affiliation{Institut f\"{u}r Experimentalphysik, University of Hamburg, Luruper Chaussee 149, D-22761 Hamburg, Germany}

\emailAdd{qixin.yu@desy.de,dieter.horns@physik.uni-hamburg.de}

\title{Searching for photon-ALPs mixing effects in AGN gamma-ray energy spectra}

\date{\today}

\abstract{
High energy gamma-rays propagating in external magnetic fields  may convert into axion-like particles (ALPs). In this case, the observed gamma-ray spectra are modified
by the resulting energy-dependent conversion probability. 
In this study, we use the energy spectra of 20 extra-galactic gamma-ray sources recorded during 10 years of \textit{Fermi}-LAT observations.
We define a test statistics based upon the likelihood ratio to test the hypothesis for a spectral model without vs. a model  
with photon-ALPs coupling. The conversion probability is calculated for fixed values of the mass and two-photon coupling of the pseudo-scalar 
particle while the external magnetic field is characterized by the additional free parameters length scale $s$ and average field strength $B$.
As a consistency check and in order to extend the analysis to include very high energy gamma-ray data, another test statistics is defined 
with the $\chi^2$ method. 
We find for 18 of the 20 sources a favorable fit, particularly for Markarian~421 and NGC~1275 a significant improvement, with the hypothesis of photon-ALPs coupling in likelihood analysis. The
test statistics of the sources are combined and the significance has been estimated  $5.3~\sigma$ (test statistics summed in local maxima of all sources) and $6.0~\sigma$ (global maxima). The significance is estimated from dedicated simulations under the null hypotheses.
The locally best-fitting values of $B$ and $s$ fall into the range that is
expected for large scale magnetic fields present in relevant astrophysical environments.}
\maketitle

\section{Introduction}\label{sec:intro}
Axions are pseudoscalars which are originally proposed as a solution to address the strong CP problem in quantum chromodynamics (QCD) \cite{Peccei:1977ur,Peccei:1977hh}. Besides the QCD axion, the existence of various axion-like particles (ALPs) has been predicted 
in the framework of extra-dimensional completions of the standard model \citep{Svrcek:2006yi,Conlon:2006tq,Cicoli:2012sz}. 
ALPs are very light pseudo-scalar bosons ($a$) characterized mainly by a two-photon coupling $g_{a\gamma\gamma}$ and its mass $m_a$. Both, the
QCD axion as well as ALPs are possible candidates for particle dark matter \citep{Preskill:1982cy,Abbott:1982af,Dine:1982ah,Jaeckel:2013uva}. 

A non-vanishing coupling of ALPs to photons leads to a rich phenomenology for photon/ALPs mixing that can be observed in the universe and
probed with laboratory experiments. 
While searches for axion/ALP type dark matter have so far only produced exclusion limits, astrophysical
searches have been considered a promising approach to 
find signatures for  photon-ALPs mixing in gamma-ray spectra
\citep[see e.g.,][]{DeAngelis:2007wiw,Hooper:2007bq,Simet:2007sa}. 
There have been several claims for indications for anomalous TeV
transparency \citep[see e.g.,][]{DeAngelis:2011id,Horns:2012kw,Rubtsov:2014uga,Libanov:2019fzq}
as well as modulation of spectra of Galactic sources \citep{Majumdar:2018sbv}. In both cases, an interpretation of the
observations has been put forward that singles out 
the mass range of neV and coupling constants  $10^{-12}\,\mathrm{GeV}^{-1}<g_{a\gamma\gamma}<10^{-10}\,\mathrm{GeV}^{-1}$ \citep{Meyer:2013pny,Kohri:2017ljt,Majumdar:2018sbv} where
the uncertainties are mainly related to the assumption of the
magnetic field present along the line of sight. The minimum value of the coupling would be accessible with the upcoming light-shining-through a wall experiment ALPS II \citep{Bahre:2013ywa}.

The upper range of preferred coupling is in tension with the upper bounds of $g_{a\gamma\gamma} < 6.6\times 10^{-11}~\mathrm{GeV}^{-1}$ (95~\% c.l.) from the CAST experiment that searches for ALPs generated in the
core of the sun and then re-converts to X-ray photons in the transversal 
magnetic field of the CAST magnet \citep{CAST:2017uph}.  However, the conversion inside the
sun may be modified,  effectively suppressing the ALPs flux emitted by the sun \citep{Pallathadka:2020vwu}. 

Here, we extend the search for spectral modulations in high energy and very high energy gamma-ray data of a sample of
high frequency peaked BL Lac type objects (HBL) and the radio galaxy NGC~1275. Different from previous studies
where a particular model for the magnetic field is used and the values of the axion-related parameters are left free, 
we instead assume a fixed mass $m_a=3.6\,\rm neV$
and coupling $g_{a\gamma\gamma}=2.3\times10^{-10}\,\rm GeV^{-1}$ motivated by \cite{Majumdar:2018sbv} and 
leave the constant magnetic field strength and its spatial extension as free parameters.

In the following sections, we present the calculation for the conversion probability in astrophysical magnetic fields (Sec.~\ref{sec:theo}), the source selection and reconstruction of energy spectra
(Sec.~\ref{sec:src_seclect}), and the results in Sec.~\ref{sec:result}.

\section{Photon-ALP oscillation model and astrophysical magnetic fields}\label{sec:theo}

The photon-ALP oscillation effect occurs in the presence of an
external magnetic field. The photon-ALP coupling is described by
the following Lagrangian \citep{Raffelt:1987im}:
\begin{equation}
\mathcal{L}=-\frac{1}{4}g_{a\gamma\gamma}aF_{\mu\nu}\tilde F^{\mu\nu}=g_{a\gamma\gamma}a\mathbf{E\cdot B},
\end{equation}
where $g_{a\gamma\gamma}$ is the coupling constant between ALPs and photons, $a$ is the ALP field, $F_{\mu\nu}$ is the electromagnetic field tensor, $\tilde F^{\mu\nu}$ is its dual tensor. $\mathbf E$ and $\mathbf B$ are the electric and magnetic fields, respectively. Considering an initially polarized photon beam propagating through a single homogeneous magnetic field domain, the propagation equation can be written in a Schr\"{o}dinger-like form: 
\begin{equation}\label{eq:mixequ}
\Bigg(i\frac{d}{dx_3}+E+\mathcal M\Bigg)\Psi(x_3)=0,
\end{equation}
with 
\begin{equation}
\Psi(x_3)=(A_1(x_3), A_2(x_3), a(x_3))^{\rm T},
\end{equation}
where $A_1(x_3)$ and $A_2(x_3)$ are the photon linear polarization states along $x_1$ and $x_2$ axis respectively, $a(x_3)$ denotes the ALP state. $\mathcal M$ represents the photon-ALP mixing matrix.

The mixing matrix could be simplified in the case where $\mathbf B$ is homogeneous. Here we use $\mathbf B_T$ the transverse magnetic field, and $B_1$ vanishes if $\mathbf B_T$ is chosen to be along the $x_2$ axis. We denote the photon polarization state parallel to the transverse magnetic field $\mathbf B_T$ direction by $A_\rVert$, and the orthogonal one by $A_\perp$. In this way $\mathcal M$ can be simplified and written as \citep{Mirizzi:2005ng,Mirizzi:2006zy,Horns:2012kw}
\begin{equation}\label{eq:matrix}
\mathcal M = 
\left( \begin{array}{ccc}
\Delta_\perp & 0 & 0\\
0 & \Delta_\rVert & \Delta_{a\gamma}\\
0 & \Delta_{a\gamma} & \Delta_a
\end{array} \right),
\end{equation}
where the terms $\Delta_\perp\equiv\Delta_{\rm pl}+\Delta^{\rm CM}_\rVert+\Delta_{\rm CMB}$, $\Delta_\rVert\equiv\Delta_{\rm pl}+\Delta_\rVert^{\rm CM}+\Delta_{\rm CMB}$, $\Delta_{a\gamma}\equiv \frac{1}{2}g_{a\gamma\gamma}B_T$ and $\Delta_a\equiv-\frac{m_a^2}{2E}$ \citep{Raffelt:1987im,Horns:2012kw}, where $m_a$ is the mass of the ALP, $\Delta_{\rm pl}$ stands for plasma effects and has the form 
\begin{equation}
\Delta_{\rm pl}\equiv-\frac{\omega_{\rm pl}^2}{2E}\simeq-1.1\times10^{-10}\times\Big(\frac{E}{\rm TeV}\Big)^{-1}\times\Big(\frac{n_e}{10^{-3}\,\rm cm^{-3}}\Big)\,\rm kpc^{-1},
\end{equation}
where $\omega_{\rm pl}=\sqrt{4\pi n_e e^2/m_e}$ is the plasma frequency and $n_e$ is the electron density in the medium (typical value of $n_e$ used here is $1.1\times10^{-2}\,\rm cm^{-3}$ \citep{Galanti:2018upl}). The terms $\Delta_{\rVert,\perp}^{\rm CM}$ (Cotton-Mouton effect) are associated with the birefringence effects of the vacuum expected from the Euler-Heisenberg Lagrangian in the presence of transverse magnetic field, and the term $\Delta_{\rm CMB}$ accounts for photon-photon dispersion \citep{Dobrynina:2014qba}. In the following, we neglect the effects of birefringence and photon-photon dispersion since they do
not affect the energy range covered with \textit{Fermi}-LAT. 
We list the relevant parameters for numerical calculation \citep{Horns:2012kw}:
\begin{equation}
\Delta_{a\gamma}\simeq7.6\times10^{-2}\times\Big(\frac{g_{a\gamma\gamma}}{5\times10^{-11}\,\rm GeV^{-1}}\Big)\times\Big(\frac{B_T}{\mu G}\Big)\,\rm kpc^{-1},
\end{equation}
\begin{equation}
\Delta_a\simeq-7.8\times10^{-3}\Big(\frac{m_a}{10\,\rm neV}\Big)^2\times\Big(\frac{E}{\rm TeV}\Big)^{-1}\,\rm kpc^{-1}.
\end{equation}
For the simplest case of a large-scale homogeneous magnetic field, the probability of a photon oscillating into an ALP (or vice versa) after traveling a distance $s$ is
\begin{equation}\label{eq:prob}
p_{\gamma\rightarrow a}=\frac{4\Delta^2_{a\gamma}}{\Delta_{\rm osc}^2}\sin^2\Big(\frac{s\,\Delta_{\rm osc}}{2}\Big),
\end{equation}
where the oscillation wave number $\Delta_{\rm osc}$ has the form
\begin{equation}
\Delta_{\rm osc}\equiv\sqrt{(\Delta_a-\Delta_{\rm pl})^2+4\Delta^2_{a\gamma}}.
\end{equation}
Furthermore, it can be seen from Eq.~\eqref{eq:prob} that the photon-ALP mixing becomes maximal and energy-independent when $E\gg E_c$ given by  
\begin{equation}\label{eq:e_crit}
E_c\equiv\frac{E|\Delta_a-\Delta_{\rm pl}|}{2\Delta_{a\gamma}}.
\end{equation}
This is similar to the resonant case, where $\Delta_a = \Delta_\mathrm{pl}$. 

 In order to take into account photon absorption, e.g., by interaction with a soft photon background field, the photon-ALP system is then described by a modified Schr\"{o}dinger-like quation similar to Eq.~\eqref{eq:mixequ}, and can be written as \citep{Dobrynina:2014qba,Csaki:2003ef,Raffelt:1987im,DeAngelis:2011id}
\begin{equation}\label{eq:mixequ2}
\Bigg(i\frac{d}{dx_3}+E+\mathcal M+iD\Bigg)\Psi(x_3)=0,
\end{equation}
with the additional matrix
\begin{displaymath}
D = 
\left( \begin{array}{ccc}
\mathcal C(x_3) & 0 & 0\\
0 & \mathcal C(x_3) & 0\\
0 & 0 & 0
\end{array} \right),
\end{displaymath}
with $\mathcal C(x_3)$  related to the
optical depth $\tau(x_3)/2=\int^{x_3}_{0}\mathcal C(x_3^\prime)dx^\prime_3$.

The formal solution of Eq.~\ref{eq:mixequ2} is then given 
for an initial condition $\Psi(0)$:
\begin{equation}
\label{eqn:solution}
    \Psi(x_3)=\exp\left(-i\int\limits^{x_3}_0(E+\mathcal M-iD)dx^\prime_3\right) \Psi(0).
\end{equation}
Then, the surviving probability of the photon in photon-ALP system can be given by \cite{DeAngelis:2011id,Horns:2012kw}:
\begin{equation}\label{eq:prr_right}
    p_{\gamma\gamma} = |A_1(x_3)|^2 + |A_2(x_3)|^2.
\end{equation}
This formalism can be readily extended to consider un-polarized initial states by introducing the density matrix formalism and
a von-Neumann type equation instead of the Schrödinger-type equation \eqref{eq:mixequ2} \citep{Raffelt:1987im,Csaki:2003ef}.
As for the magnetic fields along the propagation of photon-ALP beam, we consider three distinct regions
for conversion: the source and its vicinity, 
the intergalactic space, and the Milky Way. The magnetic field
strength and structure  present in the  Milky Way is fairly well known via 
observations of Faraday-rotation measures, the polarization
of the emission from aligned dust grains and more
indirectly through the synchrotron 
emissivity of the interstellar medium.
The magnetic field of intergalactic space is 
only constrained to be smaller than $\approx \mathrm{nG}$ \citep{Pshirkov:2015tua} and not
to be lower than $\approx 10^{-16}~\mathrm{G}$ \citep{Durrer:2013pga}. 
Finally, the magnetic field of the sources and their neighborhood is
poorly known and may differ from source to source. 

The photon-ALP mixing effect in the intergalactic magnetic field 
(IGMF) is neglected here, similar to previous studies 
\citep{Horns:2012kw}. In this case, the propagation can be separated into three regions. In the source region, we obtain 
the solution using Eq.~\eqref{eqn:solution}, neglecting absorption $\mathcal{C}=0$. In the IGMF, we do not consider the mixing, such that for the solution in Eq.~\eqref{eqn:solution}, we assume $\mathcal{M}\approx 0$. Finally, in the Milky Way, we neglect
additional absorption ($\mathcal{C}\approx 0$) 
caused by local radiation fields. 
 
The magnetic field of the source and its environment is characterized by a minimal set of parameters used here: the strength of the
transversal magnetic field $B$ and its characteristic 
coherence length $s$. The conversion in the Milky Way is calculated
using the model of the galactic magnetic field (GMF) from Ref.~\cite{Jansson:2012pc} taking into account the 
line of sight of individual sources.


\section{Source selection and data reduction} \label{sec:src_seclect}
\subsection{Source selection}
The sensitivity  for signatures of photon-ALPs conversion in 
high energy gamma-ray spectra is related to the  
uncertainties on the differential flux measurements. Conversely, the appearance of 
modulations in the gamma-ray spectra requires a sufficiently large conversion probability $p_{\gamma\rightarrow a}$
(see Eq.~\eqref{eq:prob}). Large distances with a sizeably transverse magnetic field are favorable conditions 
to search for such modulations. While in the previous study by \citep{Majumdar:2018sbv}, Galactic pulsars were used,
we extend the search to extra-galactic objects. 

Almost all extra-galactic gamma-ray sources are associated with active galactic nuclei (AGN). In order to cover a large range of energies with AGN spectra, we select objects which have a hard gamma-ray spectrum and are sufficiently bright to measure the differential flux  accurately. In order to collect our source sample, the following selection cuts are applied to the fourth \textit{Fermi}-LAT source catalogue, 4FGL \citep{Fermi-LAT:2019pir}: 

\begin{enumerate}
\item Source type (association): AGN of  BL Lac type.
\item Red shift: $z$ known or constrained $z<0.5$. 
\item TeV association: in order to potentially extend to very high energies 
(VHE: $E> 100$~GeV), we require the sources to have 
an association to known VHE sources (\textit{TeVCAT} flag).
\item Hard spectrum: photon index is smaller than 2.   
\item Signal-to-noise ratio: detection significance larger than 50 standard deviations.
\item Photon statistics: number of predicted photons ($N_\mathrm{pred}$) should exceed 1600. 
\end{enumerate}

\begin{table*}[ht!]
	\renewcommand\arraystretch{1.2}
	\centering
	\caption{\vadjust{\vspace{-2pt}}AGN sources selected for this study (in order of right ascension). The information listed are Galactic longitude ($l$) and latitude ($b$), red shift ($z$). Also, detection significance, photon index, and predicted event counts from Fermi 4FGL catalog.}\label{tab:src_collect}
	\begin{tabular*}{1.00\textwidth}{@{\extracolsep{\fill}}cccccccc}
		\hline
		\hline
		AGN name & \Centerstack{Source\\type} &$l$[$^\circ$]  &$b$[$^\circ$] & $z$ &\Centerstack{Detection\\signif. ($\sigma$)} & \Centerstack{Photon\\index} &$N_\mathrm{pred}$\\
		\hline
        $\rm 1ES\,\,0033$+595  &HBL & $120.90$ & $-3.02$   &$0.467$            &68  &1.765 &2954 \\
        $\rm 3C\,\,66A$        &IBL & $140.15$ & $-16.76$  &$0.34$             &182 &1.971 &15207\\
        $\rm PKS\,\,0301$-243  &HBL & $214.63$ & $-60.19$  &$0.2657$           &108 &1.914 &5623 \\
        $\rm NGC\,\,1275$      &Radio Galaxy & $150.58$ & $-13.26$  &$0.017559$&245 &2.114 &35561\\
        $\rm PKS\,\,0447$-439  &HBL & $248.81$ & $-39.91$  &$0.343$            &167 &1.865 &12536\\
        $\rm 1ES\,\,0502$+675  &HBL & $143.79$ & $15.89$   &$0.34$             &64  &1.601 &1718 \\
        $\rm 1ES\,\,0806$+524  &HBL & $166.25$ & $32.94$   &$0.138$            &100 &1.881 &5147 \\
        $\rm 1ES\,\,1011$+496  &HBL & $165.53$ & $52.71$   &$0.212$            &169 &1.838 &9806 \\
        $\rm Markarian\,\,421$ &HBL & $179.88$ & $65.01$   &$0.031$            &344 &1.781 &30562\\
        $\rm Markarian\,\,180$ &HBL & $131.91$ & $45.64$   &$0.045$            &50  &1.798 &1623 \\
        $\rm 1ES\,\,1215$+303  &HBL & $189.01$ & $82.05$   &$0.131$            &146 &1.933 &10779\\
        $\rm 1ES\,\,1218$+304  &HBL & $182.21$ & $82.74$   &$0.182$            &83  &1.722 &3285 \\
        $\rm PKS\,\,1440$-389  &HBL & $325.65$ & $18.71$   &$0.1385$           &78  &1.845 &3788 \\
        $\rm PG\,\,1553$+113   &HBL & $21.92$  & $43.96$   &$\lesssim0.5$       &120 &1.681 &10046\\
        $\rm Markarian\,\,501$ &HBL & $63.60$  & $38.86$   &$0.034$            &173 &1.790 &11127\\
        $\rm 1ES\,\,1727$+502  &HBL & $77.07$  & $33.54$   &$0.055$            &60  &1.790 &2251 \\
        $\rm 1ES\,\,1959$+650  &HBL & $98.00$  & $17.67$   &$0.048$            &169 &1.817 &11700\\
        $\rm PKS\,\,2005$-489  &HBL & $350.37$ & $-32.61$  &$0.071$            &70  &1.838 &3115 \\
        $\rm PKS\,\,2155$-304  &HBL & $17.74$  & $-52.25$  &$0.116$            &239 &1.850 &17766\\
        $\rm 1ES\,\,2344$+514  &HBL & $112.89$ & $-9.90$   &$0.044$            &71  &1.811 &3201 \\
		\hline
		\hline
	\end{tabular*}
\end{table*}


The 19 sources passing the selection cuts are listed 
in Table~\ref{tab:src_collect}.
Additionally, we include one more source, 
a well-known and bright radio galaxy, NGC 1275.
This source is located at the center of the 
Perseus galaxy cluster which 
most likely supports an extended \citep{Sanders:2005jx} 
as well as a turbulent magnetic field 
\citep{Taylor:2006ta}. 
This magnetized environment is favourable
for photon-ALPs mixing and has already motivated several authors to search
for spectral irregularities in the \textit{Fermi}-LAT
energy spectrum  of NGC~1275
\citep{Fermi-LAT:2016nkz,Libanov:2019fzq,Pallathadka:2020vwu}.

\subsection{\textit{Fermi}-LAT data reduction} \label{subsec:data_reduc}
In this study, we make use of 10 years of LAT data taken in
the period from Aug.~4, 2008 to Aug.~4, 2018 
in the energy range from $100\,\rm MeV$ to $500\,\rm GeV$.
We select events within a region of interest (ROI) 
defined as a cone centered on each source with a 
half-opening angle of $10^\circ$. 

The events are selected by applying a zenith cut of
$90^\circ$ to minimize the $\gamma-$ray contributions from 
the Earth's limb. We set the spatial bin size to be 
$0.1^\circ$ and distribute 48 energy bins 
(corresponds to 13 bins per decade) within the 
selected energy range for performing a binned likelihood 
analysis. Following the recommendations of the LAT instrument team, the energy dispersion is corrected 
by introducing 3 additional bins beyond the energy range analysed.

The LAT data processed in pass 8 (release 3, version 2) 
have been downloaded together with the
spacecraft file and the matching
instrumental response files (\texttt{P8R3\char`_SOURCE\char`_V2} IRFs) 
from the Fermi Science Data Center.
Subsequently, the events of class \textit{source} with conversion
in front and back part of the tracker 
are selected. The preparatory steps of the
data analysis include creation of live time cube, data cubes etc. These tasks  have been carried out using the \texttt{Fermi Science Tools ver1.2.23}\footnote{\url{https://fermi.gsfc.nasa.gov/ssc/data/analysis/documentation/}} \citep{Wood:2017yyb}.
Most of these steps of the data reduction have been
conveniently performed with the python based
\texttt{fermipy ver0.19.0}\footnote{\url{https://fermipy.readthedocs.io/en/latest/}} interface. 

The diffuse backgrounds are modeled with 
pre-processed templates of 
the Galactic diffuse emission, \texttt{gll\char`_iem\char`_v07.fits}, and the extra-galactic isotropic radiation, \texttt{iso\char`_P8R3\char`_SOUR\-CE\char`_V2\char`_v1.txt}. The energy dispersion for the background templates is already 
taken into account\footnote{\url{https://fermi.gsfc.nasa.gov/ssc/data/analysis/documentation/}}.
Point sources from the \textit{Fermi}-LAT fourth catalog (4FGL, \citep{Fermi-LAT:2019yla}), within a region of $15^\circ$, are included to the source model.
  
The resulting energy spectra are displayed as spectral energy distributions (SEDs). 
The SEDs are derived by taking the differential flux measurements and multiplying the
individual flux values in each bin by the squared geometrical mean energy of the bin. 

\section{Analysis and Results}\label{sec:result}

 \subsection{Spectral models}
 The  energy spectra 
of the sources listed in Table~\ref{tab:src_collect} are 
compared with two different models that follow from the two hypotheses considered 
here. The hypothesis $H_0(\overline{ALPS})$ 
``without" photon-ALPs mixing is the null hypothesis and
the alternative is $H_1(ALPS)$ ``with'' photon-ALPs mixing.

In our spectral analysis, the intrinsic model of any AGN is either 
described by the \textit{Logparabola} model or in a few cases by a 
single \textit{PowerLaw}, as given in Eqs.~\eqref{eq:lp} and \eqref{eq:pl} respectively.
\begin{equation}\label{eq:lp}
\left(\frac{dN}{dE}\right)_{intr.} = N_0\left(\frac{E}{E_b}\right)^{-(\alpha+\beta\ln(E/E_b))},
\end{equation}
where the free parameters $N_0$ is the 
normalization factor at scale energy $E_b$, which is usually held constant, 
$\alpha$ is the power-law index and $\beta$ the 
curvature parameter. 
\begin{equation}\label{eq:pl}
\left(\frac{dN}{dE}\right)_{intr.} = N_0\left(\frac{E}{E_b}\right)^{-\alpha},
\end{equation}
The choice of the spectral model is based upon the LAT 8-year source catalog (4FGL) \citep{Fermi-LAT:2019yla}.

The intrinsic spectrum is subsequently modified by absorption via pair-production
on the soft extra-galactic background light (EBL).
The optical depth $\tau_{\gamma\gamma}(E)$
relies on the choice of an  EBL model. Since the optical depth in the energy and red shift 
range considered here is small ($\tau_{\gamma\gamma}\ll 1$), the actual choice of the
model is not of importance for the results obtained here, but needs to be included. The model of 
\citep{Dominguez:2010bv} is used as it is conveniently integrated in the 
PhotonALPsConv package\footnote{\url{https://github.com/me-manu/PhotALPsConv}}.

Under the alternative hypothesis $H_1$ 
with photon-ALP mixing, the spectrum is multiplied with the 
photon surviving probability  $p_{\gamma\gamma}$. 
It is a function of  photon energy $E$, ALP mass $m_a$, photon-ALP coupling
$g_{a\gamma\gamma}$, transversal (constant) B-field strength $B$ 
and the distance $s$ over which the B-field is present.

In order to make the general problem of estimating the free parameters numerically tractable, we 
consider $m_a$ and $g_{a\gamma\gamma}$ fixed at values which have been found
to be favorable to explain spectral modulations present in energy spectra of Galactic
pulsars \citep{Majumdar:2018sbv}. The resulting best estimates have been found to
be $m_a=3.6\,\rm neV$, $g_{a\gamma\gamma}=2.3\times10^{-10}\,\rm GeV^{-1}$.

Therefore, the  spectra modeled in this way for the two hypotheses (with and without photon-ALP conversion)
would have the following forms: 
\begin{equation}\label{eq:woalp}
H_0:\quad \left(\frac{dN}{dE}\right)_{w/o\,ALP} = e^{-\tau_{\gamma\gamma}} \left(\frac{dN}{dE}\right)_{intr.},
\end{equation}
and
\begin{equation}\label{eq:walp}
H_1:\quad\left(\frac{dN}{dE}\right)_{w/\,ALP} = \left(\frac{dN}{dE}\right)_{intr.} p_{\gamma\gamma}(E, m_a, g_{a\gamma\gamma}, B, s),
\end{equation}
respectively, where $(dN/dE)_{intr.}$ 
is the source model referring to Eq.~\eqref{eq:lp} or Eq.~\eqref{eq:pl}, and photon survival probability $p_{\gamma\gamma}$ in Eq.~\eqref{eq:walp} is calculated with Eq.~\eqref{eq:prr_right}.


\subsection{Parameter estimates: Null hypothesis}

 We fit the experimental data with two different approaches using as test statistics separately the 
 log likelihood ratio and $\Delta\chi^2$.
 For the likelihood fitting of the SED we  use  the forward-folding method as  implemented in the fermitools. This way, we determine
 the likelihood value for the best-fitting model for both hypotheses $H_0$ and $H_1$. The effect of the survival probability
 $p_{\gamma\gamma}$ is implemented by calling the \texttt{gtlike} tool with a so-called \texttt{filefunction} model. 
 For each value
 of $B$ and $s$ chosen, we optimize the parameters of $\left(dN/dE\right)_\mathrm{intr.}$ using the likelihood fitting method. 
 
 In order to check for consistency and to be more flexible to include additional data sets (e.g. VHE spectra), 
 we also implement a $\chi^2$ fitting method with the definition for the \textit{Fermi}-LAT data
\begin{equation}
    \chi^2 = \sum_{i=1}^{N}\frac{(D_{ij}\Psi_j-\phi_i)^2}{\sigma_i^2},
\end{equation}
where $N$ is the number of energy bins ($N=18$ for all sources analyzed with $\chi^2$ method), $D_{ij}\Psi_j$ and $\phi_i$ are respectively the expected and observed $\gamma$-ray flux in bin $i$ with
a statistical uncertainty $\sigma_i$. The model flux $\Psi_j$ is corrected using the energy dispersion matrix $D_{ij}$ 
determined for the particular
observation using the tool \texttt{gtdrm} with one additional bin added to the lower and upper end of the spectrum\footnote{\url{https://fermi.gsfc.nasa.gov/ssc/data/analysis/documentation/Pass8_edisp_usage.html}}. 

In Table~\ref{tab:fitt-reswoalp}, we list the best-fitting parameters 
estimated under the null hypothesis  with the likelihood method (for $\chi^2$ estimates, see Table~\ref{tab:fitt-reswoalp-chi2} in Appendix~\ref{sec:app00}). 
For each source, we find a maximum likelihood $L_\mathrm{max}^0$ (resp. a minimum $\chi^2_\mathrm{w/o ALP}$) with 
the best-fitting normalization value $N_0$, the power-law index $\alpha$, the curvature parameter $\beta$ and the 
scaling energy $E_b$. The uncertainties  listed are calculated for a 68~\% confidence interval.
\begin{table*}[ht!]
	\renewcommand\arraystretch{1.2}
	\centering
	\caption{\vadjust{\vspace{-2pt}}Best-fitting parameters for null hypothesis with likelihood method using the modeled spectra from Eq.~\eqref{eq:woalp}, where sources with no curvature parameter are modeled with \textit{PowerLaw}, and the rest is with \textit{Logparabola}. The normalization is given in units of $10^{-12}\rm MeV^{-1}cm^{-2}s^{-1}$. The estimated uncertainties ($1\sigma$) are listed as well (except for the scaling energy $E_b$ which
    is kept fixed at the value from the catalogue).}\label{tab:fitt-reswoalp}
	\begin{tabular*}{0.70\textwidth}{@{\extracolsep{0pt}}ccccc}
	\hline
	\hline
	AGN name  & $N_0$ & $\alpha$    &\Centerstack{$\beta$\\$\times10^{-3}$} &\Centerstack{$E_b$\\[MeV]}\\
	\hline    
    $\rm 1ES\,\,0033$+595   & $0.363 (0.015 )$        &$1.68 (0.03 )$  &$-4(12)$   &3177\\
    $\rm 3C\,\,66A$         & $10.9  (0.1   )$        &$1.88 (0.01 )$  &$39(4 )$   &1211\\
    $\rm PKS\,\,0301$-243   & $5.66  (0.12  )$        &$1.83 (0.02 )$  &$31(8 )$   &954.4\\
    $\rm NGC\,\,1275$       & $56.1  (0.4   )$        &$2.04 (0.004)$  &$60(3 )$   &883.6\\
    $\rm PKS\,\,0447$-439   & $4.62  (0.07  )$        &$1.74 (0.01 )$  &$52(5 )$   &1605\\
    $\rm 1ES\,\,0502$+675   & $0.0593(0.0026)$        &$ 1.48(0.03 )$  &$-$        &6322\\
    $\rm 1ES\,\,0806$+524   & $2.31  (0.06  )$        &$1.80 (0.02 )$  &$26(8 )$   &1297\\
    $\rm 1ES\,\,1011$+496   & $7.6   (0.1   )$        &$1.75 (0.01 )$  &$33(5 )$   &1066\\
    $\rm Markarian\,\,421$  & $18.0  (0.1   )$        &$1.73 (0.005 )$ &$19(2 )$   &1286\\
    $\rm Markarian\,\,180$  & $0.164 (0.008 )$        &$ 1.77(0.03 )$  &$-$        &2679\\
    $\rm 1ES\,\,1215$+303   & $9.04  (0.14  )$        &$1.84 (0.01 )$  &$44(5 )$   &1066\\
    $\rm 1ES\,\,1218$+304   & $0.215 (0.007 )$        &$ 1.69(0.02 )$  &$-$        &4442\\
    $\rm PKS\,\,1440$-389   & $1.01  (0.03  )$        &$1.70 (0.03 )$  &$56(11)$   &2014\\
    $\rm PG\,\,1553$+113    & $3.93  (0.06  )$        &$1.56 (0.01 )$  &$38(5 )$   &1847\\
    $\rm Markarian\,\,501$  & $4.57  (0.07  )$        &$1.70 (0.01 )$  &$17(4 )$   &1478\\
    $\rm 1ES\,\,1727$+502   & $0.202 (0.008 )$        &$ 1.75(0.03 )$  &$-$        &3005\\
    $\rm 1ES\,\,1959$+650   & $3.22  (0.05  )$        &$1.76 (0.01 )$  &$23(5 )$   &1733\\
    $\rm PKS\,\,2005$-489   & $0.526 (0.016 )$        &$ 1.80(0.02 )$  &$-$        &2398\\
    $\rm PKS\,\,2155$-304   & $15.4  (0.2   )$        &$1.77 (0.01 )$  &$35(3 )$   &1136\\
    $\rm 1ES\,\,2344$+514   & $0.807 (0.03  )$        &$1.73 (0.03 )$  &$50(12)$   &1938\\
	\hline
	\hline
\end{tabular*}
\end{table*}
\subsection{Parameter estimates: ALPs hypothesis}
The alternative ($H_1$) hypothesis with photon-ALPs mixing includes two additional free parameters which relate  to the strength $B$ of the
magnetic field and the distance $s$ over which the photons can mix in the constant external magnetic field. For each pair of
$B$, $s$, we maximize the likelihood $L_\mathrm{max}^1$ (or minimize  $\chi^2_\mathrm{w/ ALP}$). We carry out this procedure for a discrete set
of pairs of $B$ and $s$ located on a logarithmic grid with
 $(150\times150)$ steps where $10^{-3}\mu G\leq B\leq 1\,\rm\mu G-10^3\,\rm\mu G$ and $10^{-2}\,\rm kpc-1\,\rm kpc\leq s\leq10^3\,\rm kpc-10^4\,\rm kpc$. Units!! 
 The ranges are chosen such that the critical energy $E_c$ could fall into the analyzed energy range, 
 and to include the best-fitting parameters of $(B, s)$  under the $H_1$ hypothesis.
 In case of multiple local maxima found, we choose the
 combination which minimizes the
 total energy present in the magnetic field given by 
 $\propto s^3 B^2$. The same criterion is used for the
 grid with the $\chi^2$ values.

In order to test the significance of the alternative hypothesis against the null hypothesis, we introduce the test statistics (TS) based upon the likelihood ratio:
\begin{equation}\label{eq:TS}
TS(B, s) = -2\times(\ln(L_{max}^0) - \ln(L_{max}^1(B,s))).
\end{equation}

For example, in the left panel of Fig.~\ref{fig:mkn421scan} (for figures of other sources see Figs.~\ref{fig:es0033}-\ref{fig:es2344} in Appendix~\ref{sec:app00}) 
we show the resulting values of $TS(B,s)$ on the grid for  Markarian~421 (Mkn~421). 

The value of the TS varies in a characteristic way for different values of $B$ and $s$. For small values of $B$ and $s$, 
the two hypotheses are not distinguishable as the survival probability $p_{\gamma\rightarrow a}$ is too small in comparison
with the measurement uncertainties. For high values of $B$ and $s$ a large part of the parameter space is excluded. Notably,
a repetitive pattern of local maxima occurs which are aligned along increasing values of $B$ and $s$. The local maxima
correspond to the case where  $s\cdot \Delta_\mathrm{osc}>2\pi$ and therefore multiple oscillations occur. For increasing 
values of $B$, the critical energy $E_\mathrm{crit}$ decreases therefore, a wider part of the energy spectrum is affected. 

On the search grid, we locate the global maximum of 
$TS(\hat{B},\hat{s})=18.5$ for $\hat B=21.0~\mathrm{nG}$ and $\hat s =216.4~\mathrm{kpc}$ (see Table~\ref{tab:best-fitt-reswalp} in Appendix~\ref{sec:app00} for global maxima of other sources), which is marked with a white triangle error bar. In this particular case, the global maximum $(\hat B, \hat s)$ is located at a local maximum which corresponds
to the parameters with the smallest value of $B^2\cdot s^3$ which is proportional to the total energy required
to build up the magnetic field. 

We mark down the chosen local maximum $TS(\hat{B_0},\hat{s_0})$ with the smallest value of $B^2\cdot s^3$, which in this case is identical with the global best-fitting results, and is shown as a black point error bar in Fig.~\ref{fig:mkn421scan}. 

In a consistent way, we obtain the best-fitting parameters for ($\hat B_0,\hat s_0$) of chosen local maxima  under the ALP hypothesis for the remaining sources listed in Table~\ref{tab:fitt-reswalp}.
 
Similar to the approach used for the $TS$ defined by the likelihood ratio, 
\begin{equation}\label{eq:TS_chi2}
    \Delta\chi^2 = \chi^2_{\rm w/o\,ALP} - \chi^2_{\rm w/\,ALP},
\end{equation}
is also calculated for the same grid and
the best-fitting parameters which maximise the $\Delta \chi^2$ are obtained\footnote{Note, the definition of $\Delta \chi^2$ and the sign is chosen
such that we can use comparable value of $\Delta \chi^2$ and $TS$.}. As an example, the right panel of Fig.~\ref{fig:mkn421scan} shows
the corresponding grid of $\Delta \chi^2$ values.
When comparing the $\Delta \chi^2$ values on the same grid as the $TS$ values, the same patterns emerge and similar best-fitting values
for ($\hat B$, $\hat s$) are found, as well as the values for ($\hat B_0$, $\hat s_0$).  There are however some differences which relate to the fact that the $\Delta \chi^2$ method is based upon a coarser
binning of the energy spectra and therefore the oscillation features remain in some cases under-sampled. 
In the same way, we have obtained the best-fitting combinations of ($\hat B$, $\hat s$) and ($\hat B_0$, $\hat s_0$) from the
$\chi^2$ fit which are listed in Tables~\ref{tab:fitt-reswalp-chi2} and \ref{tab:best-fitt-reswalp-chi2} (in Appendix~\ref{sec:app00}) respectively.

The best-fitting distance $\hat s_0$ range from $\approx 0.1$~kpc (1ES 1218+304) up to $\approx 262$~kpc (Markarian 180). The bulk of the source spectra favor a conversion within a distance range of 1~kpc to 200~kpc with a magnetic field strength between 10~nG and 10~$\mu$G.

\begin{figure}[ht!]
\begin{minipage}[t]{0.48\linewidth}
\centering
\includegraphics[width=1.0\textwidth]{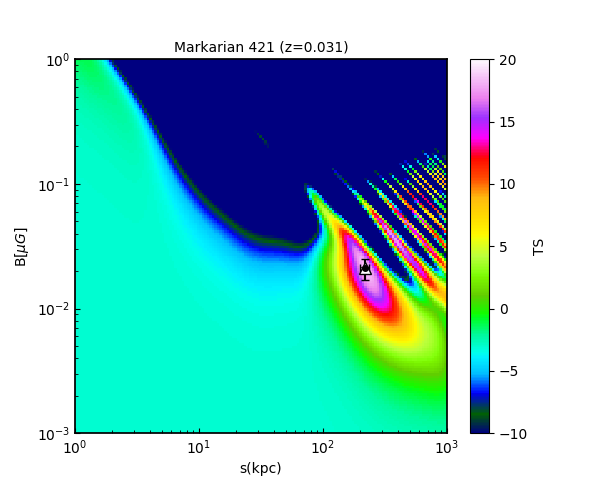}
\end{minipage}%
\begin{minipage}[t]{0.48\linewidth}
\centering
\includegraphics[width=1.0\textwidth]{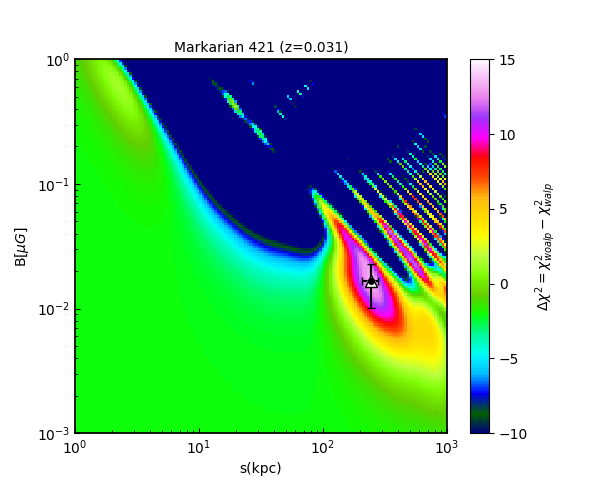}
\end{minipage}
\caption{Left panel: $(B, s)$ grid map where the color bar indicates the test statistics $TS$ which is twice the difference of log-likelihood values between null and ALP hypotheses. Right panel: $(B, s)$ grid map where the color bar indicates the difference of $\chi^2$ values fitted in null and ALP hypotheses. The black
and white marker correspond to the local and 
global best-fitting parameters respectively. }\label{fig:mkn421scan}
\end{figure}
\begin{table*}[ht!]
	\renewcommand\arraystretch{1.2}
	\centering
	\caption{\vadjust{\vspace{-2pt}}Best-fitting parameters of local maxima for ALP hypothesis with likelihood method using the modeled spectra from Eq.~\eqref{eq:walp}. $B$ and $s$ are additional free parameters relating to the strength and length scale for the
    external magnetic field that is responsible for photo-ALP mixing effects. The normalisation is given in units of $10^{-12}\rm MeV^{-1}cm^{-2}s^{-1}$. Parameters uncertainties ($1\sigma$) are included.}\label{tab:fitt-reswalp}
        \scalebox{0.9}{
	\begin{tabular*}{1.03\textwidth}{@{\extracolsep{0pt}}ccccccc}
	\hline
	\hline
	AGN name &$N_0$ &$\alpha$ &\Centerstack{$\beta$\\$\times10^{-3}$} &\Centerstack{$E_b$\\[MeV]} &\Centerstack{$\hat B_0$\\[nG]} &\Centerstack{$\hat s_0$\\[kpc]}\\
	\hline
    $\rm 1ES\,\,0033$+595   & $0.593 (0.023)$  & $1.54 (0.03)$   &$36(12)$    &3177   &54.9  (27.9 )&81.8 (28.4)\\    
    $\rm 3C\,\,66A$         & $12.5  (0.2  )$  & $1.80 (0.01)$   &$43 (6 )$   &1211   &322.7 (46.1) &7.2  (0.9 )\\
    $\rm PKS\,\,0301$-243   & $10.4  (2.8  )$  & $1.78 (0.05)$   &$41 (10)$   &954.4  &24396.3(7912.3)&0.2 (0.1 )\\
    $\rm NGC\,\,1275$       & $103   (5    )$  & $1.99 (0.01)$   &$82 (3 )$   &883.6  &26268.2(1142.6)&0.2 (0.01)\\
    $\rm PKS\,\,0447$-439   & $7.20  (0.14 )$  & $1.55 (0.01)$   &$87 (6 )$   &1605   &1675.6(157.3) &2.0  (0.1 )\\
    $\rm 1ES\,\,0502$+675   &$0.0731(0.0047)$  & $1.49 (0.03)$   &$-$         &6322   &820.5 (74.1 )&12.2 (0.7)\\
    $\rm 1ES\,\,0806$+524   & $2.32  (0.06 )$  & $1.78 (0.02)$   &$11 (10)$   &1297   &143.8 (21.2 )&43.5 (3.8 )\\
    $\rm 1ES\,\,1011$+496   & $7.67  (0.12 )$  & $1.75 (0.01)$   &$25 (6 )$   &1066   &27.1  (5.7  )&206.2(24.1)\\
    $\rm Markarian\,\,421$  & $19.1  (0.2  )$  & $1.69 (0.005)$  &$13 (2 )$   &1286   &21.0  (4.1  )&216.4(18.4)\\
    $\rm Markarian\,\,180$  & $0.177 (0.008)$  & $1.72 (0.03)$   &$-$         &2679   &16.7  (11.9 )&262.8(84.4)\\
    $\rm 1ES\,\,1215$+303   & $17.1  (3.0 )$  & $1.77 (0.06)$   &$61 (5 )$   &1066   &18574.9(4545.8)&0.3(0.1)\\
    $\rm 1ES\,\,1218$+304   & $0.426 (0.015)$  & $1.69 (0.02)$   &$-$         &4442   &32031.9(7981.4)&0.1(0.04)\\
    $\rm PKS\,\,1440$-389   & $1.76  (0.09 )$  & $1.50 (0.03)$   &$105 (12)$  &2014   &2013.3(629.7)&2.2  (0.4 )\\
    $\rm PG\,\,1553$+113    & $4.96  (0.10 )$  & $1.44 (0.01)$   &$43 (6 )$   &1847   &846.0 (36.5 )&11.1(0.3)\\
    $\rm Markarian\,\,501$  & $7.88  (1.40 )$  & $1.67 (0.02)$   &$24 (4 )$   &1478   &29047.0(7747.1)&0.2(0.03)\\
    $\rm 1ES\,\,1727$+502   & $0.260 (0.012)$  & $1.72 (0.02)$   &$-$         &3005   &1987.1(65.4) &9.3 (0.2)\\
    $\rm 1ES\,\,1959$+650   & $3.41  (0.06 )$  & $1.71 (0.01)$   &$15 (6 )$   &1733   &63.8  (29.5 )&29.6 (8.1 )\\
    $\rm PKS\,\,2005$-489   & $0.707 (0.048)$  & $1.73 (0.02)$   &$-$         &2398   &6400.6(374.4)&1.6 (0.1 )\\
    $\rm PKS\,\,2155$-304   & $30.8  (0.3  )$  & $1.76 (0.01)$   &$42 (3 )$  &1136   &30009.3(6113.0)&0.1(0.01)\\              
    $\rm 1ES\,\,2344$+514   & $0.870 (0.037)$  & $1.62 (0.04)$   &$33 (16)$   &1938   &244.2 (71.1 )&15.0 (2.9 )\\
	\hline
	\hline
\end{tabular*}}
\end{table*}

\begin{figure}[ht!]
\begin{minipage}[t]{0.48\linewidth}
\centering
\includegraphics[width=1.0\textwidth]{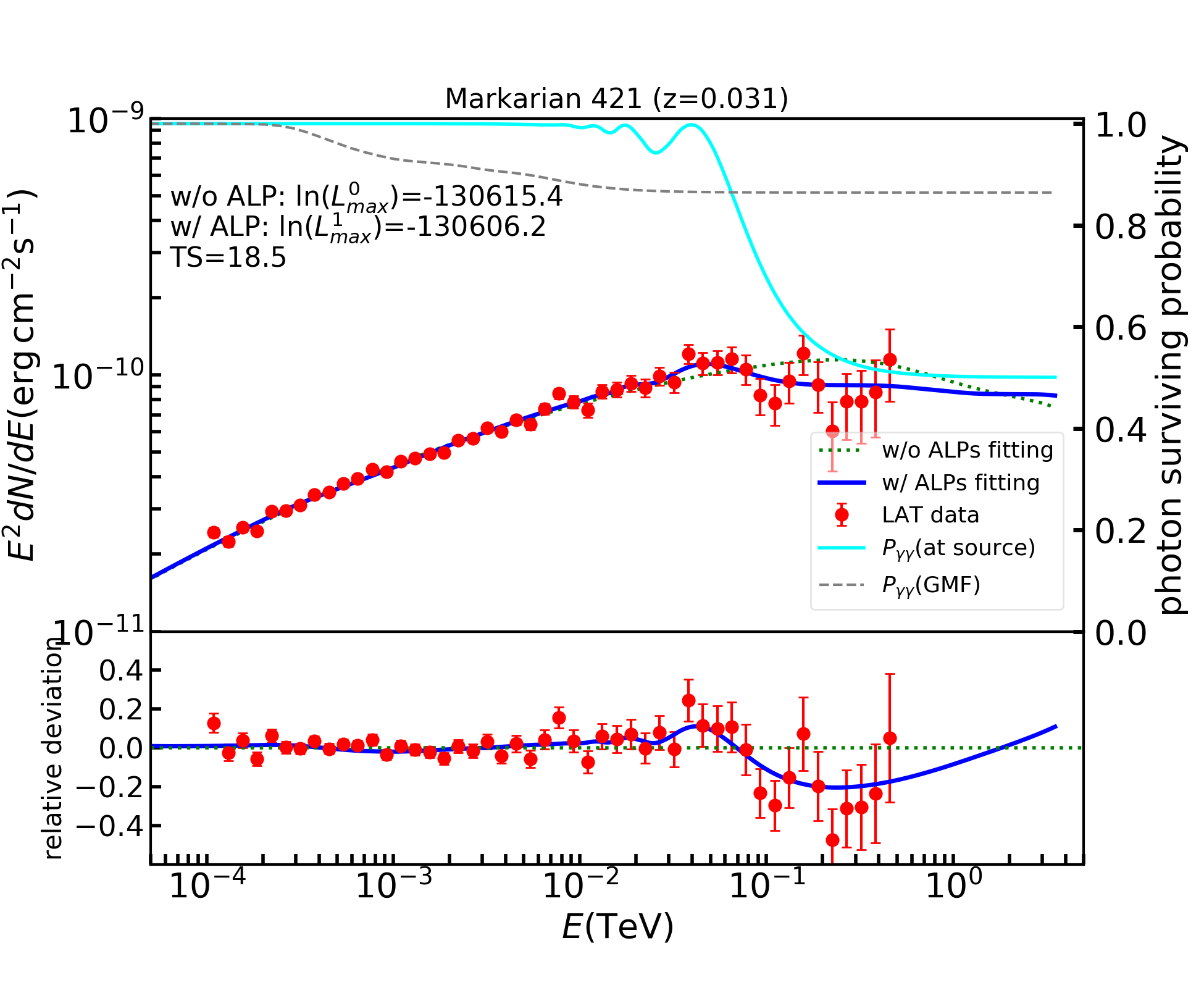}
\end{minipage}
\begin{minipage}[t]{0.48\linewidth}
\centering
\includegraphics[width=1.0\textwidth]{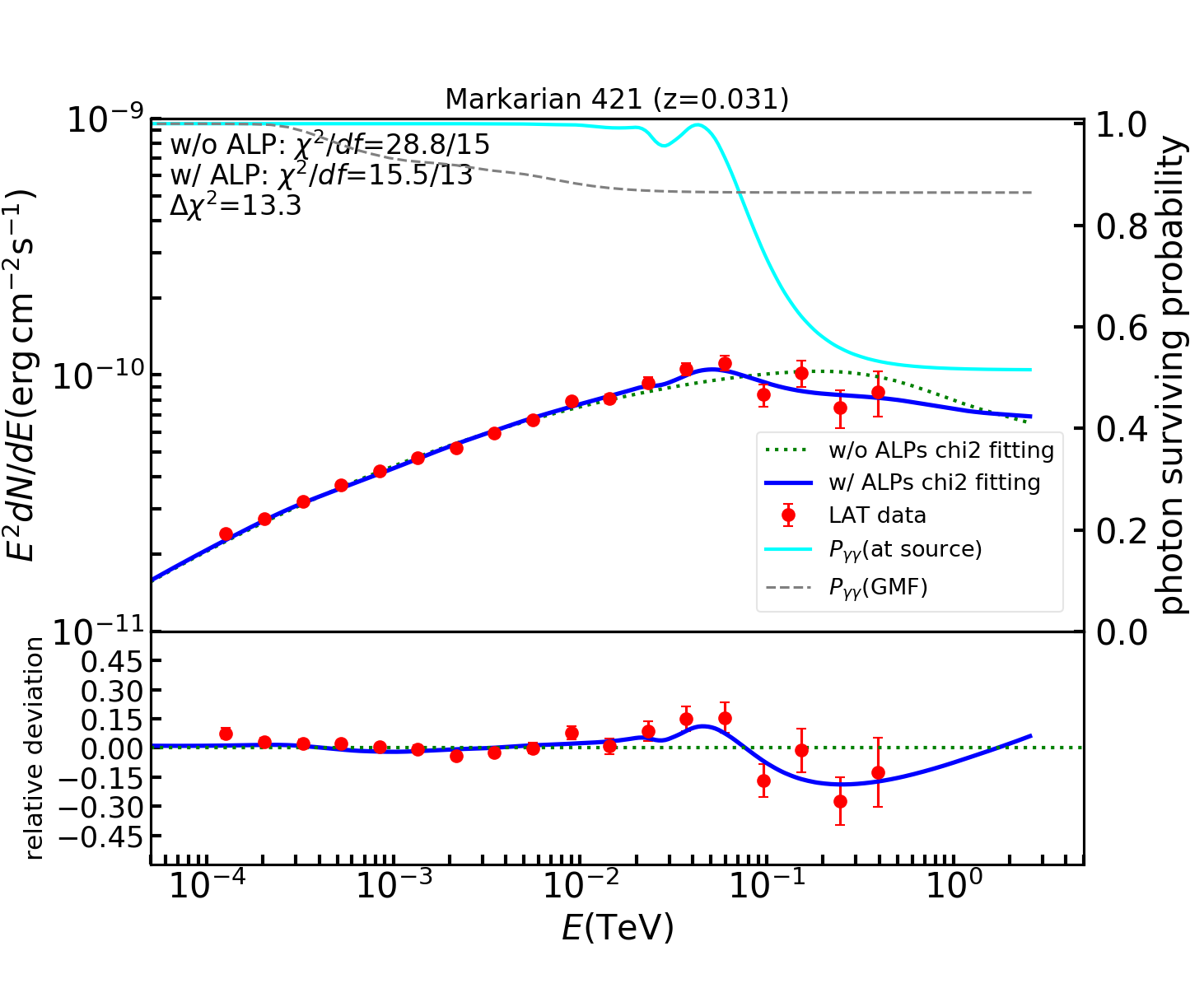}
\end{minipage}
\caption{Left panel: the spectral energy distribution for source Mkn~421 with likelihood fitting method. The red data points are collected from a 10-yrs LAT observation. The blue straight line is the best-fitting model with photon-ALP mixing effects included, and the green dashed line is the best-fitting model without the assumption of photon-ALP mixing. The cyan solid line is the photon surviving probability at source and the gray dashed line is the photon surviving probability at Milky Way. In the lower panel, we show the relative deviations of the flux points and ``w/ ALP" scenario from the baseline (``w/o ALP"). Right panel: SED for source Mkn~421 with $\chi^2$ fitting method.}\label{fig:mkn421sed}
\end{figure}

As an illustration of the best-fitting SED for the two hypotheses, we show in Fig.~\ref{fig:mkn421sed},
the observed SED data points of Mkn~421 (see Figs.~\ref{fig:es0033sed}-\ref{fig:es2344sed} in Appendix~\ref{sec:app00} for other sources) together with the model curves. 
In the left panel, the SED data points are calculated with the likelihood binning while in the right figure, the SED points
are calculated for a coarser binning. The best-fitting curve for the null hypothesis is shown as a green dashed line and 
is for both fitting methods very similar. For the case of photon-ALPs mixing, the resulting conversion probability leads to modifications of the spectrum 
mainly between 50 GeV and 500~GeV (shown as blue solid line). The relative amplitude of the modulation is about $15~\%$. 

\subsection{Hypotheses testing for the joined \textit{Fermi}-LAT spectra}
\label{subsection:hypotheses}
The hypotheses testing is performed first on the individual 
energy spectra and subsequently in a analysis of the joined
test statistics of all spectra. 

For either rejection or acceptance of null hypothesis, we  estimate the distribution of $TS$ under the 
null hypothesis following a similar procedure as described in \cite{Fermi-LAT:2016nkz}. 
We generate as pseudoexperiments (PE) 400 sets of  simulated gamma-ray spectra for each source under the
null hypothesis.
The simulation of PE data sets
is done through Gaussian sampling of the expected event numbers in a counts cube generated for the gamma-ray sources and diffuse
emission present in the region of interest \citep{Wood:2017yyb}. The resulting sim data sets are then subject to the same
data analysis procedure as outlined above. For each source, this results in two distributions with 400 values of TS according to
Eq.~\eqref{eq:TS} and $\Delta \chi^2$ as defined in Eq.~\eqref{eq:TS_chi2}, respectively. 

In the case of nested hypotheses 
the distribution of the test statistic should asymptotically approach a $\chi^2$ 
distribution (in this case a non-central $\chi^2$ distribution under null hypothesis) 
if the number of simulations is sufficiently high  \citep{Wilks:1938dza}.

\begin{figure}[ht!]
\begin{minipage}[t]{0.48\linewidth}
\centering
\includegraphics[width=1.0\textwidth]{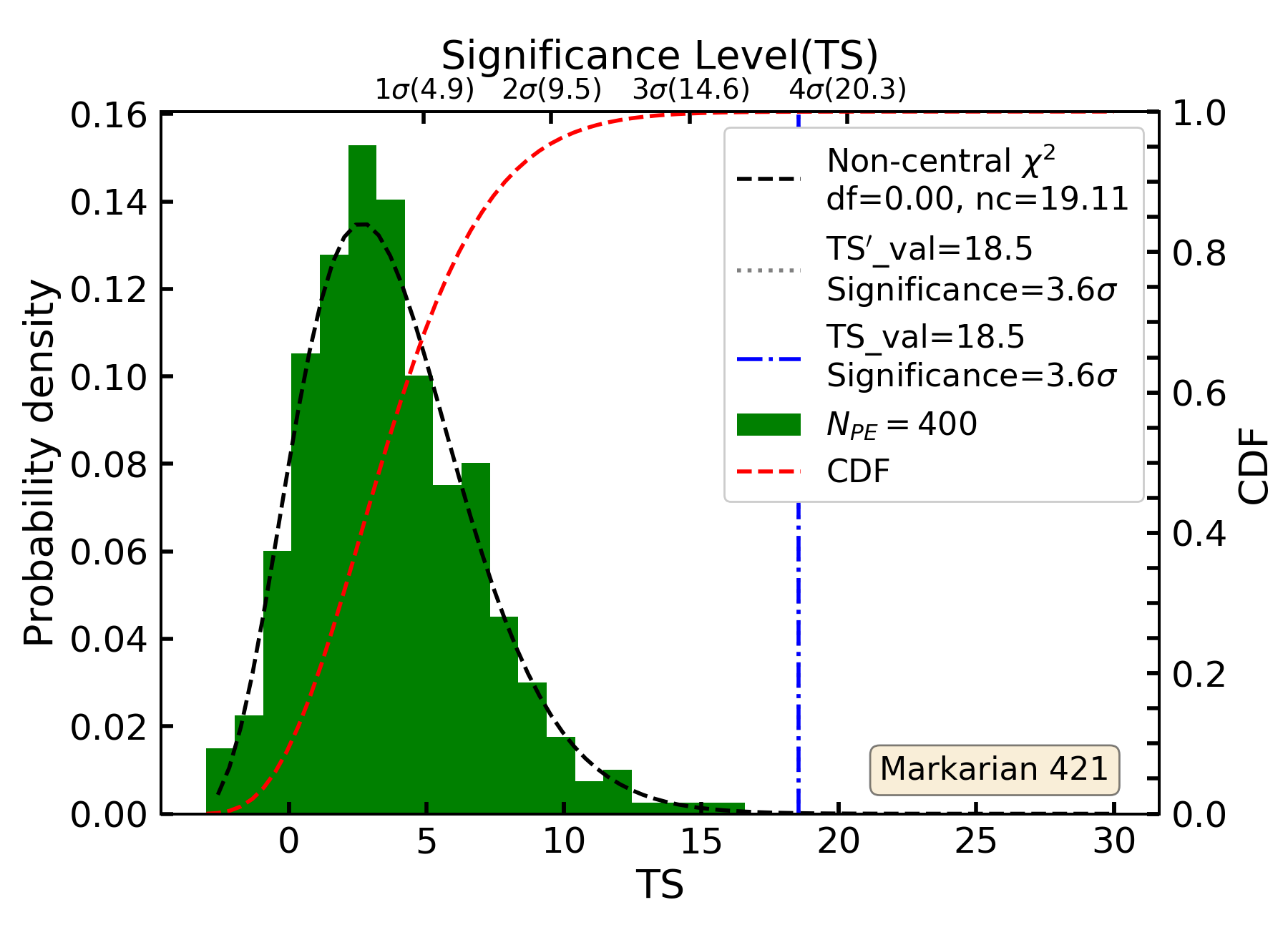}
\end{minipage}%
\begin{minipage}[t]{0.48\linewidth}
\centering
\includegraphics[width=1.0\textwidth]{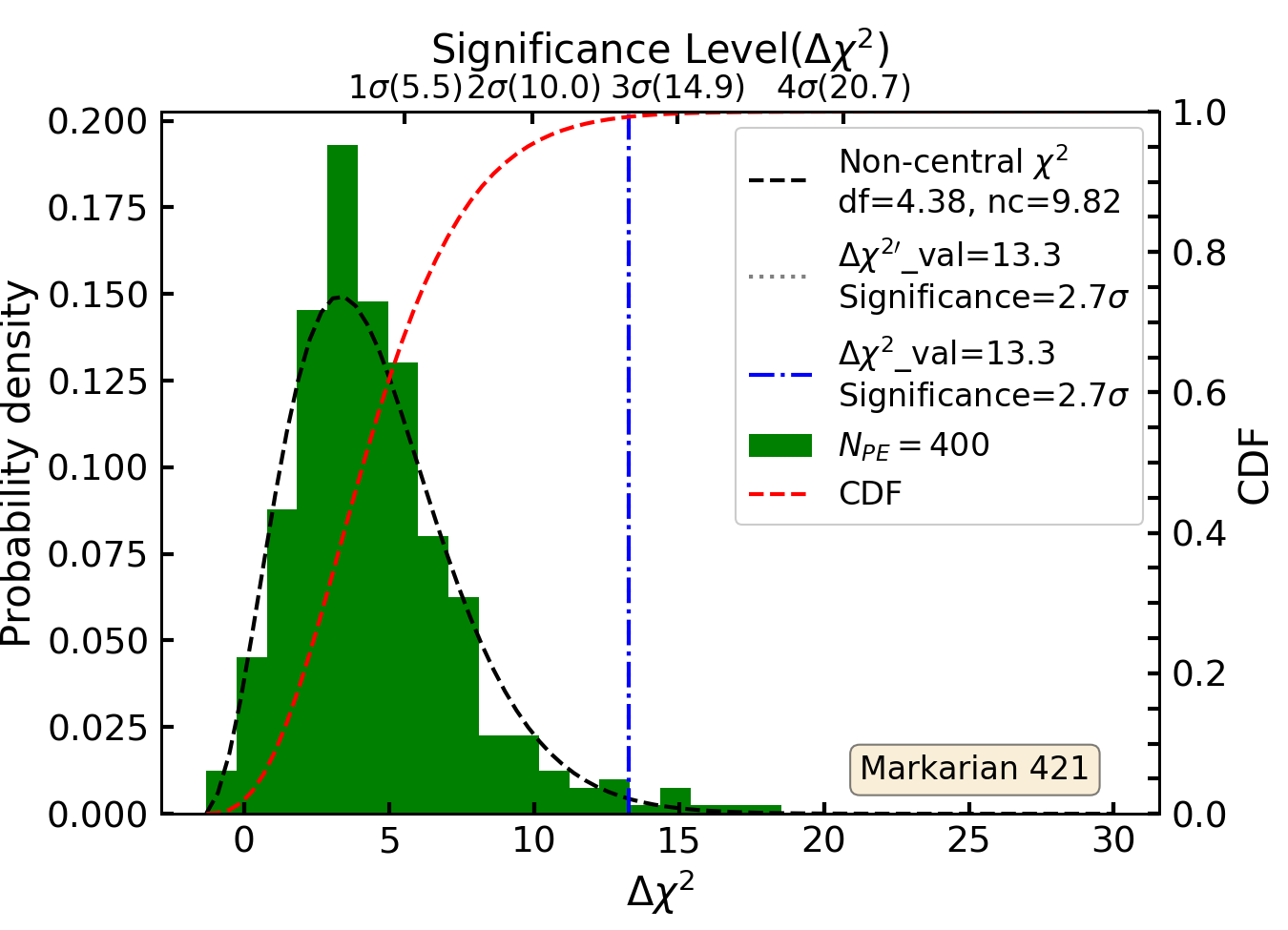}
\end{minipage}
\caption{Left panel: Simulated null distribution for Mrk~421 from likelihood ratio test. Right panel: Simulated null distribution from $\Delta\chi^2$ test for the same source. 
The black dashed line indicates  a fit to the distribution with a non-central $\chi^2$ function. The red solid line represents the resulting cumulative distribution function (CDF).
The TS($\Delta\chi^2$) value derived from the local maxima of original data is marked as a blue (dot-dash) vertical line, while the TS($\Delta\chi^2$) value obtained from the global maxima is marked as a gray dotted line (in this case, the blue line coincides with the gray line).}\label{fig:mkn421sim}
\end{figure}

As an example, we present in  Fig.~\ref{fig:mkn421sim}  the distributions of TS for Mkn~421 with likelihood ratio test.  The TS distribution 
is best approximated with a non-central $\chi^2$ distribution (NCD) with about 0 degree of freedom ($df$) 
and non-centrality ($nc$) parameter $nc=19.11$. 
With the accumulated NCD, we derive the probability to find a value of $TS$ larger than the one found in data ($TS=18.5$)
to be $p(TS>18.5; df=0.00, nc=19.11)=2.79\times10^{-4}$, corresponding to a significance level of $3.6~\sigma$. The  result on Mkn~421 for the 
$\chi^2$ fit is slightly less significant with $p(\Delta \chi^2>13.3;df=4.38,nc=9.82)=7.39\times10^{-3}$, corresponding to a significance level of $2.7~\sigma$.
Upon closer inspection, the binning for the $\chi^2$-fit is under-sampling the modulation predicted for the 
spectrum when photon-ALPs oscillation is considered.
We can conclude from both tests, that 
in this case, the photon-ALP hypothesis is preferred over the null hypothesis.

The goodness of fit for the hypothesis $H_1$ is acceptable for
12 of the 20 spectra. Particularly, for 1ES~0502+675 and 
1ES 1727+502, the resulting values of $\chi^2(df)=38.2(14)$
and $\chi^2(df)=32.7(14)$ are too large to be acceptable. The
corresponding probability to obtain a larger value of $\chi^2$
is $p(\chi^2>38.2, df=14)=4.8\times 10^{-4}$
and $p(\chi^2>25.9, df=14)=3.2\times 10^{-3}$, indicating
a poor fit in both cases. Upon inspecting the SEDs and
the residuals in Figs.~\ref{fig:es0502sed} and \ref{fig:es1727sed}, additional features in the spectrum are
present which are not well described by the model. 


Following the same approach,
we present the obtained TS values for all the sources in Table~\ref{tab:TS}, as well as their corresponding significance levels derived from null distribution for each source. As is evident from Table~\ref{tab:TS}, the other sources show a similar preference
for hypotheses $H_1$ with photon-ALPs mixing. 

In order to test the overall preference of the joint data sets, we combine 
the TS for the individual sources:
\begin{equation}
    TS_\mathrm{tot} = \sum_i TS_i, 
\end{equation}
where $TS_i$ is the test statistics for each individual source. Similarly, we combine the PE results
from the individual sources in a bootstrapping approach.
In order to do so, we take $10^7$ sequences  of 20 uniform random deviates $n_1,\ldots,n_{20}$ 
 in order to 
combine the sources in a random way: 
\begin{equation}\label{eq:PE1}
 PE=\{ (TS_{n_1},\ldots,TS_{n_{20}})|  n_1,\ldots,n_{20} \in \{1,\ldots,400\} \}.
\end{equation}
This way, we calculate a distribution of $10^7$ values of $TS^{PE}$ derived from the PE:
\begin{equation}\label{eq:PE2}
    TS^{PE} = \sum_i TS_{n_i}.
\end{equation}
With this approach we benefit from the combinatorial factor of $400^{20}\approx 10^{52}$ different possibilities to combine the 
simulated data sets. 
The combined analysis of the $\Delta\chi^2$ test is done in a similar way:  we  add up the individual $\Delta\chi^2$ values to obtain $\Delta\chi^2_{tot}$, and generate $10^7$ values of 
$(\Delta\chi^2)^{PE}$ similar to the procedure outlined in  Eqs.~\eqref{eq:PE1}, \eqref{eq:PE2}.

\begin{figure}[ht!]
\begin{minipage}[t]{0.48\linewidth}
\centering
\includegraphics[width=1.0\textwidth]{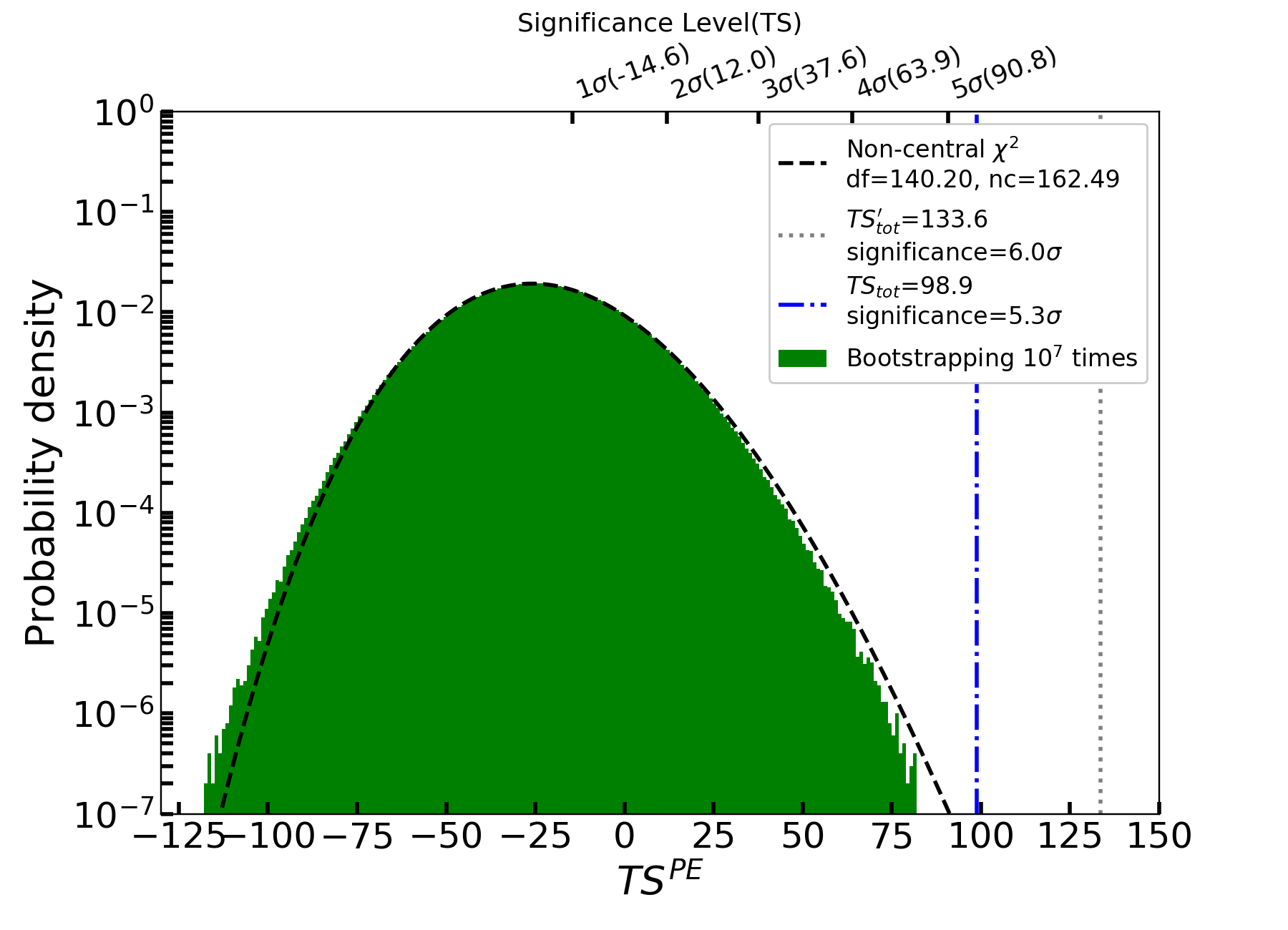}
\end{minipage}%
\begin{minipage}[t]{0.48\linewidth}
\centering
\includegraphics[width=1.0\textwidth]{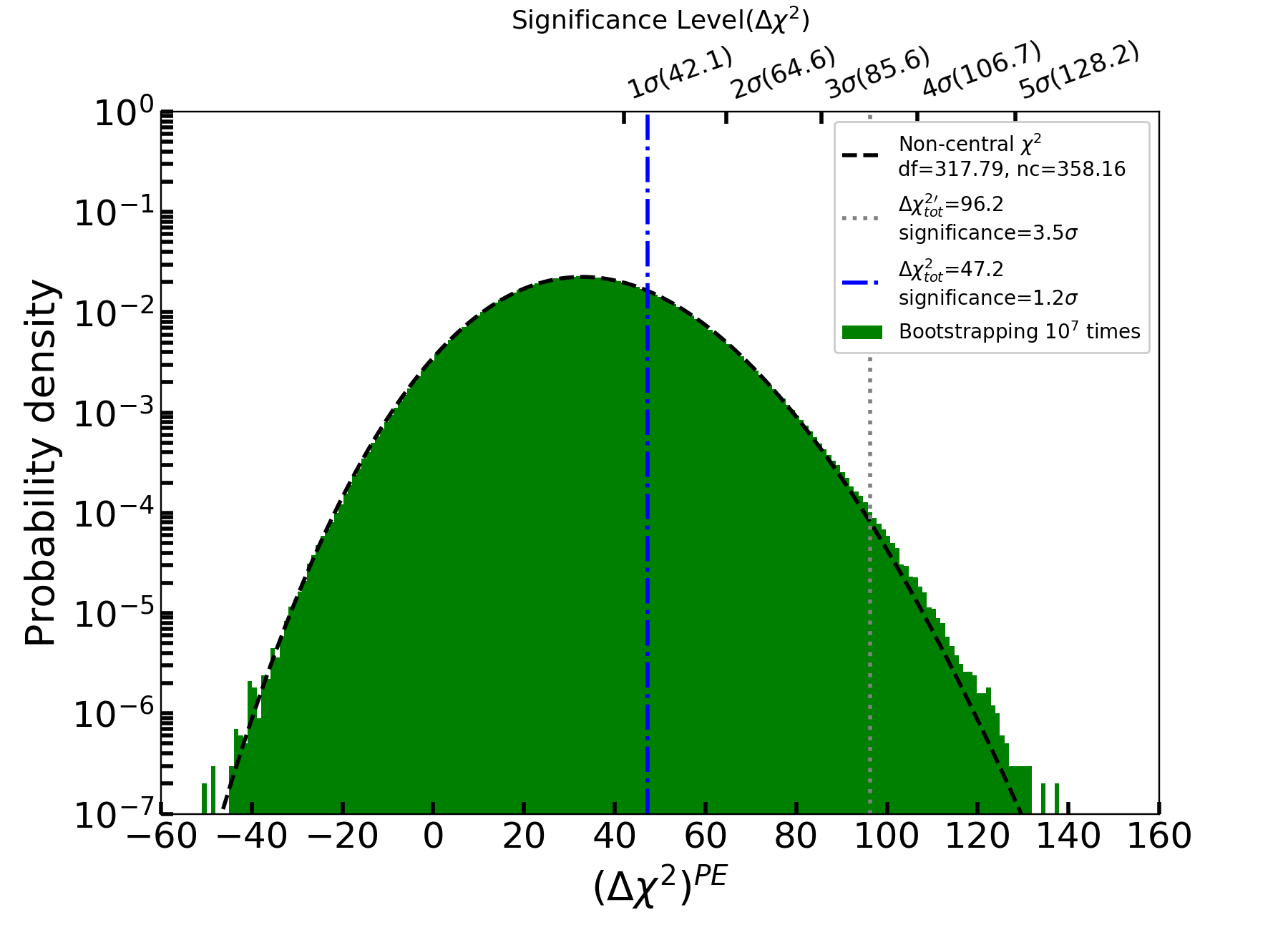}
\end{minipage}
\caption{Left panel: Combined full null TS distribution from likelihood ratio test using bootstrapping method. Right panel: Combined full null $\Delta\chi^2$ distribution from $\Delta\chi^2$ test using bootstrapping method. Black dashed line denotes fitting to the histogram of $TS^{PE} ((\Delta\chi^2)^{PE})$ values with non-central $\chi^2$ (NCD) function. Blue dotted-dashed line indicates the position of $TS_{tot} (\Delta\chi^2_{tot})$. Gray dotted line stands for the position of $TS^{\prime}_{tot} (\Delta\chi^{2\prime}_{tot})$ summed over all global maximal values on $(B, s)$ grid maps (see  App.~\ref{sec:app00} the white triangle on each grid map).}\label{fig:combine_sim}
\end{figure}

The resulting distributions of $TS^{PE}$ and $(\Delta \chi^2)^{PE}$ are shown in Fig.~\ref{fig:combine_sim}. The probability density function can be approximated
by a NCD, similar to the distributions for the individual sources. The distribution of $TS^{PE}$ is well fit by the NCD and the probability to find a value of $TS^{PE}>TS_{tot}$ 
can be estimated from the best-fitting NCD\footnote{A similar value is obtained by counting the number of entries in the simulated distribution with $TS^{PE}>TS_{tot}$.} to be
$p(TS^{PE}>TS_{tot}=98.9; df=140.20, nc=162.49)=1.22\times10^{-7}$, corresponding to a $z$-score of $5.3$. For the $\Delta \chi^2$ based hypotheses test, we find the NCD fit a poor description
of the underlying simulated distribution for small values of $\Delta \chi^2$. For larger values of $\Delta \chi^2$ the fit matches closely the 
distribution. We estimate the $z$-score to be smaller than the value found for the $TS$-based distribution
at $1.4$.  This is consistent with the findings from the individual sources.

\begin{table*}[ht!]
	\renewcommand\arraystretch{1.2}
	\centering
	\caption{\vadjust{\vspace{-2pt}}Best-fitting log-likelihood and $\chi^2$ values of the local maxima for null ($H_0$) and ALP hypotheses ($H_1$). TS values are calculated with Eqs.~\eqref{eq:TS} and \eqref{eq:TS_chi2} for likelihood ratio test and $\Delta\chi^2$ test respectively. Corresponding significance levels for both tests are listed as well.}\label{tab:TS}
	\scalebox{0.8}{
	\begin{tabular*}{1.15\textwidth}{@{\extracolsep{\fill}}ccccccccc}
		\hline
		\hline
		AGN name &\Centerstack{$H_0$\\$\ln(L^0_{max})$} &\Centerstack{$H_1$\\$\ln(L^1_{max})$}  &$TS$ &\Centerstack{$z$-score\\($H_1/H_0$)} &\Centerstack{$H_0$\\$\chi^2_{\rm w/o\,ALP}/df$} &\Centerstack{$H_1$\\$\chi^2_{\rm w/\,ALP}/df$} &$\Delta\chi^2$ &\Centerstack{$z$-score\\($H_1/H_0$)}\\
        \hline
        $\rm 1ES\,\,0033$+595   & $712253.0$      & $712257.2 $   &$8.4 $ &$2.1$ &$22.2/15$ &$14.2/13$ &$8.1 $ &$1.9$\\
        $\rm 3C\,\,66A$         & $-122858.4$     & $-122856.5$   &$3.8 $ &$2.0$ &$20.6/15$ &$16.3/13$ &$4.3 $ &$2.3$\\
        $\rm PKS\,\,0301$-243   & $-154508.6$     & $-154508.2$   &$0.7 $ &$0.5$ &$16.3/15$ &$16.4/13$ &$-0.2$ &$0.5$\\
        $\rm NGC\,\,1275$       & $26767.7	$     & $26777.6  $   &$19.8$ &$3.6$ &$25.6/15$ &$22.6/13$ &$3.1$  &$1.8$\\
        $\rm PKS\,\,0447$-439   & $-146200.2$     & $-146199.2$   &$2.0$  &$1.9$ &$26.2/15$ &$25.8/13$ &$0.4 $ &$1.5$\\
        $\rm 1ES\,\,0502$+675   & $-44659.7$      & $-44656.4 $   &$6.7 $ &$1.3$ &$41.9/16$ &$38.2/14$ &$3.6 $ &$0.4$\\
        $\rm 1ES\,\,0806$+524   & $-159751.7$     & $-159750.4$   &$2.6 $ &$0.2$ &$20.5/15$ &$16.6/13$ &$3.9 $ &$0.5$\\
        $\rm 1ES\,\,1011$+496   & $-153504.9$     & $-153502.4$   &$5.0 $ &$0.8$ &$12.7/15$ &$9.2 /13$ &$3.5 $ &$0.4$\\
        $\rm Markarian\,\,421$  & &   & & & & &\\
        \multicolumn{1}{r}{\tiny{10yrs LAT}}&$-130615.4$&$-130606.2$&$18.5$&$3.6$&$28.8/15$ &$15.5/13$ &$13.3$ &$2.7$\\
        \multicolumn{1}{r}{\tiny{simul. LAT+MAGIC}}&$-$&$-$&$-$&$-$&$33.4/15$&$20.7/13$&$12.6$&$1.8$\\
        $\rm Markarian\,\,180$  & $-132628.5$     & $-132627.9$   &$1.0$  &$0.01$ &$23.8/16$&$20.5/14$ &$3.3 $ &$0.4$\\
        $\rm 1ES\,\,1215$+303   & $-143988.2$     & $-143984.9$   &$6.6$  &$2.0$ &$15.5/15$ &$13.4/13$ &$2.1 $ &$1.2$\\
        $\rm 1ES\,\,1218$+304   & $-145577.4$     & $-145577.9$   &$-1.0$ &$0.9$ &$19.0/16$ &$20.2/14$ &$-1.2$ &$1.1$\\
        $\rm PKS\,\,1440$-389   & $60781.3$       & $60783.1  $   &$3.7$  &$1.3$ &$18.9/15$ &$17.0/13$ &$3.5 $ &$0.6$\\
        $\rm PG\,\,1553$+113    & $-152448.4$     & $-152450.6$   &$-4.2$ &$0.02$&$17.8/15$ &$27.7/13$ &$-9.9$ &$0.0$\\
        $\rm Markarian\,\,501$  & $-95747.1$      & $-95746.2 $   &$1.7$  &$0.4$ &$18.5/15$ &$17.2/13$ &$1.4 $ &$0.3$\\
        $\rm 1ES\,\,1727$+502   & $-154208.5$     & $-154207.4$   &$2.4$  &$0.4$ &$26.7/16$ &$32.7/14$ &$-6.1$ &$0.0$\\
        $\rm 1ES\,\,1959$+650   & $-49456.2$      & $-49453.9 $   &$4.6$  &$0.7$ &$21.6/15$ &$16.3/13$ &$5.3 $ &$0.9$\\
        $\rm PKS\,\,2005$-489   & $-166314.9$     & $-166309.0$   &$11.4$ &$2.1$ &$33.9/16$ &$20.9/14$ &$12.9$ &$2.1$\\
        $\rm PKS\,\,2155$-304   &     &   & & & & & &\\
        \multicolumn{1}{r}{\tiny{10yrs LAT}}&$-144503.6$&$-144503.1$&$0.9$&$1.5$&$15.8/15$&$19.5/13$&-3.6&0.5\\
        \multicolumn{1}{r}{\tiny{simul. LAT+H.E.S.S.}}&$-$&$-$&$-$&$-$&$33.3/18$&$29.2/16$&$4.1$&1.6\\
        $\rm 1ES\,\,2344$+514   & $-61475.6$      & $-61472.2$    &$6.7$  &$1.2$ &$15.2/15$ &$14.0/13$ &$1.2$  &$0.5$\\
		\hline
		\hline
	\end{tabular*}}
\end{table*}

\subsection{Combined HE and VHE spectra}
 The large collection area of ground-based instruments
extends  the  high energy (HE)
range accessible with \textit{Fermi}-LAT 
towards very high energies (VHE), where photon statistics limit the sensitivity
for space based instruments. The downside of the ground-based technique
is a limited field of view. Therefore, the VHE 
spectrum is in most cases recorded during flaring states  whereas the HE spectrum is recorded quasi-continuously with the
all-sky instrument of \textit{Fermi}-LAT.
The flare-selected observation of AGN with ground based instruments   
introduces
a bias in the observed energy spectrum towards a high flux-state
which is not necessarily representative of a truly time-averaged spectrum.

 In the following, we consider examples for the combination
 of HE and VHE data taken from PKS~2155-304 ($z=0.116$) and 
 Mkn~421 ($z=0.031$), where HE and VHE data are recorded contemporaneously
with \textit{Fermi}-LAT and ground-based instruments. 
\subsubsection{Combined spectrum of PKS~2155-304} 
The nearby X-ray selected AGN PKS~2155-304 
is the first extra-galactic very high energy gamma-ray source 
discovered in the southern sky \citep{Chadwick:1999ile}. 
It has been closely monitored, both during 
periods of quiescence as well as during flares \citep{HESS:2010tfm}.

We consider a quasi-simultaneous observation  to avoid the combination of data sets averaged over different
flux states. Non-simultaneous spectral data could lead to an apparent spectral break or irregularities
close to the transition energy of the two measurements.  The constraint on  available contemporaneous observation time leads to 
larger statistical uncertainties on the detected photon numbers which in turn reduce the sensitivity for spectral features. 
During  contemporaneous 
observations of PKS~2155-304 with H.E.S.S.-II and \textit{Fermi}-LAT in 2013, 
a spectral break between the HE and VHE data is observed \citep{HESS:2016btr}. The H.E.S.S. Phase II observations achieved 
a reduced energy threshold in comparison with the previous measurements 
recorded with the smaller H.E.S.S. Phase I instruments \citep{HESS:2010tfm}. 
The lower threshold of H.E.S.S. II observations
improves the overlap in the energy range covered with space and ground based instruments. We re-analyse the contemporaneous \textit{Fermi}-LAT data set
used by \citep{HESS:2016btr} with identical energy bins to combine the two measurements. 

\begin{figure}[ht!]
  \centering
  \includegraphics[width=0.60\textwidth]{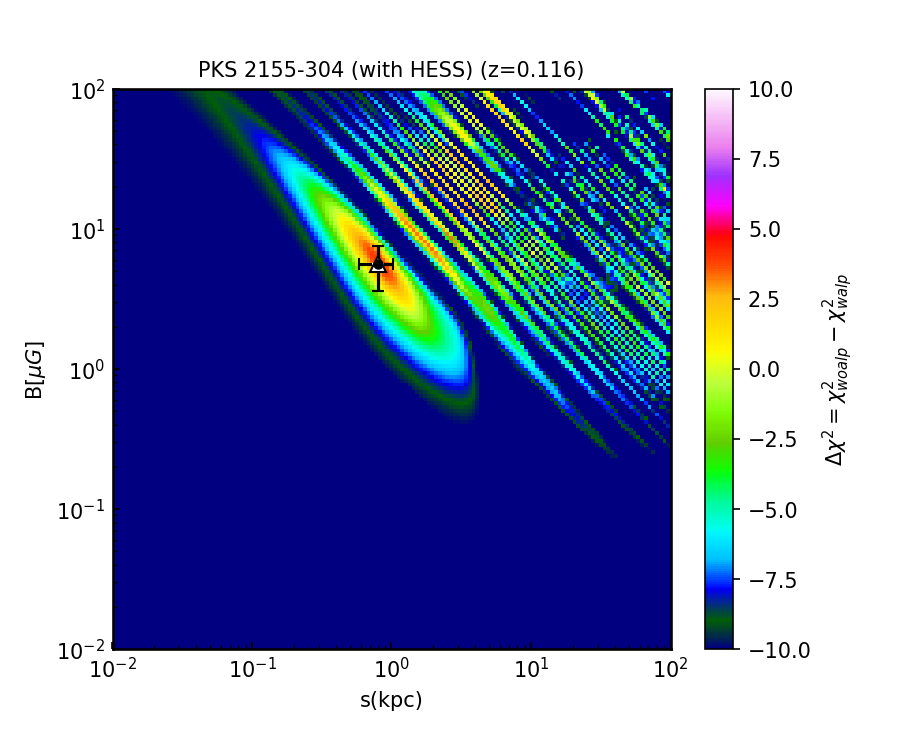}\\
  \caption{$\Delta\chi^2$ for a grid of values of B-field strength $B$ and distance $s$. 
  The color bar indicates the $\Delta\chi^2$ values when fitting the combined contemporaneous LAT and H.E.S.S. data in 2013. 
  The black point marker indicates the local maximum of $\Delta \chi^2$ derived from the fit of the SED to the
  combined spectrum, while the white triangle marker represents the global best-fitting parameters.}\label{fig:pks2155scan}
\end{figure}

We  present in Fig.~\ref{fig:pks2155scan} the scan of $\Delta \chi^2(B,s)$ 
from the combined energy spectrum for the ALP hypothesis $H_1$. The global best-fitting parameters are found to be at $\hat B=5.5\,\mu G$, with $\hat s=0.2~$kpc, where $\Delta\chi^2=4.1$ is obtained. 
As can be seen from Fig.~\ref{fig:pks2155scan},
the local maximum $(\hat B_0, \hat s_0)$ (indicated with a black point) 
coincides with the global maximum (indicated with a white triangle). 

\begin{figure}[ht!]
\begin{minipage}[t]{0.48\linewidth}
\centering
\includegraphics[width=1.0\textwidth]{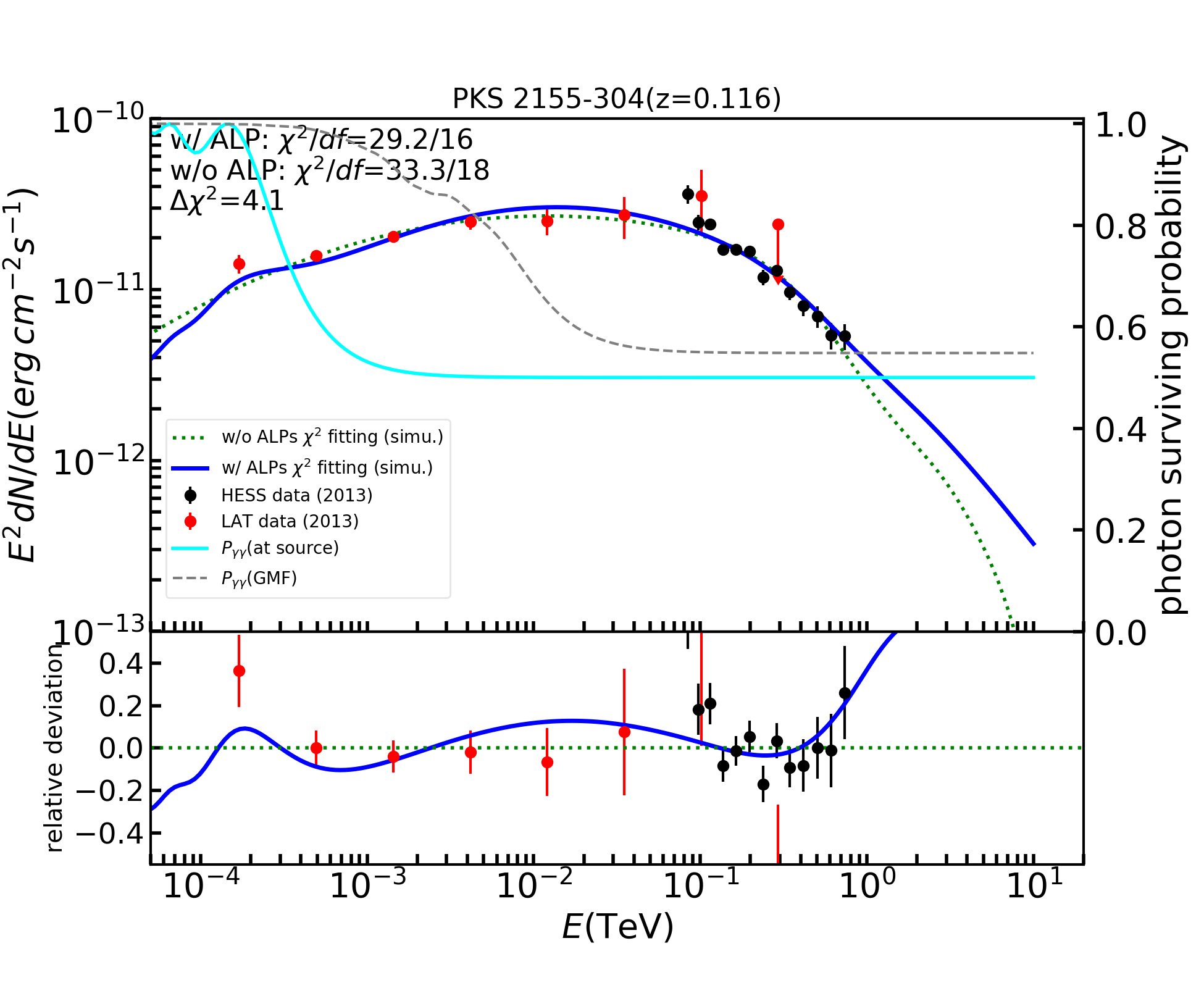}
\end{minipage}%
\begin{minipage}[t]{0.49\linewidth}
\includegraphics[width=1.0\textwidth]{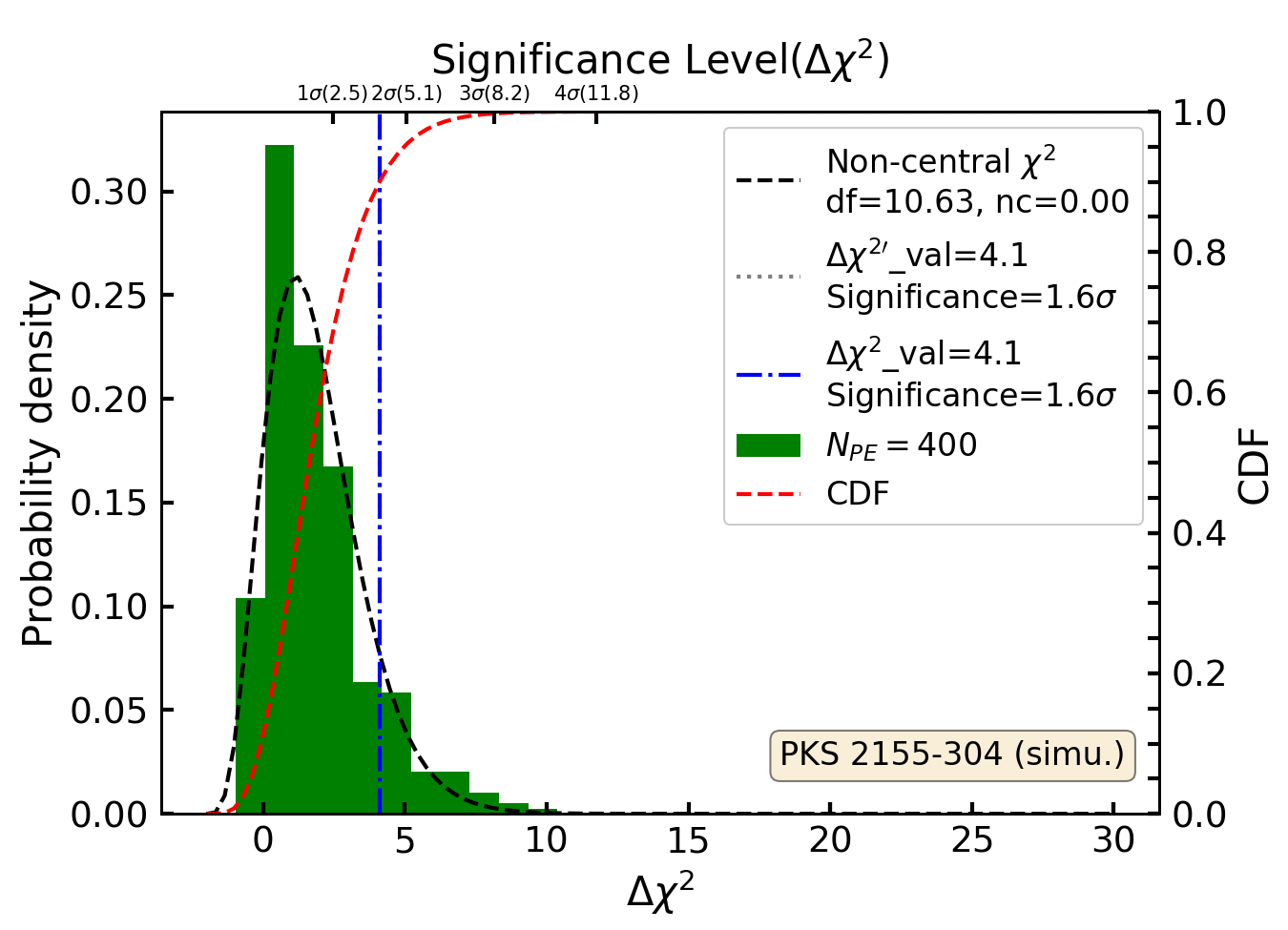}
\centering
\end{minipage}
\caption{Left panel: the spectral energy distribution for  $\rm PKS\,\,2155$-304 during
contemporaneous observations with H.E.S.S.-II and \textit{Fermi}-{LAT} in 2013. The red data points represent the 2013 LAT observations, and the black data points are from H.E.S.S during the same year. The blue solid and green dashed lines are the best-fitting models under $H_0$ and $H_1$ hypotheses respectively. The cyan solid and gray dashed lines stand for the photon surviving probabilities in different regions along the line of sight.
Right panel: simulated null distribution from $\Delta\chi^2$ test for 2013 H.E.S.S and LAT observations. The black dashed line indicates a fit to the distribution with a non-central $\chi^2$ function. The red solid line represents the resulting cumulative distribution function (CDF). The $\Delta\chi^2$ value derived from the original data is marked as a blue (dot-dash) vertical line.}\label{fig:pks2155sed_simu} 
\end{figure}

The resulting spectral energy distribution is shown 
in Fig.~\ref{fig:pks2155sed_simu}. The spectral break is observed at an energy of $(48\pm12)$~GeV when fitting a
broken-power law to the combined SED. The flux measurements
in the overlapping energy range between 80 GeV and 300 GeV are consistent between the two instruments.  The $H_0$ hypothesis is not providing
a good description of the data while the $H_1$ hypothesis improves slightly the fit by $\Delta \chi^2=4.1$.
Using mock data sets, we estimate the
significance in a similar way as before. The resulting distribution for $\Delta \chi^2$
and a NCD fit function is shown in Fig.~\ref{fig:pks2155sed_simu} (right panel). The
z-score for the improvement is estimated to be $\approx 1.6$.



\subsubsection{Combined spectrum from Mkn~421}
The northern, nearby AGN Mkn~421 ($z=0.031$) is a highly variable BL Lac type
object that has been closely monitored since the
discovery of its VHE emission \citep{Punch:1992xw}. 
While a number of simultaneous multi-wavelength observations have been 
carried out for this source, we select the result reported by 
\citep{LAT:2011mmt} on a simultaneous observation campaign
with Fermi-LAT and the MAGIC telescopes from January to June 2009. 
During this campaign, the combined energy spectrum from the two
instruments covers a very broad energy range with substantial overlap between the two instruments. 

\begin{figure}[ht!]
  \centering
  \includegraphics[width=0.70\textwidth]{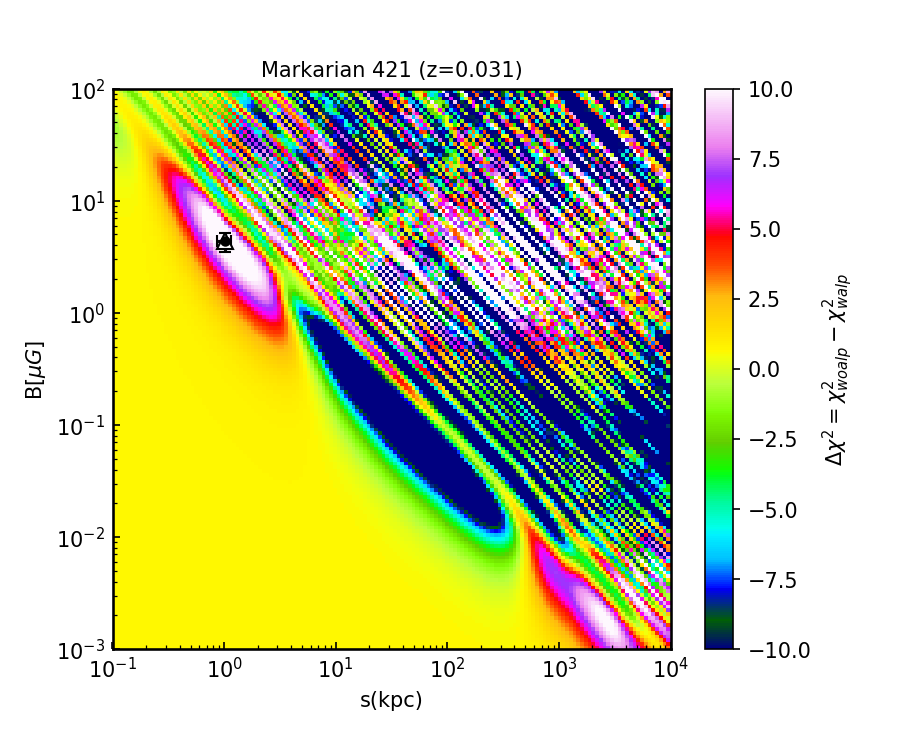}\\
  \caption{$\Delta\chi^2$ distribution as functions of B-field strength $B$ and distance $s$. 
  The color bar indicates the $\Delta\chi^2$ values when fitting the combined time-averaged LAT and MAGIC data \citep{LAT:2011mmt}. 
  The black point marker indicates the local maximum of $\Delta \chi^2$ derived from the fit of the SED to the
  time averaged spectrum, while the white triangle marker stands for the global maximum of $\Delta \chi^2$.}\label{fig:mkn421scan_simu}
\end{figure}

The scan of the parameters $s$ and $B$ for the combined spectrum shows several maxima which would favor either a large magnetic field of several $\mu$G on kpc scales or a very weak magnetic field of several nG over Mpc distances (see Fig.~\ref{fig:mkn421scan_simu}). The chosen the local maximum, corresponding to a minimum energy, is marked with a black cross in Fig.~\ref{fig:mkn421scan_simu} where the resulting critical energy $\approx 100$ MeV (see Eq.~\eqref{eq:e_crit}).  

\begin{figure}[ht!]
\begin{minipage}[t]{0.48\linewidth}
\centering
\includegraphics[width=1.0\textwidth]{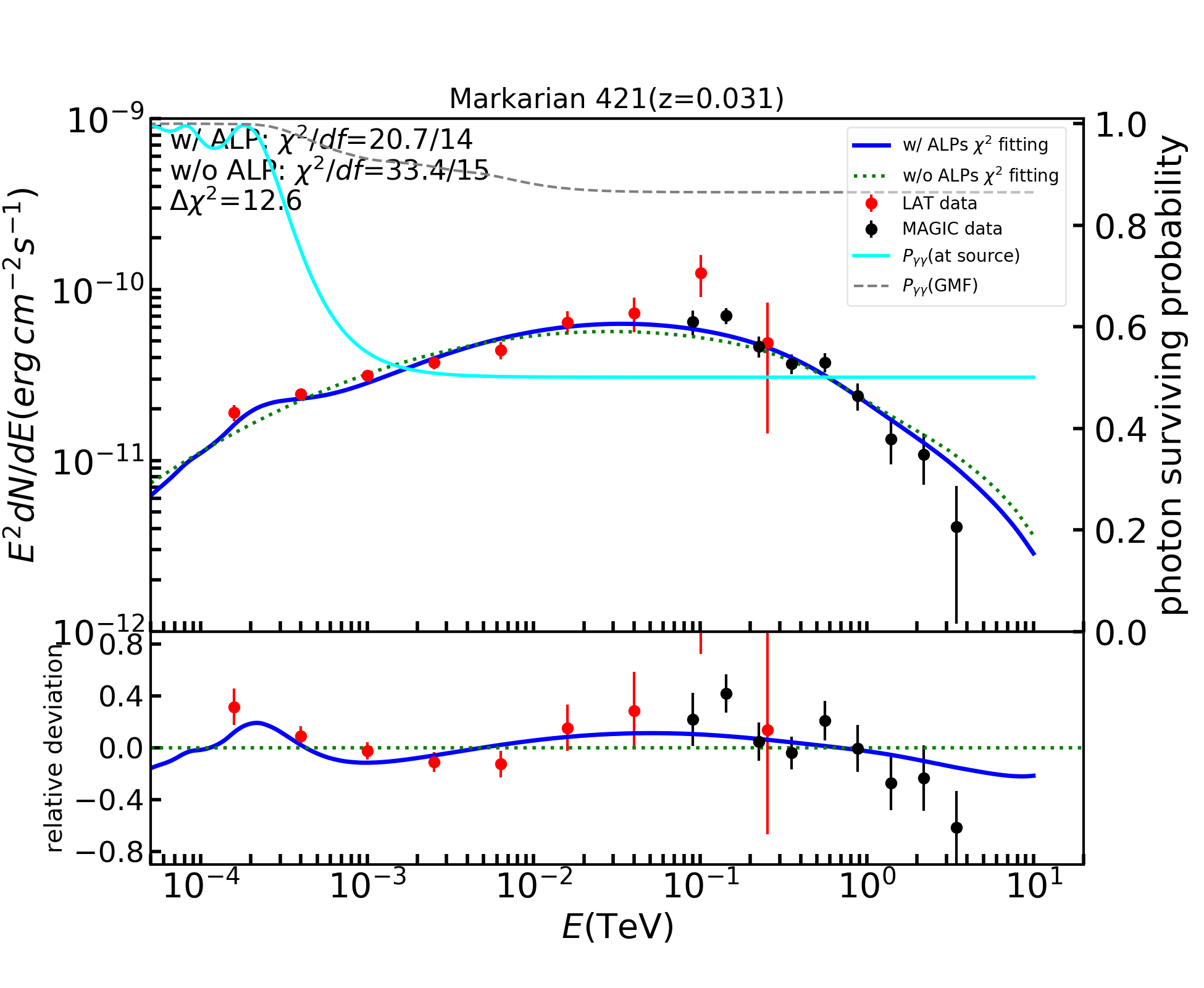}
\end{minipage}%
\begin{minipage}[t]{0.48\linewidth}
\includegraphics[width=1.0\textwidth]{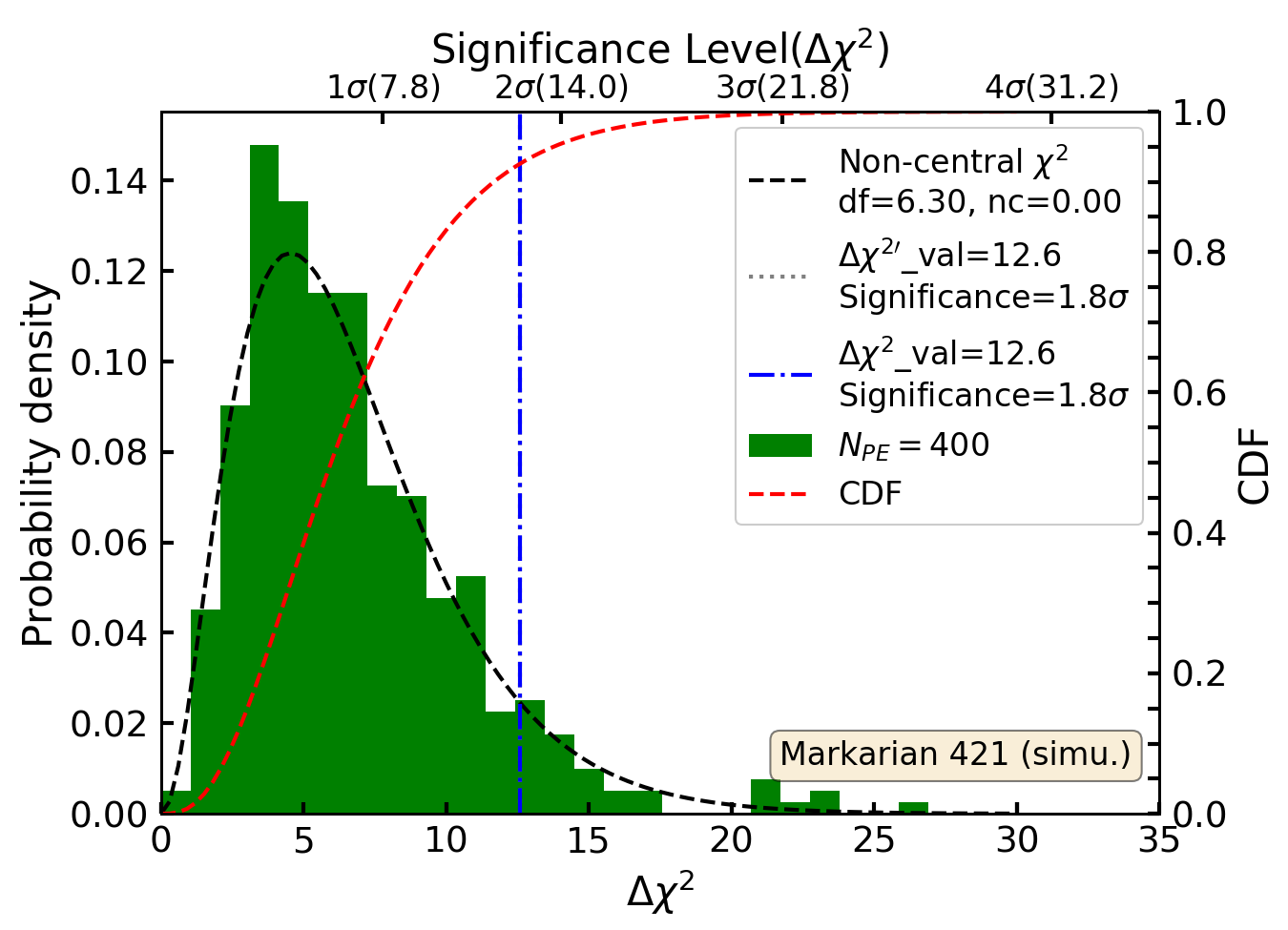}
\centering
\end{minipage}
\caption{Left panel: the spectral energy distribution for Markarian~421 during
contemporaneous observations with MAGIC and \textit{Fermi}-LAT in 2009. 
The red data points represent the 2009 LAT observations, 
and the black data points are from MAGIC \citep{LAT:2011mmt} during 
the same year. 
The blue solid and green dashed lines are the best-fitting models under $H_0$ and $H_1$ hypotheses respectively. The cyan solid and gray dashed lines stand for the photon surviving probabilities in different regions along the line of sight. Right panel: simulated null distribution from $\Delta\chi^2$ test for 2009 MAGIC and LAT observations. The black dashed line indicates a fit to the distribution with a non-central $\chi^2$ function. The red solid line represents the resulting cumulative distribution function (CDF). The $\Delta\chi^2$ value derived from the original data is marked as a blue (dot-dash) vertical line.}\label{fig:mkn421sed_simu} 
\end{figure}

The SED for Mkn~421 is obtained by reanalysing \textit{Fermi}-LAT data from
the observation season covered with MAGIC from January to June 2009 \citep{LAT:2011mmt}. The resulting SED is displayed in Fig.~\ref{fig:mkn421sed_simu}.
The spectrum shows a softening just below TeV energies, 
deviating noticeably from the  log-parabola shape (green dashed line in Fig.~\ref{fig:mkn421sed_simu}). The fit under the alternative $H_1$ hypothesis improves the 
goodness of fit by $\Delta \chi^2=12.6$ such that the resulting $\chi^2_{H1}=20.7$
for 13 degrees of freedom. This value is slightly larger than expected due to two
flux points between 100 and 200 GeV which deviate by more than two standard deviations from the fit. 

The z-score of the improvement is estimated to be $1.8$ (see Fig.~\ref{fig:mkn421sed_simu}, right panel).

 
\section{Discussion}

The search for spectral modulations in extra-galactic
energy spectra has been carried out for 20 objects that have
been selected to provide an optimized coverage in energy. 
For 18 sources (see Table~\ref{tab:TS}), 
the fits  show a consistent improvement  when including a photon-ALPs conversion (hypothesis $H_1$) in comparison to the  null hypothesis ($H_0$). 
When inspecting the individual spectra (see Figs.~\ref{fig:es0033sed}-\ref{fig:es2344sed}), the photon-ALPs conversion in the GMF and the
magnetic field intrinsic to the source leads to a rich phenomenology of spectral shapes. 
The resulting breaks, dips, and bumps occur predominantly at the 
critical energy (see Eq.~\ref{eq:e_crit}) specific to the GMF and the source-intrinsic magnetic field. 

The most significant improvement can be seen 
as expected in the spectra which have the largest signal-to-noise ratio, 
i.e. Mkn~421 (Fig.~\ref{fig:mkn421sed}) with $TS=18.5$ and NGC1275 (Fig.~\ref{fig:ngc1275sed}) with $TS=19.8$. Another source with a well-measured
spectrum is PG~1553+113 (Fig.~\ref{fig:pg1553sed}). The spectrum 
has apparently several features that are not predicted in our model. Subsequently, the resulting $TS=-4.2$ would  favor the $H_0$ hypothesis.
However, the source could be embedded in a cluster environment where 
weak mixing in the turbulent magnetic field provides a better description
of the observed spectrum. 

The findings obtained with the likelihood-method have been largely confirmed
when fitting the SED with a $\chi^2$-based approach. The wider binning leads to 
an under-sampling of the spectral features that are visible in the spectra 
obtained for the likelihood analysis. The resulting significance in the $\chi^2$-based
approach is therefore smaller than for the likelihood approach.
The $\chi^2$-based approach is however useful when combining data sets from \textit{Fermi}-LAT with
ground-based measurements. We have demonstrated that for contemporaneous data-sets
on PKS2155-304 and Mkn~421, the energy range can be extended in a meaningful 
way. However, no additional features are observed in the wider energy range. 
In the case of PKS~2155-304, the re-conversion of ALPs leads to an enhanced
flux at energies exceeding a few TeV which is slightly favored by the data. 

The combination of the results obtained with the relevant likelihood-analysis
has been carried out for all 20 sources. When combining the 
likelihood results for the local maxima in the $B$-$s$ plane, we find 
a total TS value of 
$TS_\mathrm{tot}=98.9$. The local maxima is chosen to minimize the energy requirement
to sustain a magnetic field with energy density $\propto B^2$ over a volume $\propto s^3$. 
This is considerably smaller than the value found when combining the global maxima of all sources with $TS'_\mathrm{tot}=133.6$.

A bootstrap-type combination of the same analyses carried out
on mock data-sets that have been simulated under the null hypothesis are
used to estimate the significance. We estimate the  
 chance probability to find a $TS_\mathrm{tot}$ value larger than $98.9$ to be 
$1.2\times 10^{-7}$ corresponding to a significance of $5.3~\sigma$.
For the global maximum, the estimated significance reaches $6~\sigma$.



\begin{figure}[ht!]
  \centering
  \includegraphics[width=0.80\textwidth]{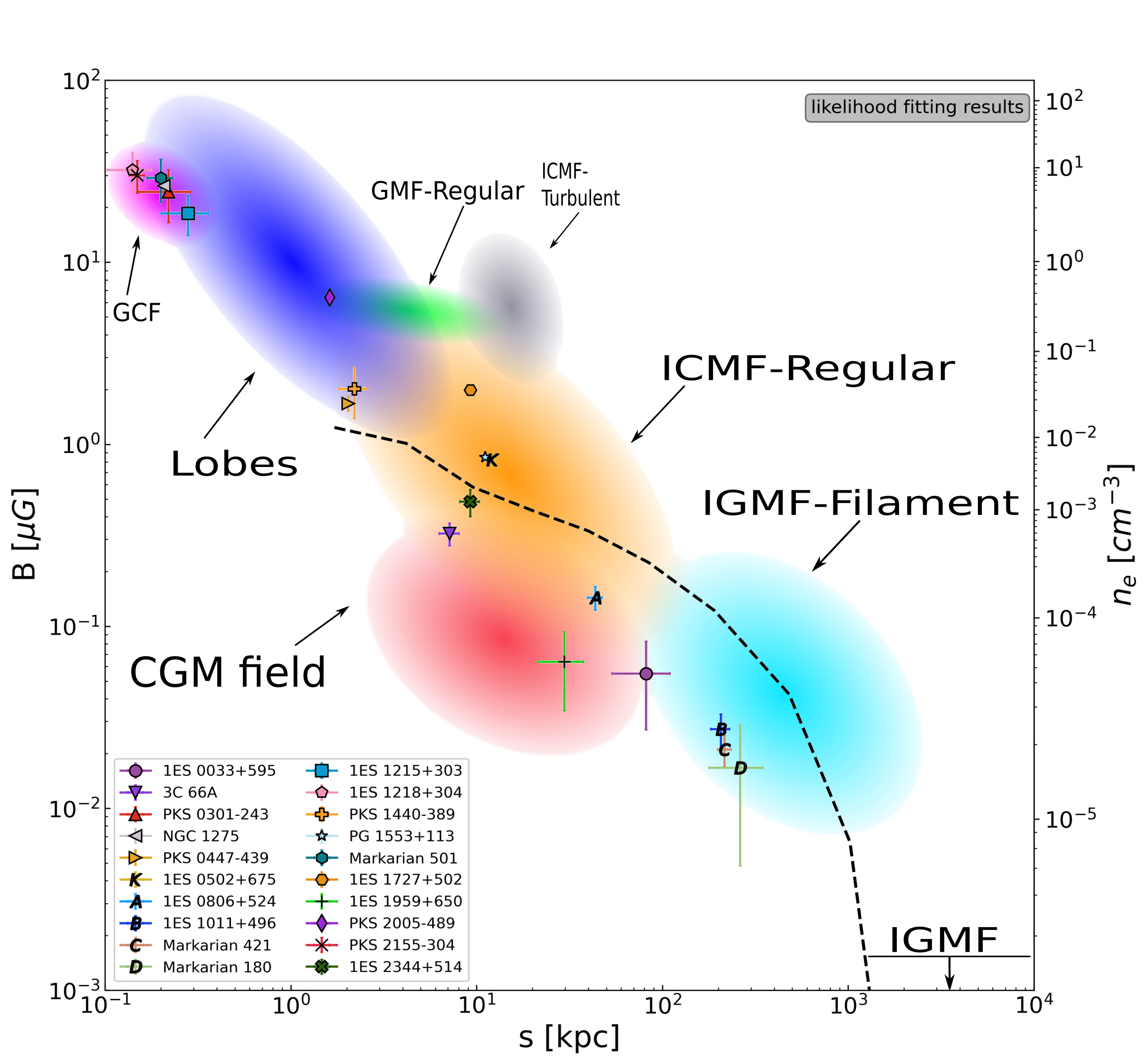}\\
  \caption{Likelihood best-fitting results of $(\hat B_0, \hat s_0)$ from Table~\ref{tab:fitt-reswalp}: The black dashed line is the cosmological magnetohydrodynamical simulation for radio haloes and magnetic fields in galaxy clusters which relate to the
  average electron density $n_e$ \cite{Marinacci:2017wew, Springel:2017tpz}. }\label{fig:b_bounds}
\end{figure}

 The required values of average transversal magnetic
field $\hat B_0$ and extension $\hat s_0$ found for the individual sources
fall into a wide range 
covering several orders of magnitude as shown in Fig.~\ref{fig:b_bounds}. 

We compare in the same figure
the values found in the fitting procedure with the range
of values for magnetic fields possibly present in the vicinity of the considered sources. This includes 
the magnetic field in the outer regions of the jet (lobes) \citep{Meyer:2014gta,Tavecchio:2014yoa,Feain:2009rf} as well as the magnetic field present in the host galaxy. 
In the wider vicinity of the source, we can expect 
that some objects are located in galaxy groups or galaxy clusters which are known to support an intra-cluster magnetic field (ICMF) with a turbulent and large scale component \citep{Kim:1990,Carilli:2001hj,Govoni:2004as,Sanders:2005jx,Subramanian:2005hf,Taylor:2006ta,Akahori:2010ym,Libanov:2019fzq,Fermi-LAT:2016nkz}. 
Recently, low-frequency radio-observations have revealed the presence
of a large-scale magnetic field in the circumgalactic medium (CGM) \cite{2023A&A...670L..23H}. 

Finally, the IGMF in filaments \citep{Vernstrom:2017jvh,Brown:2017dwx,Vacca:2018rta,OSullivan:2018shr,Locatelli:2021byc,Vernstrom:2021hru,Carretti:2022tbj}
along the line of sight could contribute additional
conversion regions. The IGMF \citep{Durrer:2013pga} in voids, however, is not relevant.

We note that the estimated values for $\hat B$ and $\hat s$ found for the 20 sources are nicely aligned with the astrophysically known magnetic fields.  There is a noticeable
cluster of six sources with $B=20~\mu\mathrm{G}-30~\mu\mathrm{G}$ for a spatial scale
between $150~\mathrm{pc}-300~\mathrm{pc}$. Similar large-scale fields are present
in the central 200~pc of the Milky Way (galactic center field, GCF) \cite{2013ApJ...762...33Y}, suggesting that the
photon-ALPs conversion takes place in a similar environment in these sources. 

In addition, we also indicate the  prediction for formation of magnetic fields
from magnetohydrodynamical simulation ~\cite{Marinacci:2017wew,Springel:2017tpz}\footnote{The converted B-field values shown here are from Illustris TNG-300 simulation setup in Ref.~\cite{Marinacci:2017wew}}. 
The result of the simulation traces well the observed values indicated
in Fig.~\ref{fig:b_bounds}. 

The result obtained here have been found under the assumption of fixed values for
mass $m_a=3.6~\mathrm{neV}$ and coupling $g_{a\gamma\gamma}=2.3\times 10^{-10}~\mathrm{GeV}^{-1}$ \cite{Majumdar:2018sbv},
that are not consistent with the bounds
provided by CAST \cite{CAST:2017uph} (see however Ref.~\cite{2021JCAP...11..036G} 
for a consistent interpretation of the  two results). It is important to note, that 
the features visible in the energy spectra can also be fit with values for the coupling
consistent with the bound from CAST ($g_{a\gamma\gamma}<6.6\times 10^{-11}~\mathrm{GeV}^{-1}$, 95~\% c.l.). 
In this case, the product of required magnetic field and length scale would need 
to be increased
by a factor of $\approx 4$, which would not lead to unreasonable values of these
parameters. However, more recent constraints for $g_{a\gamma\gamma}<5.4\times 10^{-12}~\mathrm{GeV}^{-1}$ (95~\% c.l.) have been derived 
from a re-analysis of the polarization 
from magnetized white dwarfs \cite{2022PhRvD.105j3034D}. In this case, the
necessary increase of the product of $B$ and $s$ to be $\approx 40$ would push the best-fitting values to the
upper end of astrophysically motivated values (see Fig.~\ref{fig:b_bounds}). 
\section{Summary}
\begin{itemize}
   \item The fits to high-energy spectra of 20 extra-galactic gamma-ray sources improves significantly with $5.3(6.0)\,\sigma$ when including photon-ALPs mixing in a homogeneous magnetic field $\hat B_0(\hat B)$ with spatial extension $\hat s_0(\hat s)$ left free to vary for a fixed $m_a$ and 
   $g_{a\gamma\gamma}$. The values $\hat s_0$, $\hat B_0$ relate to the local maximum in the TS with 
   the additional condition that the total energy in the magnetic field $\propto \hat B_0^2~\hat s_0^3$ is minimized.
   The value in the parentheses corresponds to the global maximum. 
   \item The individual sources with strong indications
   for additional spectral features present in
    the \textit{Fermi}-LAT data are NGC1275 with $3.6(3.6)~\sigma$ and Mkn~421 with also $3.6(3.6)~\sigma$. 
 \item The range of values found for $\hat B_0$  is
   consistent with expected and plausible
   values for the magnetic field strength found
   in astrophysical environments characterized by
   the length
   scales of $\hat s_0$  from several 100 pc to several
   100 kpc (see Fig.~\ref{fig:b_bounds}). 
   \item The observations of spectral modulations
   in AGN establish 
   the disappearance channel of
   photon-ALPs mixing which is complementary to the 
   appearance channel (anomalous transparency) that has provided the first indications for photon-ALPs mixing in gamma-ray spectra.  
   \item The result shown here is degenerate for choosing a particular 
   combination of coupling $g_{a\gamma\gamma}$ and mass
   $m_a$ to achieve the required conversion probability $p_{\gamma\rightarrow a}\propto g_{a\gamma\gamma}^2B^2s^2$
above
   the critical energy $E_c\propto m_a^2/(g_{a\gamma\gamma} B)$. In this regards, the result presented here 
   remains valid for 
   different choices of $g_{a\gamma\gamma}$ and $m_a$ 
   which would be consistent with other exclusion limits and
   astrophysical expectations for magnetic field strength.
\end{itemize}
\begin{acknowledgments}
Part of this work has made use of data and software provided by the Fermi Science Support Center. Q.Y. acknowledges the support from China Scholarship Council and the Deutsche Forschungsgemeinschaft (DFG, German Research Foundation) under Germany’s Excellence Strategy – EXC 2121 ``Quantum Universe" – 390833306. 
\end{acknowledgments}

\appendix
\section{Additional fit-results}\label{sec:app00}
In Table~\ref{tab:fitt-reswoalp-chi2} we list the best-fitting parameters estimated
using the $\chi^2$-minimization under the null-hypothesis
(see Table~\ref{tab:fitt-reswoalp} for the corresponding
likelihood-fit results). The fit-results obtained with 
the $\chi^2$-method under the alternative hypothesis and
evaluated at the local minimum are listed in Table~\ref{tab:fitt-reswalp-chi2} (see Table~\ref{tab:fitt-reswalp} for
the likelihood-fit results).
In Tables~\ref{tab:best-fitt-reswalp-chi2} and \ref{tab:best-fitt-reswalp} we present the global best-fitting parameters with the $\chi^2$ and likelihood methods respectively 
(supplementary to the information presented in 
Table~\ref{tab:best-TS}). Finally, the Table~\ref{tab:best-TS} summarizes the result of the hypotheses
tests when considering the global extrema (compare with
Table~\ref{tab:TS} for the local extrema).

\begin{table*}[ht!]
	\renewcommand\arraystretch{1.2}
	\centering                                                                                     
	\caption{\vadjust{\vspace{-2pt}}Best-fitting parameters for 
	the null hypothesis with the $\chi^2$ method using the modeled spectra from Eq.~\eqref{eq:woalp}. The estimated uncertainties ($1\sigma$) for each fitting parameters are listed as well.
	The value of $E_b=10^5$~MeV is fixed, the normalisation is given in units of $10^{-15}\rm MeV^{-1}cm^{-2}s^{-1}$.}\label{tab:fitt-reswoalp-chi2}
	\begin{tabular*}{0.57\textwidth}{@{\extracolsep{0pt}}cccc}
	\hline
	\hline
    AGN name  &$N_0$     & $\alpha$        &\Centerstack{$\beta$\\$\times10^{-3}$}\\
    \hline    
    $\rm 1ES\,\,0033$+595   & $0.983(0.115)$        &$1.71(0.08)$  &$1 (13)$   \\
    $\rm 3C\,\,66A$         & $1.10 (0.09 )$        &$2.28(0.04)$  &$45(5 )$   \\
    $\rm PKS\,\,0301$-243   & $0.509(0.064)$        &$2.17(0.06)$  &$36(8 )$   \\
    $\rm NGC\,\,1275$       & $0.802(0.046)$        &$2.70(0.02)$  &$70(3 )$   \\
    $\rm PKS\,\,0447$-439   & $1.29 (0.10 )$        &$2.20(0.04)$  &$55(5 )$   \\
    $\rm 1ES\,\,0502$+675   & $0.86 (0.07 )$        &$2.50(0.03)$  &$-$        \\
    $\rm 1ES\,\,0806$+524   & $0.46 (0.05 )$        &$2.11(0.06)$  &$33(8 )$   \\
    $\rm 1ES\,\,1011$+496   & $1.15 (0.09 )$        &$2.06(0.04)$  &$32(5 )$   \\
    $\rm Markarian\,\,421$  & & &  \\
    \multicolumn{1}{r}{\tiny{10yrs LAT}}&$6.34(0.20)$&$1.92(0.02)$&$22(2)$   \\
    \multicolumn{1}{r}{\tiny{simul. LAT+MAGIC}}&3.28(0.18)&$2.12(0.02)$ &$50(5)$    \\
    $\rm Markarian\,\,180$  & $0.226(0.026)$        &$2.19(0.03)$  &$-$        \\
    $\rm 1ES\,\,1215$+303   & $0.772(0.069)$        &$2.28(0.04)$  &$48(5 )$   \\
    $\rm 1ES\,\,1218$+304   & $1.02 (0.07 )$        &$2.29(0.02)$  &$-$        \\
    $\rm PKS\,\,1440$-389   & $0.484(0.06 )$        &$2.20(0.07)$  &$63(11)$   \\
    $\rm PG\,\,1553$+113    & $3.96 (0.21 )$        &$1.87(0.03)$  &$37(5 )$   \\
    $\rm Markarian\,\,501$  & $2.41 (0.13 )$        &$1.86(0.03)$  &$17(4 )$   \\
    $\rm 1ES\,\,1727$+502   & $0.344(0.033)$        &$2.20(0.02)$  &$-$        \\
    $\rm 1ES\,\,1959$+650   & $1.60 (0.11 )$        &$1.98(0.04)$  &$26(5 )$   \\
    $\rm PKS\,\,2005$-489   & $0.572(0.043)$        &$2.18(0.02)$  &$-$        \\
    PKS\,\,2155-304 & & &  \\
    \multicolumn{1}{r}{\tiny{10yrs LAT}}&2.44(0.13)&$2.13(0.03)$ &$40(3)$    \\
    \multicolumn{1}{r}{\tiny{simul. LAT+H.E.S.S.}}&1.32(0.04)&$2.22(0.03)$ &$53(7)$    \\
    $\rm 1ES\,\,2344$+514   & $0.321(0.06 )$        &$2.23(0.11)$  &$61(16)$   \\
	\hline
	\hline
    \end{tabular*}
\end{table*}
\begin{table*}[ht!]
	\renewcommand\arraystretch{1.2}
	\centering
	\caption{\vadjust{\vspace{-2pt}}Best-fitting parameters of local maxima for ALP hypothesis with $\chi^2$ method. Parameters uncertainties ($1\sigma$) are included. The value of $E_b=10^5$~MeV is fixed, the normalisation is given in units of $10^{-15}\rm MeV^{-1}cm^{-2}s^{-1}$.}\label{tab:fitt-reswalp-chi2}
        \scalebox{0.9}{
	\begin{tabular*}{0.94\textwidth}{@{\extracolsep{0pt}}cccccc}
	\hline
	\hline
	AGN name &$N_0$ &$\alpha$ &\Centerstack{$\beta$\\$\times10^{-3}$} &\Centerstack{$\hat B_0$\\[nG]} &\Centerstack{$\hat s_0$\\[kpc]}\\
    \hline                                                                                                
    $\rm 1ES\,\,0033$+595   & $1.74 (0.20 )$  & $1.78(0.08)$   &$18 8(12)$   &$7547.5(3281.5)$ &$1.5  (0.6)$ \\
    $\rm 3C\,\,66A$         & $1.89 (0.26 )$  & $2.20(0.05)$   &$46  (5 )$   &$374.7(62.6)$ &$6.9  (0.8)$ \\
    $\rm PKS\,\,0301$-243   & $0.950(0.145)$  & $2.18(0.12)$   &$46  (13)$   &$21672.5(2918.8)$ &$0.3  (0.01)$ \\
    $\rm NGC\,\,1275$       & $1.47 (0.02 )$  & $2.74(0.01)$   &$76  (1 )$   &$39621.7(317.6)$ &$0.1  (1.6)$ \\
    $\rm PKS\,\,0447$-439   & $2.46 (0.22 )$  & $2.30(0.07)$   &$90   9$      &$1731.8(283.0)$ &$2.0  (0.2)$ \\
    $\rm 1ES\,\,0502$+675   & $1.35 (0.21 )$  & $2.59(0.05)$   &$-$          &$104.4(54.0)$ &$53.5 (17.4)$ \\
    $\rm 1ES\,\,0806$+524   & $0.581(0.075)$  & $2.00(0.07)$   &$21  (9 )$   &$25.5(10.2)$ &$202.6(36.2)$ \\
    $\rm 1ES\,\,1011$+496   & $1.42 (0.14 )$  & $1.98(0.05)$   &$24   (6)$      &$35.7(7.3)$ &$170.5(17.0)$ \\
    $\rm Markarian\,\,421$  &    &     &    &   &  \\
    \multicolumn{1}{r}{\tiny{10yrs LAT}}& $8.71 (0.43 )$  & $1.84(0.02)$   &$16(3)$   &$16.5    (6.4)$ &$242.0(35.1)$ \\
    \multicolumn{1}{r}{\tiny{simul. LAT+MAGIC}}&$7.25(0.37)$&$2.15(0.02)$&$71(5)$&$4344.2(845.6)$&$1.0(0.1)$\\   
    $\rm Markarian\,\,180$  & $0.308(0.040)$  & $2.25(0.03)$   &$-$          &$20.2(11.0)$ &$226.2(72.9)$ \\
    $\rm 1ES\,\,1215$+303   & $1.44 (0.16 )$  & $2.36(0.07)$   &$69  (12)$   &$7385.9(5817.8)$ &$0.4  (0.3)$ \\
    $\rm 1ES\,\,1218$+304   & $1.98 (0.14 )$  & $2.29(0.02)$   &$-$          &$295842.8(65007.2)$    &$0.0  (0.01)$ \\
    $\rm PKS\,\,1440$-389   & $0.859(0.121)$  & $2.38(0.08)$   &$110 (12)$   &$2326.0(1404.3)$ &$1.9  (0.5)$ \\
    $\rm PG\,\,1553$+113    & $7.40 (0.40 )$  & $1.78(0.04)$   &$40  (5 )$   &$840.1(88.9)$ &$10.9 (1.0)$ \\
    $\rm Markarian\,\,501$  & $4.07 (0.38 )$  & $1.87(0.02)$   &$24  (3 )$   &$29607.5(3114.0)$ &$0.2  (0.02)$ \\
    $\rm 1ES\,\,1727$+502   & $0.479(0.044)$  & $2.23(0.02)$   &$-$          &$1355.6(341.8)$ &$6.8  (1.3)$ \\
    $\rm 1ES\,\,1959$+650   & $2.33 (0.31 )$  & $1.89(0.05)$   &$22  (5)$   &$137.7(67.8)$ &$11.1 (3.4)$ \\
    $\rm PKS\,\,2005$-489   & $1.01 (0.09 )$  & $2.24(0.02)$   &$-$          &$8025.7(2126.2)$ &$1.3  (0.3)$ \\
    $\rm PKS\,\,2155$-304   &    &     &      &   &   \\
    \multicolumn{1}{r}{\tiny{10yrs LAT}}  & $4.69 (0.24 )$  & $2.16(0.03)$&43(3)&$60241.1 (10007.3)$ &$0.1  (0.01)$ \\ 
    \multicolumn{1}{r}{\tiny{simul. LAT+H.E.S.S.}} &$2.69(0.09)$&$2.34(0.02)$&$86(7)$&$5589.7(1975.4)$&$0.8(0.2)$\\
    $\rm 1ES\,\,2344$+514   & $0.731(0.132)$  & $2.15(0.12)$   &$75  (17)$   &$619.9(204.6)$ &$7.5  (1.5)$ \\
	\hline
	\hline
\end{tabular*}}
\end{table*}

\begin{table*}[ht!]
	\renewcommand\arraystretch{1.2}
	\centering
	\caption{\vadjust{\vspace{-2pt}}Best-fitting parameters for ALP hypothesis with $\chi^2$ method using the modeled spectra from Eq.~\eqref{eq:walp}. Parameters uncertainties ($1\sigma$) are included. The value of $E_b=10^5$~MeV is fixed, the normalisation is given in units of $10^{-15}\rm MeV^{-1}cm^{-2}s^{-1}$.}\label{tab:best-fitt-reswalp-chi2}
        \scalebox{0.9}{
	\begin{tabular*}{0.93\textwidth}{@{\extracolsep{0pt}}cccccc}
	\hline
	\hline
	AGN name &$N_0$ &$\alpha$ &\Centerstack{$\beta$\\$\times10^{-3}$} &\Centerstack{$\hat B$\\[nG]} &\Centerstack{$\hat s$\\[kpc]}\\
    \hline                                                                                                
$\rm 1ES\,\,0033$+595       & $1.77 (0.20 )$  & $1.75(0.08)$  & $13(13)$    & $17047.7 (2689.8)$  & $3.8     (0.6  )$ \\  
$\rm 3C\,\,66A$             & $1.89 (0.26 )$  & $2.20(0.05)$  & $46(5 )$    & $374.8   (62.6  )$  & $6.9     (0.8  )$ \\ 
$\rm PKS\,\,0301$-243       & $0.955(0.078)$  & $2.17(0.05)$  & $46(5 )$    & $88403.9 (532.1 )$  & $1.8     (0.01 )$ \\
$\rm NGC\,\,1275$           & $1.43 (0.13 )$  & $2.75(0.03)$  & $78(4 )$    & $33752.1 (9973.8)$  & $0.1     (0.03 )$ \\
$\rm PKS\,\,0447$-439       & $2.50 (0.19 )$  & $2.23(0.04)$  & $74(6 )$    & $8922.0  (54.8  )$  & $57.7    (0.4  )$ \\  
$\rm 1ES\,\,0502$+675       & $1.42 (0.17 )$  & $2.57(0.03)$  & $-$         & $910.7   (6.9   )$  & $272.8   (1.8  )$ \\  
$\rm 1ES\,\,0806$+524       & $0.617(0.085)$  & $1.98(0.07)$  & $19(10)$    & $51.8    (7.6   )$  & $278.5   (28.4 )$ \\  
$\rm 1ES\,\,1011$+496       & $1.79 (0.16 )$  & $1.89(0.04)$  & $16(6 )$    & $120.9   (1.6   )$  & $855.7   (8.9  )$ \\  
$\rm Markarian\,\,421$      &   &    &    &    &    \\  
\multicolumn{1}{r}{\tiny{10yrs LAT}}& $8.71 (0.44 )$  & $1.84(0.02)$  & $16(3 )$    & $16.5 (6.6)$  & $242.2 (37.4)$  \\
\multicolumn{1}{r}{\tiny{simul. LAT+MAGIC}}&$7.25(0.37)$&$2.15(0.02)$&$71(5)$&$4344.2(845.6)$&$1.0(0.1)$\\                 
$\rm 1ES\,\,1215$+303       & $1.44 (0.16 )$  & $2.36(0.07)$  & $69(12)$    & $7392.7  (5743.8)$  & $0.4     (0.3  )$  \\ 
$\rm 1ES\,\,1218$+304       & $2.00 (0.14 )$  & $2.30(0.02)$  & $-$         & $308906.8(516.5 )$  & $7.3     (0.01 )$  \\ 
$\rm PKS\,\,1440$-389       & $0.931(0.119)$  & $2.28(0.08)$  & $91(12)$    & $30894.1 (26.7  )$  & $200.5   (0.2  )$  \\ 
$\rm PG\,\,1553$+113        & $7.29 (0.16 )$  & $1.88(0.01)$  & $53(1 )$    & $2647.4  (0.0   )$  & $226832.3(0.4  )$  \\ 
$\rm Markarian\,\,501$      & $4.40 (0.25 )$  & $1.85(0.03)$  & $15(5 )$    & $59644.3 (1303.1)$  & $2.5     (0.1  )$  \\ 
$\rm 1ES\,\,1727$+502       & $0.446(0.042)$  & $2.18(0.02)$  & $-$         & $61936.7 (16.3  )$  & $395.7   (0.1  )$  \\ 
$\rm 1ES\,\,1959$+650       & $2.75 (0.29 )$  & $1.81(0.05)$  & $13(7 )$    & $180.6   (1.8   )$  & $627.7   (3.5  )$  \\ 
$\rm PKS\,\,2005$-489       & $0.895(0.076)$  & $2.22(0.02)$  & $-$         & $11432.8 (151.3 )$  & $20.4    (0.3  )$  \\ 
$\rm PKS\,\,2155$-304       &    &   &     &   &    \\ 
 \multicolumn{1}{r}{\tiny{10yrs LAT}}  & $4.51 (0.24 )$  & $2.11(0.03)$  & $49(4 )$ & $4912.4  (1.4)$ & $925.6(0.2  )$  \\
 \multicolumn{1}{r}{\tiny{simul. LAT+H.E.S.S.}} &$2.69(0.09)$&$2.34(0.02)$&$86(7)$&$5589.7(1975.4)$&$0.8(0.2)$\\
$\rm 1ES\,\,2344$+514       & $0.690(0.152)$  & $2.21(0.12)$  & $83(17)$    & $1159.7  (390.0 )$  & $11.9    (3.3  )$  \\ 
	\hline
	\hline
\end{tabular*}}
\end{table*}

\begin{table*}[ht!]
	\renewcommand\arraystretch{1.2}
	\centering
	\caption{\vadjust{\vspace{-2pt}}Best-fitting parameters for ALP hypothesis with likelihood method using the modeled spectra from Eq.~\eqref{eq:walp}. The normalisation is given in units of $10^{-12}\rm MeV^{-1}cm^{-2}s^{-1}$. Parameters uncertainties ($1\sigma$) are included.}\label{tab:best-fitt-reswalp}
        \scalebox{0.9}{
	\begin{tabular*}{1.02\textwidth}{@{\extracolsep{0pt}}ccccccc}
	\hline
	\hline
	AGN name &$N_0$ &$\alpha$ &\Centerstack{$\beta$\\$\times10^{-3}$} &\Centerstack{$E_b$\\[MeV]} &\Centerstack{$\hat B$\\[nG]} &\Centerstack{$\hat s$\\[kpc]}\\
	\hline
$\rm 1ES\,\,0033$+595       & $0.629  (0.025)$  & $1.54 (0.03)$  & $42 (12)$    & $3177 $  & $911.9   (1.5  )$  & $911.4  (1.3 )$   \\ 
$\rm 3C\,\,66A$             & $12.5   (0.2 )$   & $1.80 (0.01)$  & $43 (6 )$    & $1211 $  & $311.0   (44.4 )$  & $7.3    (1.0 )$   \\
$\rm PKS\,\,0301$-243       & $11.1   (0.4  )$  & $1.81 (0.02)$  & $38 (10)$    & $954.4$  & $50660.9 (314.0)$  & $1.3    (0.01 )$  \\
$\rm NGC\,\,1275$           & $102    (2    )$  & $1.98 (0.01)$  & $85 (3 )$    & $883.6$  & $14007.6 (694.6)$  & $0.2    (0.01 )$  \\
$\rm PKS\,\,0447$-439       & $7.09   (0.13 )$  & $1.56 (0.01)$  & $83 (6 )$    & $1605 $  & $2151.5  (41.6 )$  & $5.5    (0.1 )$   \\
$\rm 1ES\,\,0502$+675       & $0.083  (0.004)$  & $ 1.38(0.03)$  & $-$          & $6322 $  & $830.6   (0.7  )$  & $1000.0 (0.7  )$  \\
$\rm 1ES\,\,0806$+524       & $2.31   (0.06 )$  & $1.79 (0.02)$  & $12 (10)$    & $1297 $  & $74.6    (2.4  )$  & $435.4  (12.4)$   \\
$\rm 1ES\,\,1011$+496       & $7.68   (0.10 )$  & $1.73 (0.01)$  & $17 (5 )$    & $1066 $  & $118.6   (0.9  )$  & $723.0  (5.0 )$   \\
$\rm Markarian\,\,421$      & $19.1   (0.2  )$  & $1.69 (0.005)$ & $13 (2 )$    & $1286 $  & $21.1    (4.2  )$  & $216.3  (18.4)$   \\
$\rm Markarian\,\,180$      & $0.182  (0.008)$  & $ 1.70(0.03)$  & $-$          & $2679 $  & $98.5    (1.5  )$  & $870.1  (11.9)$   \\
$\rm 1ES\,\,1215$+303       & $17.9   (0.4  )$  & $1.81 (0.01)$  & $53 (7 )$    & $1066 $  & $47661.7 (202.6)$  & $1.4    (0.01 )$  \\ 
$\rm 1ES\,\,1218$+304       & $0.418  (0.015)$  & $ 1.68(0.02)$  & $-$          & $4442 $  & $241288.9(42.7 )$  & $9.4    (0.002)$  \\
$\rm PKS\,\,1440$-389       & $1.75   (0.06 )$  & $1.51 (0.03)$  & $100(13)$    & $2014 $  & $3331.2  (38.9 )$  & $11.7   (0.1  )$  \\
$\rm PG\,\,1553$+113        & $5.66   (0.09 )$  & $1.45 (0.01)$  & $49 (5 )$    & $1847 $  & $3496.4  (0.0  )$  & $98466.5(0.1  )$  \\
$\rm Markarian\,\,501$      & $8.35   (0.34 )$  & $1.70 (0.01)$  & $17 (5 )$    & $1478 $  & $85669.0 (71.4 )$  & $3.2    (0.003)$  \\
$\rm 1ES\,\,1727$+502       & $0.297  (0.017)$  & $ 1.75(0.03)$  & $-$          & $3005 $  & $226832.0(0.6  )$  & $690.1  (0.002)$  \\
$\rm 1ES\,\,1959$+650       & $3.39   (0.06 )$  & $1.70 (0.01)$  & $9  (5 )$    & $1733 $  & $89.5    (0.9  )$  & $831.8  (10.0 )$  \\
$\rm PKS\,\,2005$-489       & $0.747  (0.050)$  & $ 1.76(0.02)$  & $-$          & $2398 $  & $21312.2 (78.1 )$  & $5.6    (0.02 )$  \\
$\rm PKS\,\,2155$-304       & $30.9   (0.4  )$  & $1.76 (0.01)$  & $40 (5 )$    & $1136 $  & $344288.2(57.1 )$  & $5.4    (0.001)$  \\
$\rm 1ES\,\,2344$+514       & $1.07   (0.05 )$  & $1.56 (0.03)$  & $61 (14)$    & $1938 $  & $482.4   (82.1 )$  & $9.2    (1.1  )$  \\
	\hline
	\hline
\end{tabular*}}
\end{table*}

\begin{table*}[ht!]
	\renewcommand\arraystretch{1.2}
	\centering
	\caption{\vadjust{\vspace{-2pt}}Best-fitting log-likelihood and $\chi^2$ values for null ($H_0$) and ALP hypotheses ($H_1$). TS values are calculated with Eqs.~\eqref{eq:TS} and \eqref{eq:TS_chi2} for likelihood ratio test and $\Delta\chi^2$ test respectively. Corresponding significance levels for both tests are listed as well.}\label{tab:best-TS}
	\scalebox{0.8}{
	\begin{tabular*}{1.15\textwidth}{@{\extracolsep{\fill}}ccccccccc}
		\hline
		\hline
		AGN name &\Centerstack{$H_0$\\$\ln(L^0_{max})$} &\Centerstack{$H_1$\\$\ln(L^1_{max})$}  &$TS$ &\Centerstack{$z$-score\\($H_1/H_0$)} &\Centerstack{$H_0$\\$\chi^2_{\rm w/o\,ALP}/df$} &\Centerstack{$H_1$\\$\chi^2_{\rm w/\,ALP}/df$} &$\Delta\chi^2$ &\Centerstack{$z$-score\\($H_1/H_0$)}\\
        \hline
        $\rm 1ES\,\,0033$+595   & $712253.0$      & $712257.8$    &$9.6$  &$2.3$ &$22.2/15$ &$13.5/13$ &$8.8 $ &$2.0$\\
        $\rm 3C\,\,66A$         & $-122858.4$     & $-122856.5$   &$3.8$  &$2.0$ &$20.6/15$ &$16.3/13$ &$4.3 $ &$2.3$\\
        $\rm PKS\,\,0301$-243   & $-154508.6$     & $-154507.9$   &$1.4$  &$0.6$ &$16.3/15$ &$16.3/13$ &$-0.1$ &$0.5$\\
        $\rm NGC\,\,1275$       & $26767.7	$     & $26777.6$     &$19.8$ &$3.6$ &$25.6/15$ &$22.4/13$ &$3.2$  &$1.8$\\
        $\rm PKS\,\,0447$-439   & $-146200.2$     & $-146199.1$   &$2.3$  &$2.0$ &$26.2/15$ &$22.6/13$ &$3.6$  &$2.1$\\
        $\rm 1ES\,\,0502$+675   & $-44659.7$      & $-44656.2$    &$7.0$  &$1.3$ &$41.9/16$ &$36.2/14$ &$5.7 $ &$1.0$\\
        $\rm 1ES\,\,0806$+524   & $-159751.7$     & $-159750.3$   &$2.9$  &$0.2$ &$20.5/15$ &$16.4/13$ &$4.1 $ &$0.5$\\
        $\rm 1ES\,\,1011$+496   & $-153505.0$     & $-153500.7$   &$8.6$  &$1.6$ &$12.7/15$ &$8.0 /13$ &$4.7 $ &$0.7$\\
        $\rm Markarian\,\,421$  & &   & & & & &\\
        \multicolumn{1}{r}{\tiny{10yrs LAT}}&$-130615.4$&$-130606.2$&$18.5$&$3.6$&$28.8/15$ &$15.5/13$ &$13.3$ &$2.7$\\
        \multicolumn{1}{r}{\tiny{simul. LAT+MAGIC}}&$-$&$-$&$-$&$-$&$33.4/15$&$15.2/13$&$18.2$&$1.8$\\
        $\rm Markarian\,\,180$  & $-132628.4$     & $-132627.7$   &$1.4$  &$0.01$ &$23.8/16$&$19.8/14$ &$4.0 $ &$0.5$\\
        $\rm 1ES\,\,1215$+303   & $-143988.2$     & $-143984.0$   &$8.4$  &$2.2$ &$15.5/15$ &$13.4/13$ &$2.1 $ &$1.2$\\
        $\rm 1ES\,\,1218$+304   & $-145577.3$     & $-145576.5$   &$1.6$  &$1.3$ &$19.0/16$ &$19.9/14$ &$-0.9$ &$1.1$\\
        $\rm PKS\,\,1440$-389   & $60781.3$       & $60783.8$     &$5.0$  &$1.5$ &$18.9/15$ &$16.1/13$ &$2.7 $ &$0.8$\\
        $\rm PG\,\,1553$+113    & $-152448.5$     & $-152447.6$   &$1.7$  &$0.4$ &$17.8/15$ &$14.7/13$ &$3.1 $ &$0.5$\\
        $\rm Markarian\,\,501$  & $-95747.1$      & $-95742.8$    &$8.7$  &$1.7$ &$18.5/15$ &$12.8/13$ &$5.7 $ &$1.0$\\
        $\rm 1ES\,\,1727$+502   & $-154208.5$     & $-154205.2$   &$6.8$  &$1.4$ &$26.7/16$ &$22.7/14$ &$3.9 $ &$0.3$\\
        $\rm 1ES\,\,1959$+650   & $-49456.2$      & $-49453.1$    &$6.3$  &$1.1$ &$21.6/15$ &$13.5/13$ &$8.1$  &$1.5$\\
        $\rm PKS\,\,2005$-489   & $-166314.9$     & $-166307.9$   &$13.6$ &$2.5$ &$33.9/16$ &$17.2/14$ &$16.7$ &$2.7$\\
        $\rm PKS\,\,2155$-304   &     &   & & & & & &\\
        \multicolumn{1}{r}{\tiny{10yrs LAT}}&$-144503.6$&$-144502.7$&$1.7$&$1.5$&$15.8/15$&$14.3/13$&1.5&1.1\\
        \multicolumn{1}{r}{\tiny{simul. LAT+H.E.S.S.}}&$-$&$-$&$-$&$-$&$33.3/18$&$29.2/16$&$4.1$&1.6\\
        $\rm 1ES\,\,2344$+514   & $-61482.2$      & $-61480.1$    &$4.3$  &$1.2$ &$15.2/15$ &$13.6/13$  &$1.6 $ &$0.6$\\
		\hline
		\hline
	\end{tabular*}}
\end{table*}

\begin{figure}[ht!]
\begin{minipage}[t]{0.455\linewidth}
\centering
\includegraphics[width=0.9\textwidth]{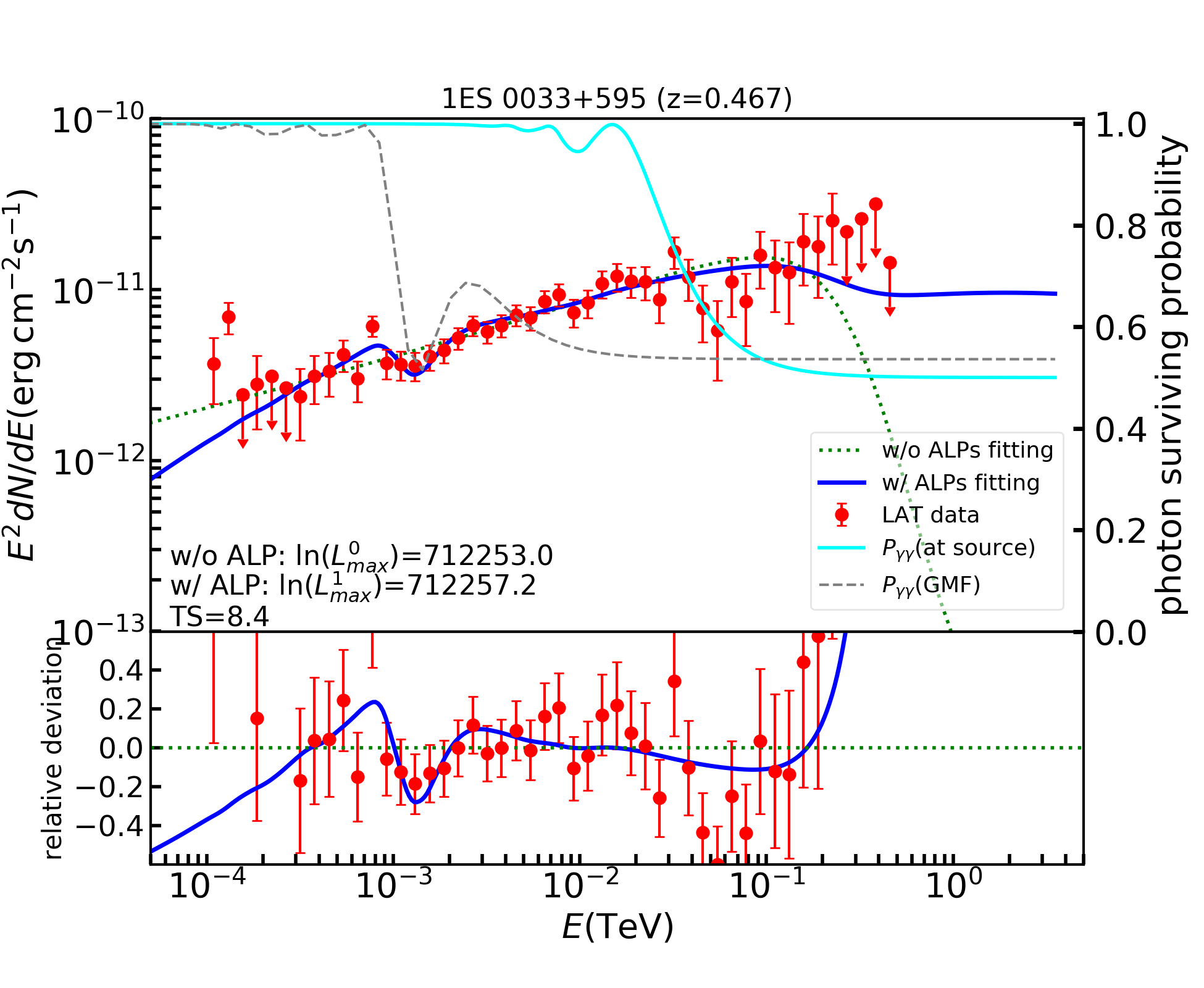}
\end{minipage}%
\begin{minipage}[t]{0.455\linewidth}\centering
\includegraphics[width=0.9\textwidth]{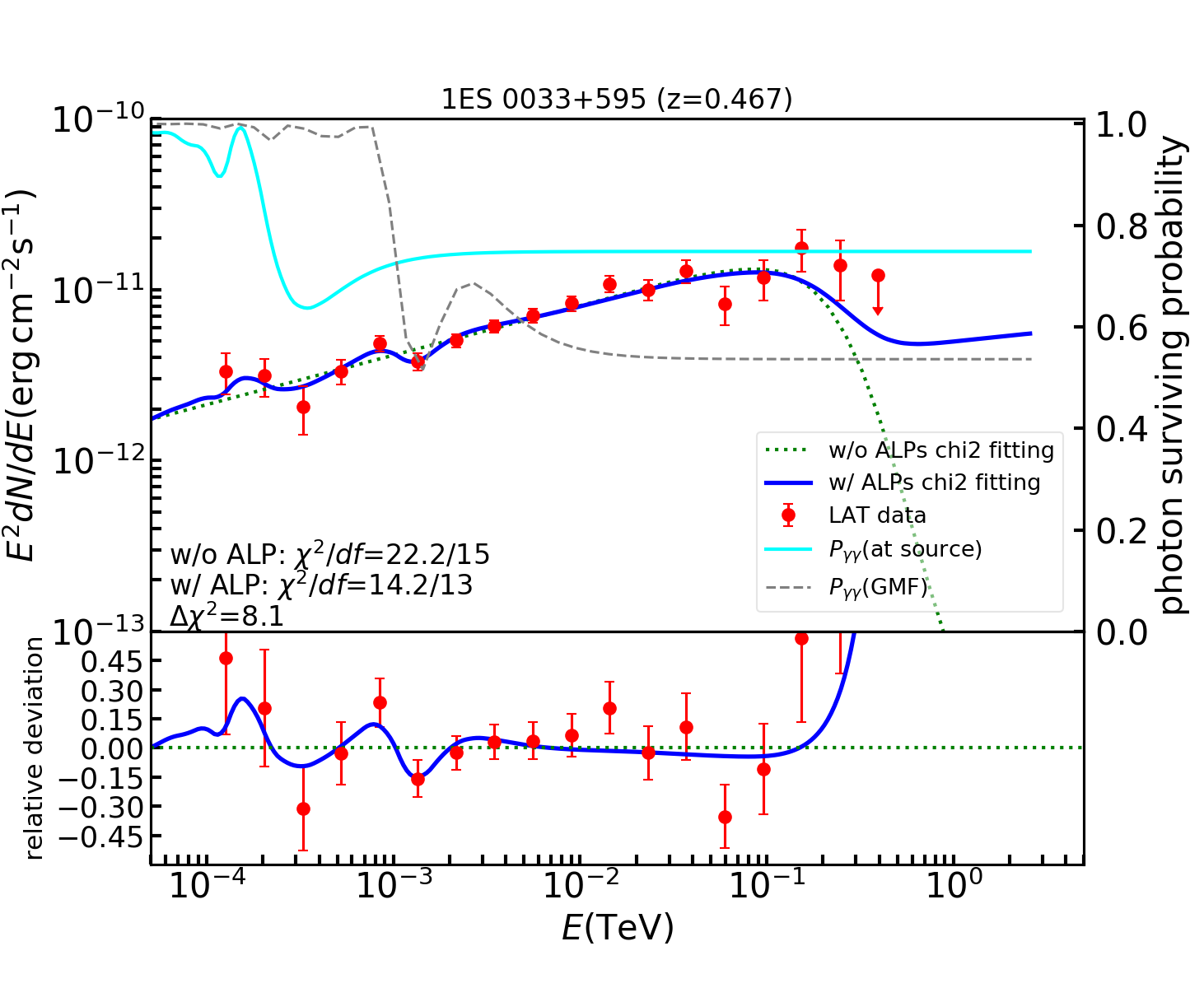}
\end{minipage}
\caption{1ES~0033+595, same as Fig.~\ref{fig:mkn421sed}.}\label{fig:es0033sed}
\end{figure}

\begin{figure}[ht!]
\begin{minipage}[t]{0.455\linewidth}
\centering
\includegraphics[width=0.9\textwidth]{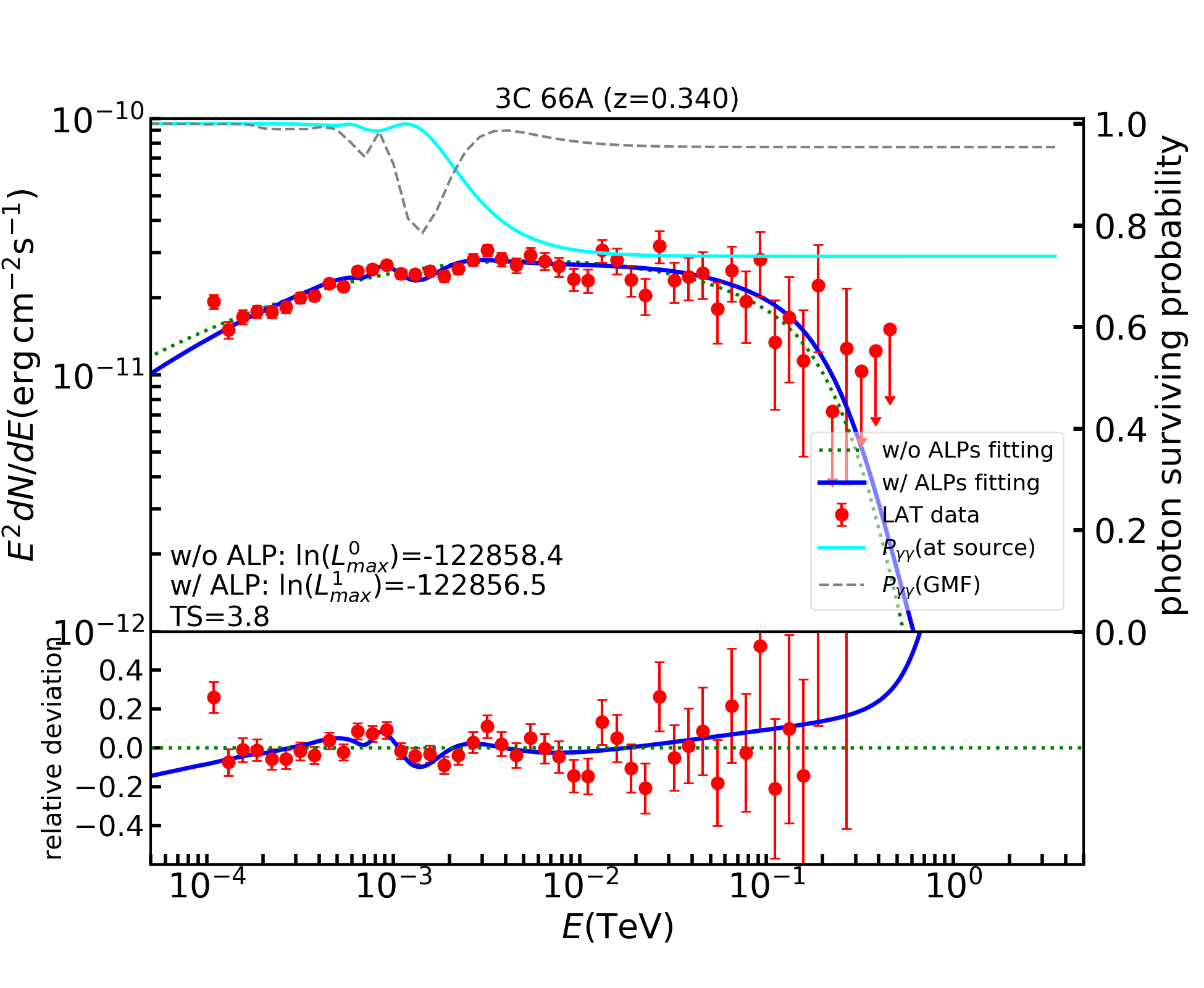}
\end{minipage}%
\begin{minipage}[t]{0.455\linewidth}
\centering
\includegraphics[width=0.9\textwidth]{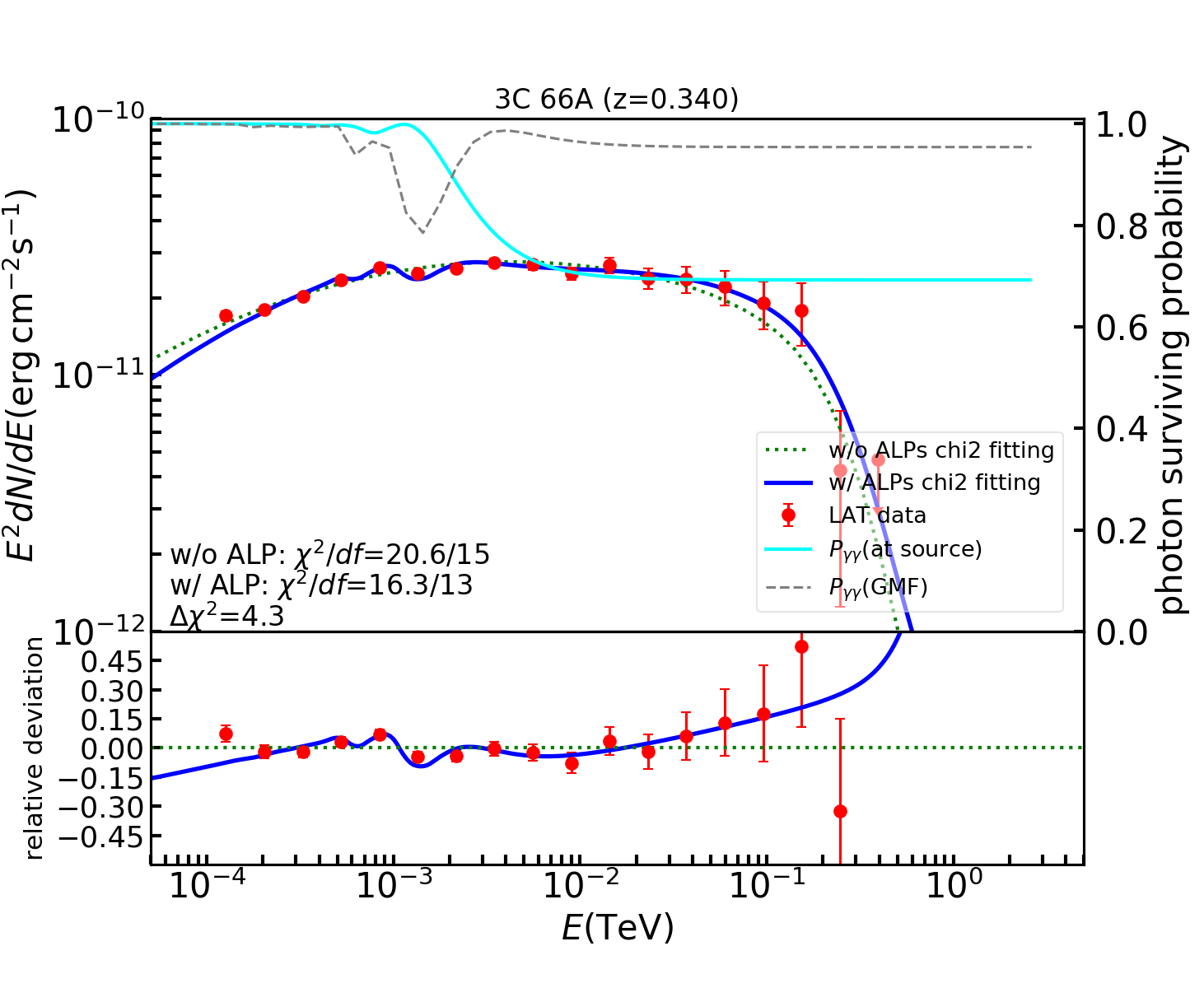}
\end{minipage}
\caption{3C~66A, same as Fig.~\ref{fig:mkn421sed}.}\label{fig:3c66ased}
\end{figure}

\begin{figure}[ht!]
\begin{minipage}[t]{0.455\linewidth}
\centering
\includegraphics[width=0.9\textwidth]{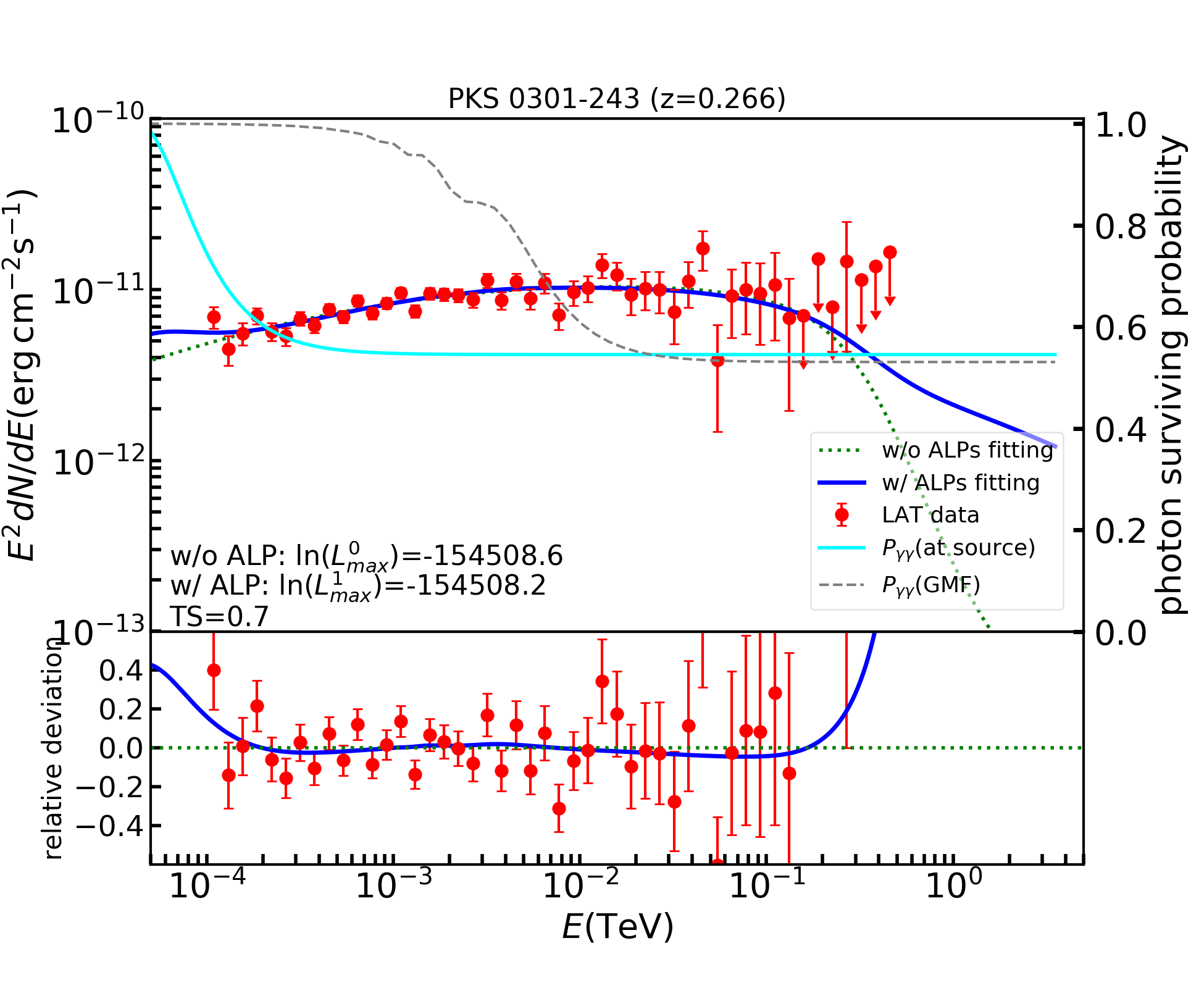}
\end{minipage}%
\begin{minipage}[t]{0.455\linewidth}
\centering
\includegraphics[width=0.9\textwidth]{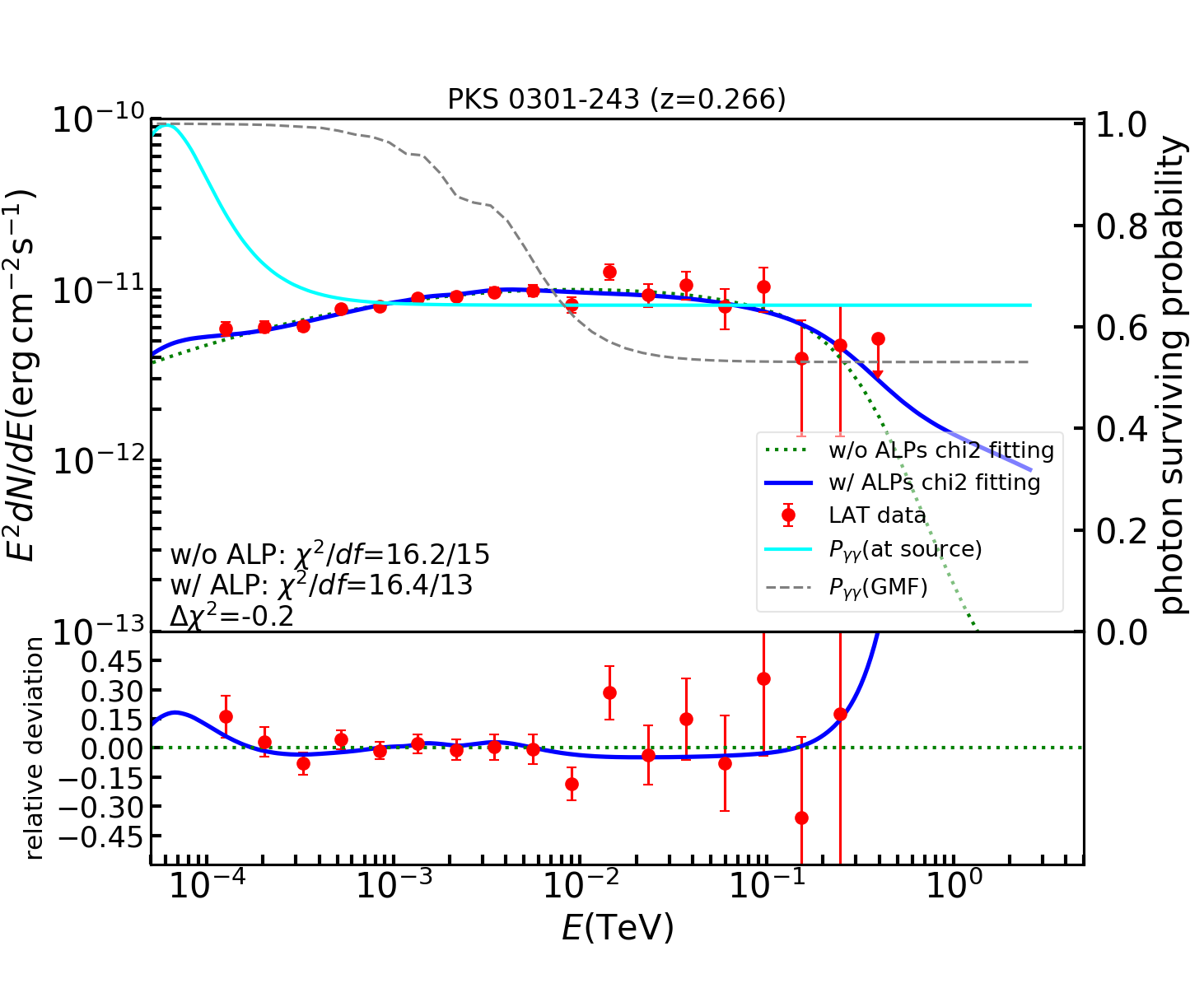}
\end{minipage}
\caption{PKS~0301-243, same as Fig.~\ref{fig:mkn421sed}.}\label{fig:pks0301sed}
\end{figure}

\begin{figure}[ht!]
\begin{minipage}[t]{0.455\linewidth}
\centering
\includegraphics[width=0.9\textwidth]{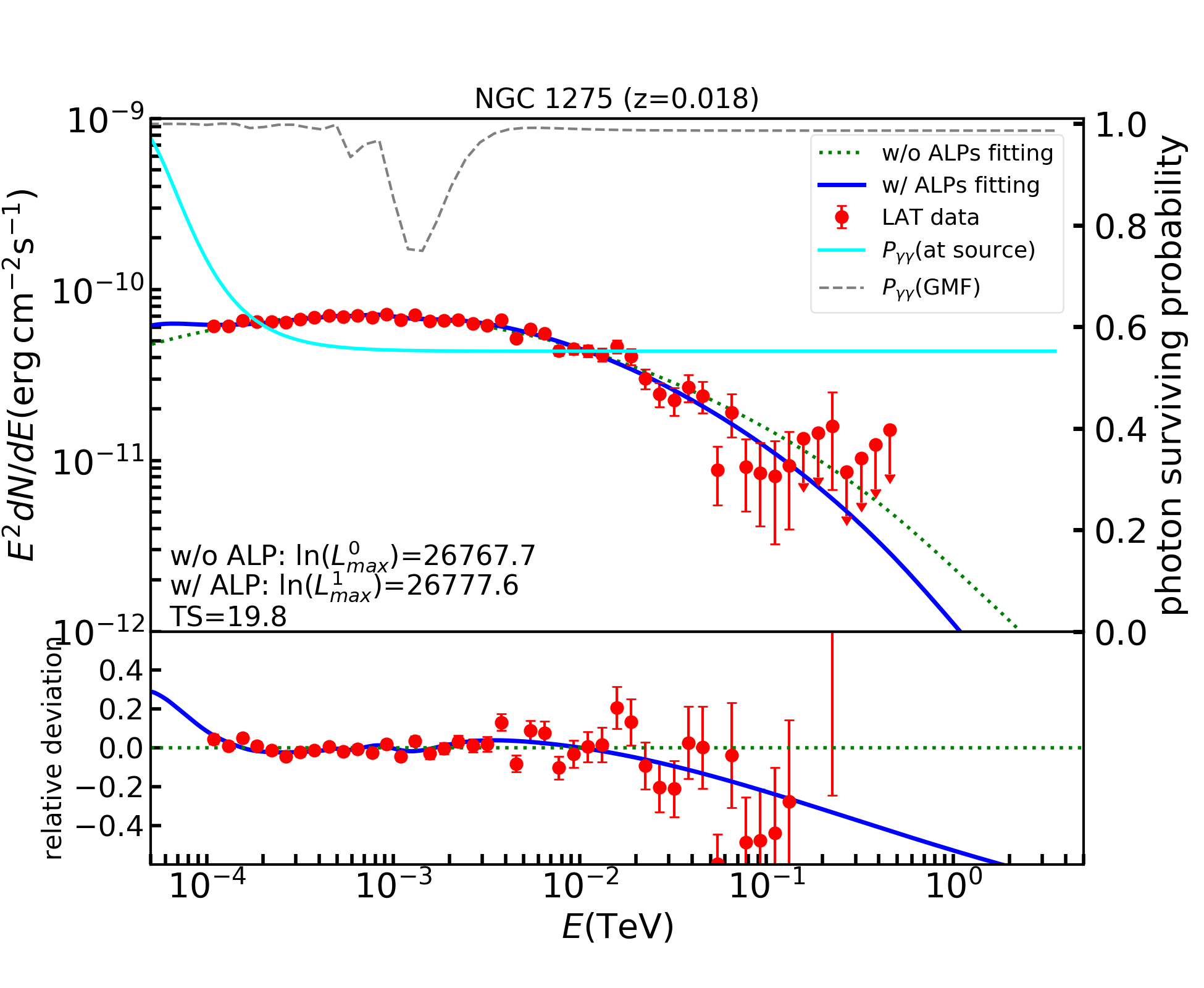}
\end{minipage}%
\begin{minipage}[t]{0.455\linewidth}
\centering
\includegraphics[width=0.9\textwidth]{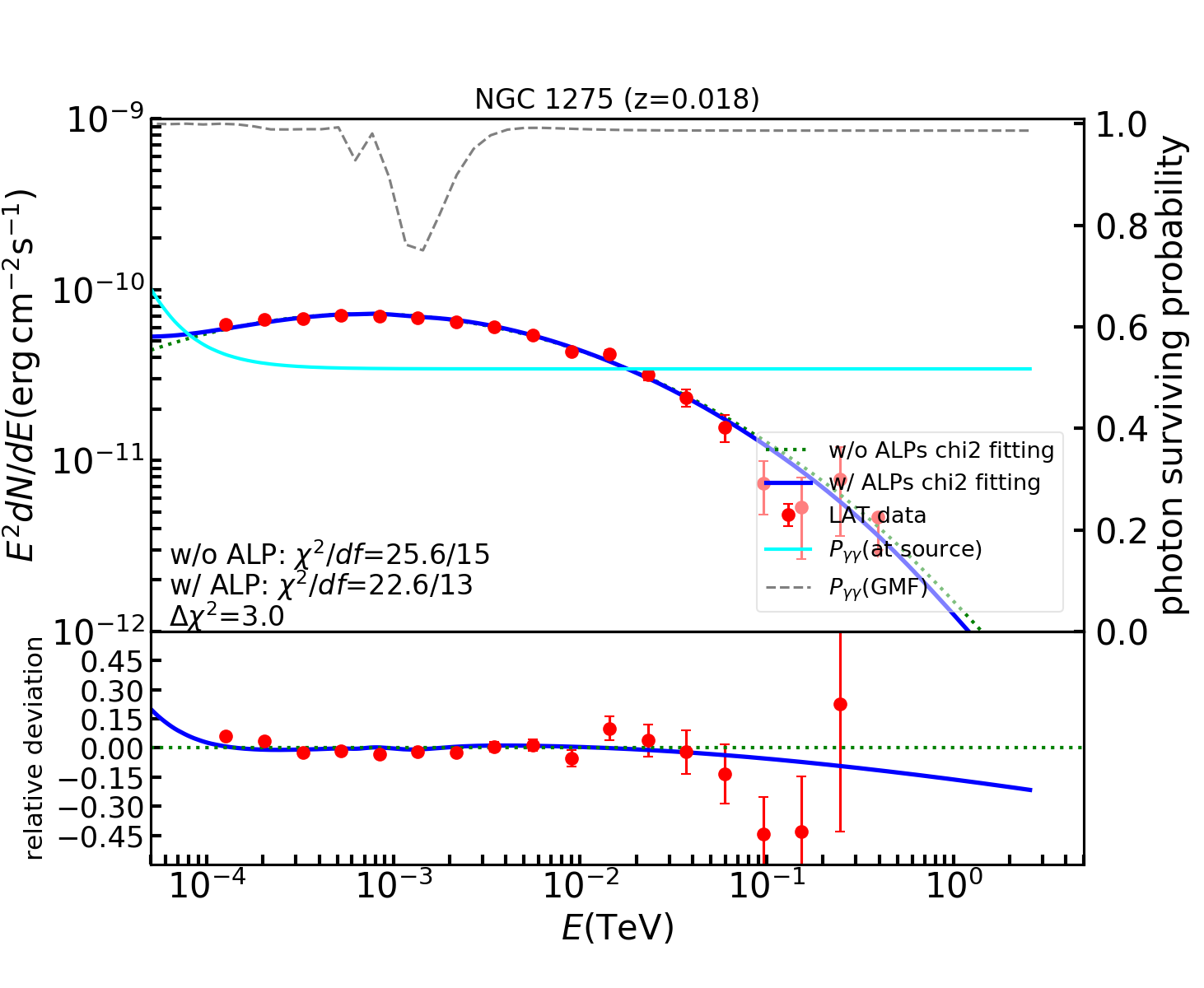}
\end{minipage}
\caption{NGC~1275, same as Fig.~\ref{fig:mkn421sed}.}\label{fig:ngc1275sed}
\end{figure}

\begin{figure}[ht!]
\begin{minipage}[t]{0.455\linewidth}
\centering
\includegraphics[width=0.9\textwidth]{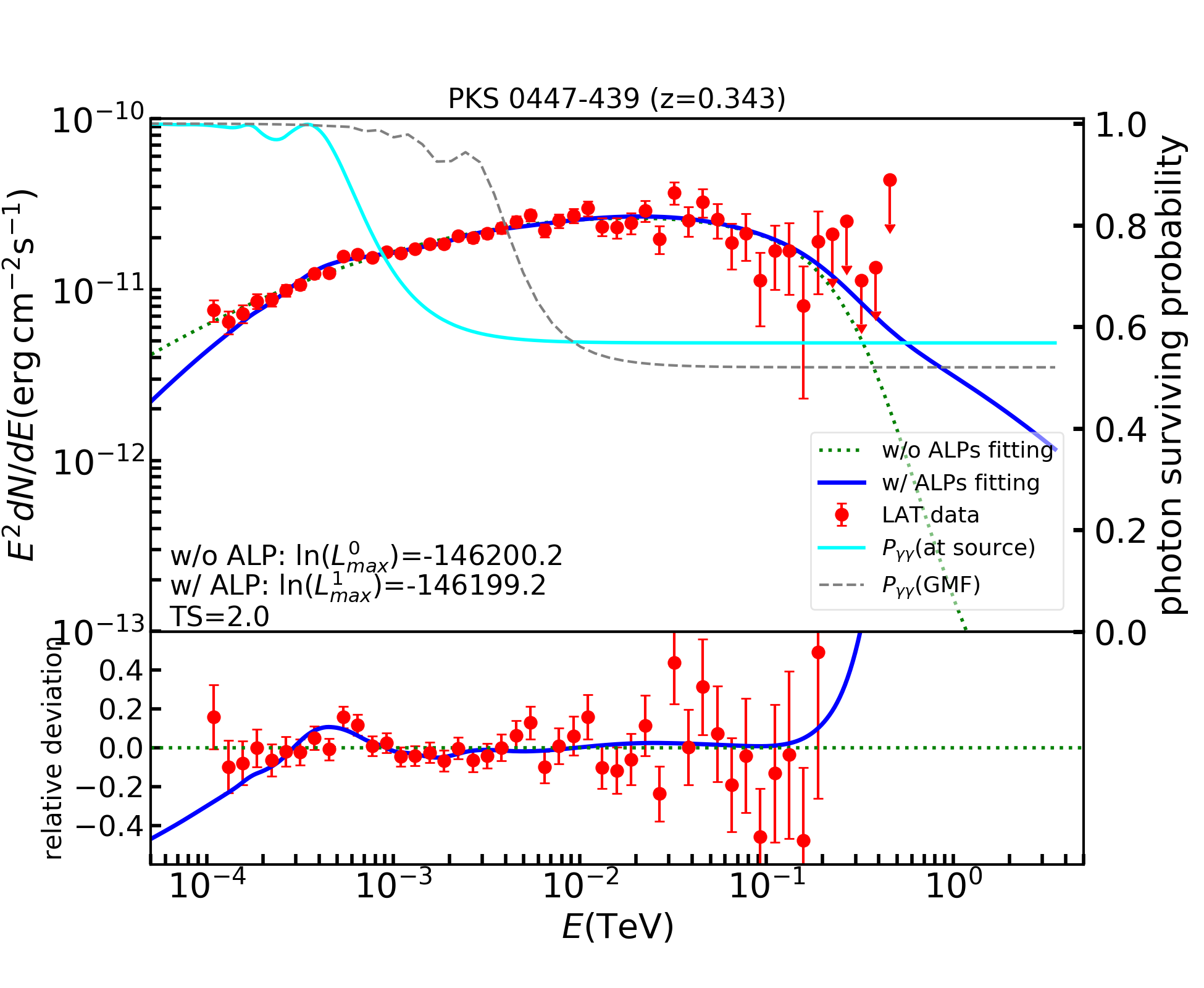}
\end{minipage}%
\begin{minipage}[t]{0.455\linewidth}
\centering
\includegraphics[width=0.9\textwidth]{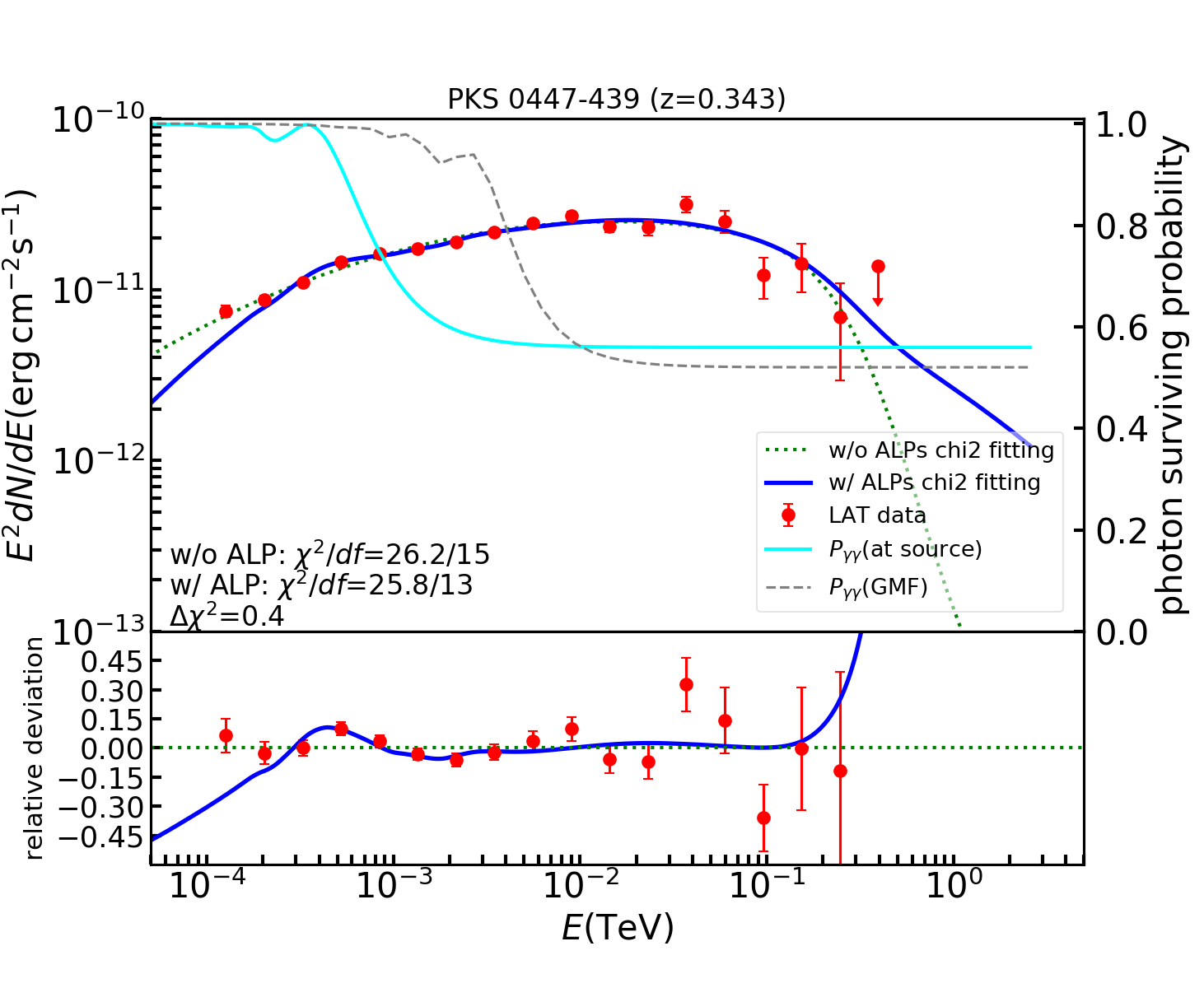}
\end{minipage}
\caption{PKS~0447-439, same as Fig.~\ref{fig:mkn421sed}.}\label{fig:pks0447sed}
\end{figure}

\begin{figure}[ht!]
\begin{minipage}[t]{0.455\linewidth}
\centering
\includegraphics[width=0.9\textwidth]{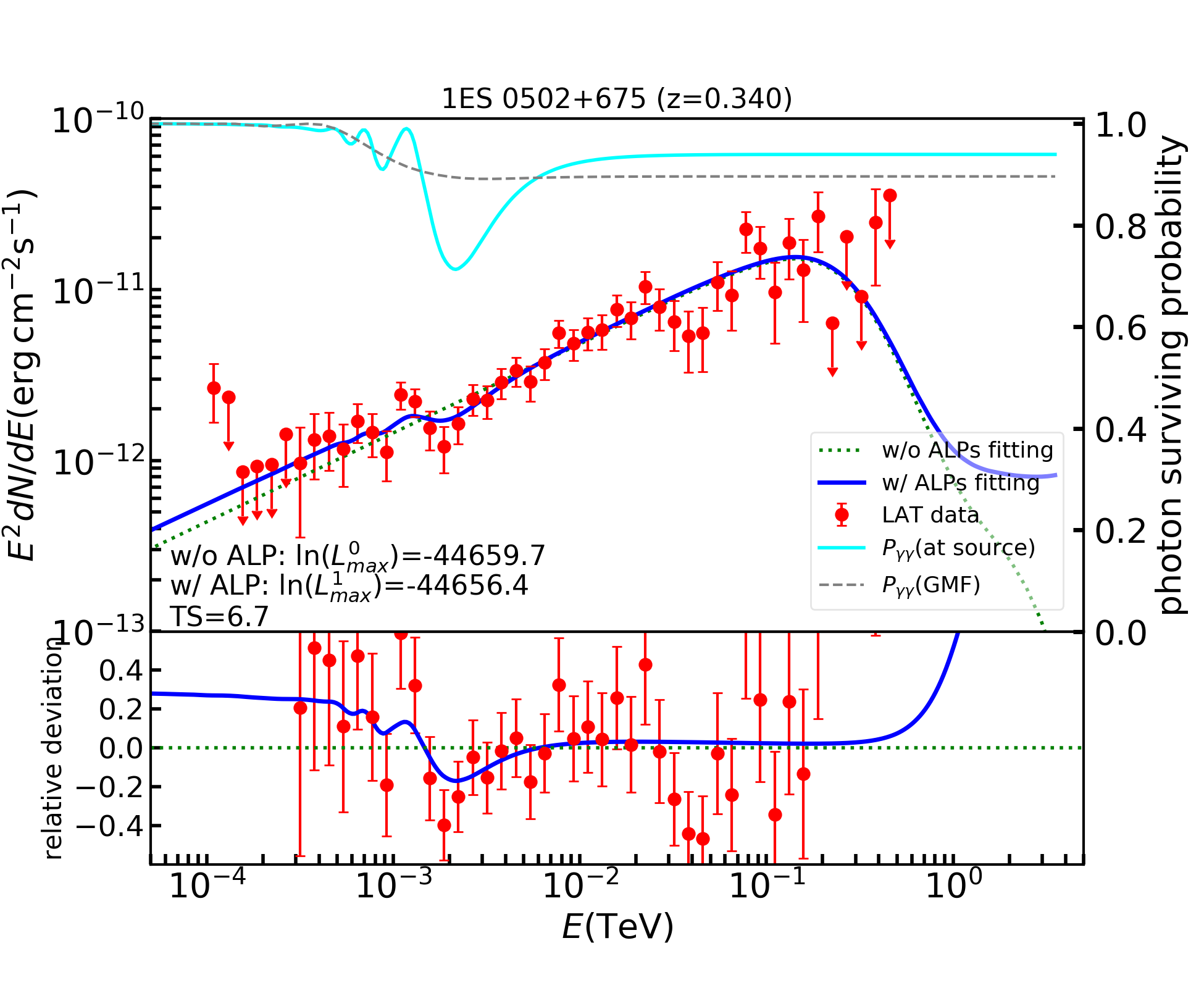}
\end{minipage}%
\begin{minipage}[t]{0.455\linewidth}
\centering
\includegraphics[width=0.9\textwidth]{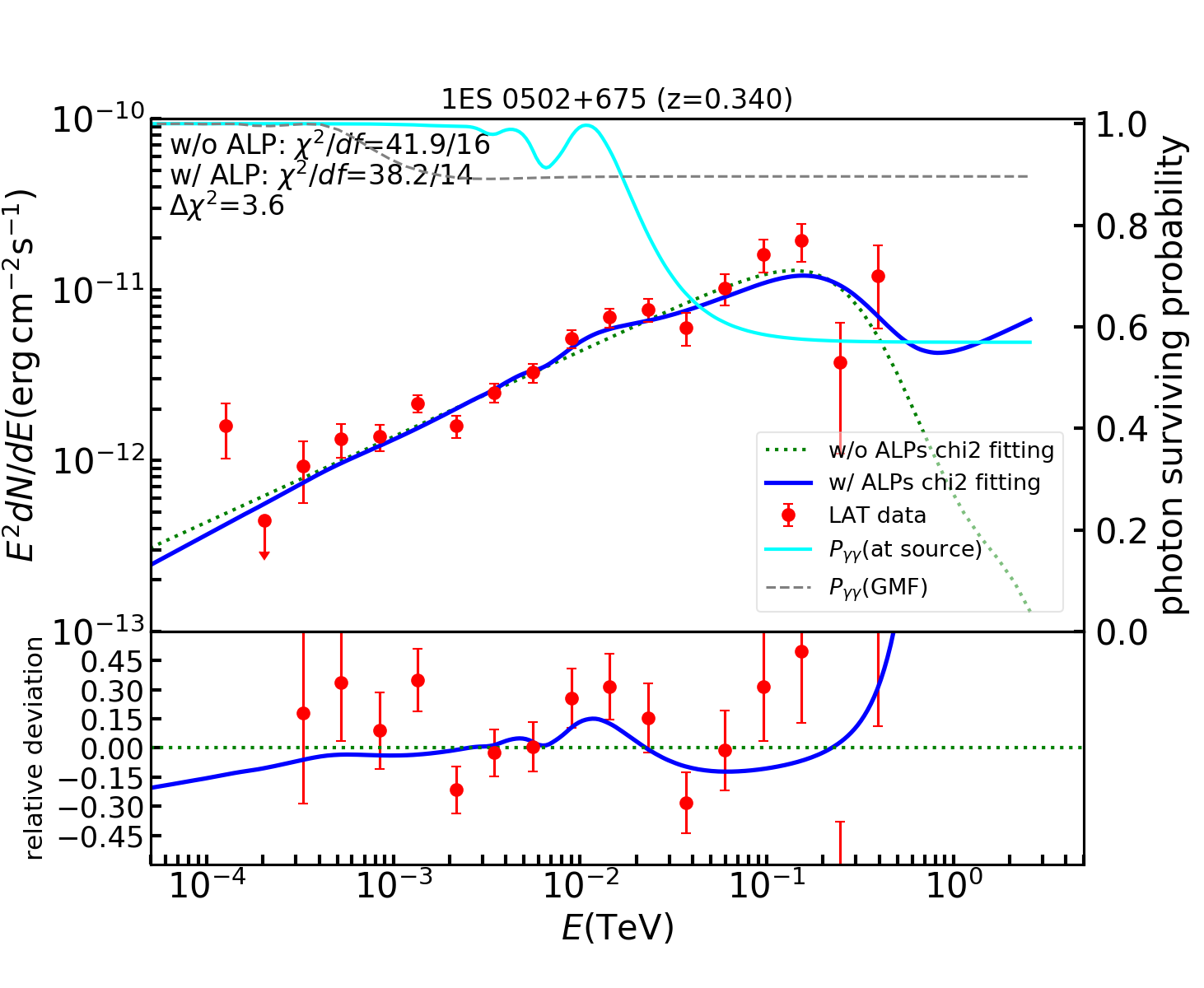}
\end{minipage}
\caption{1ES~0502+675, same as Fig.~\ref{fig:mkn421sed}.}\label{fig:es0502sed}
\end{figure}

\begin{figure}[ht!]
\begin{minipage}[t]{0.455\linewidth}
\centering
\includegraphics[width=0.9\textwidth]{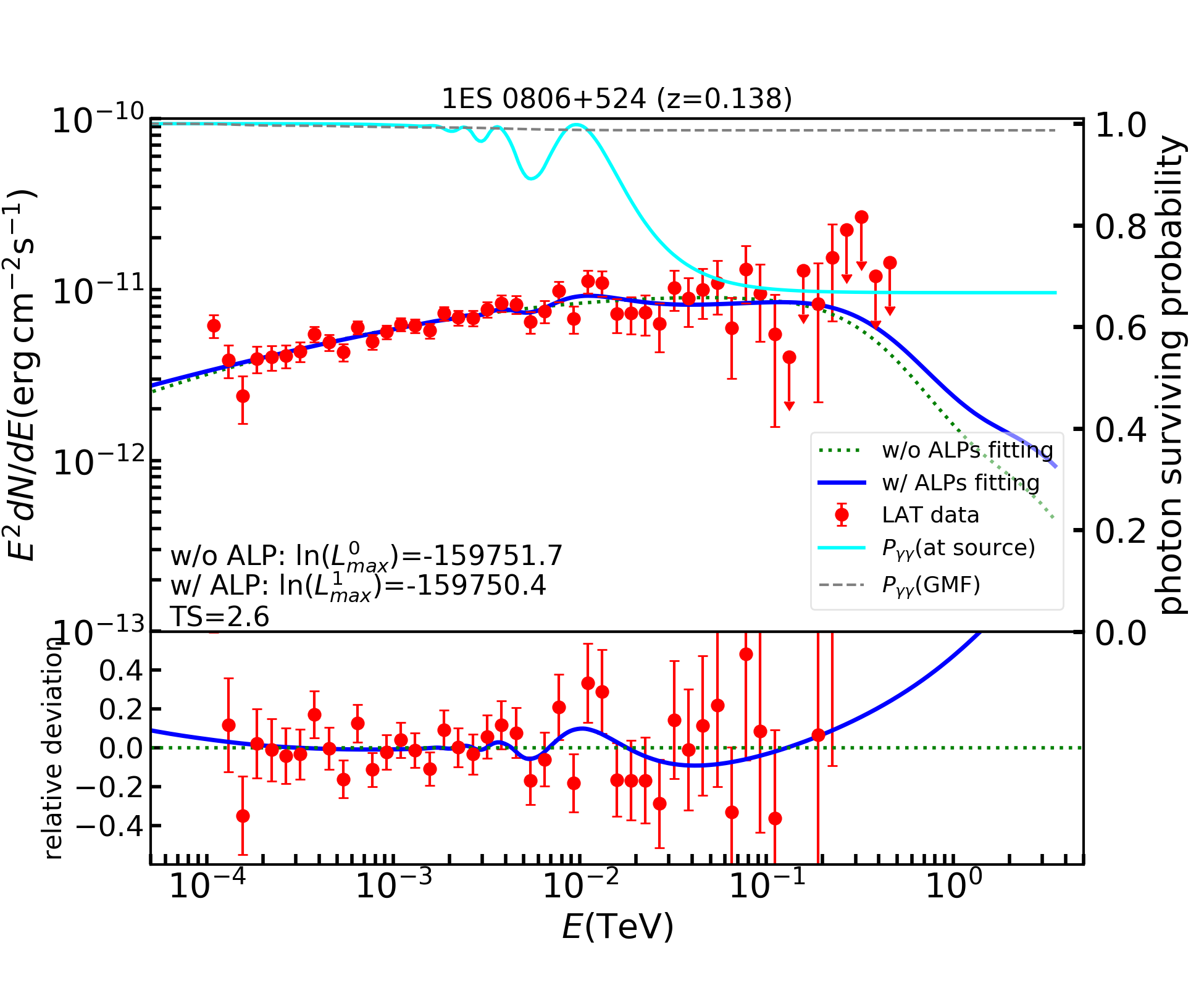}
\end{minipage}%
\begin{minipage}[t]{0.455\linewidth}
\centering
\includegraphics[width=0.9\textwidth]{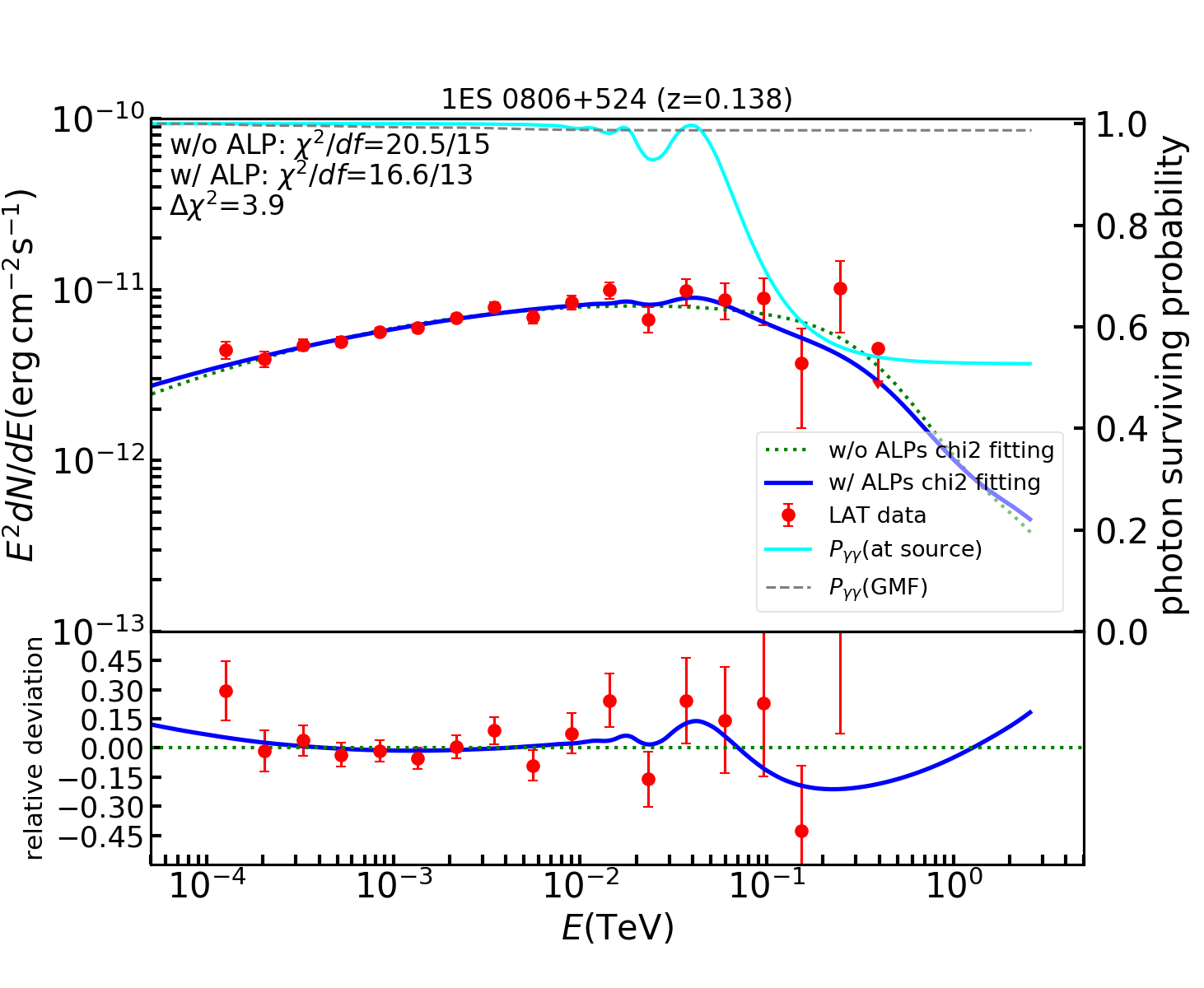}
\end{minipage}
\caption{1ES~0806+524, same as Fig.~\ref{fig:mkn421sed}.}\label{fig:es0806sed}
\end{figure}

\begin{figure}[ht!]
\begin{minipage}[t]{0.455\linewidth}
\centering
\includegraphics[width=0.9\textwidth]{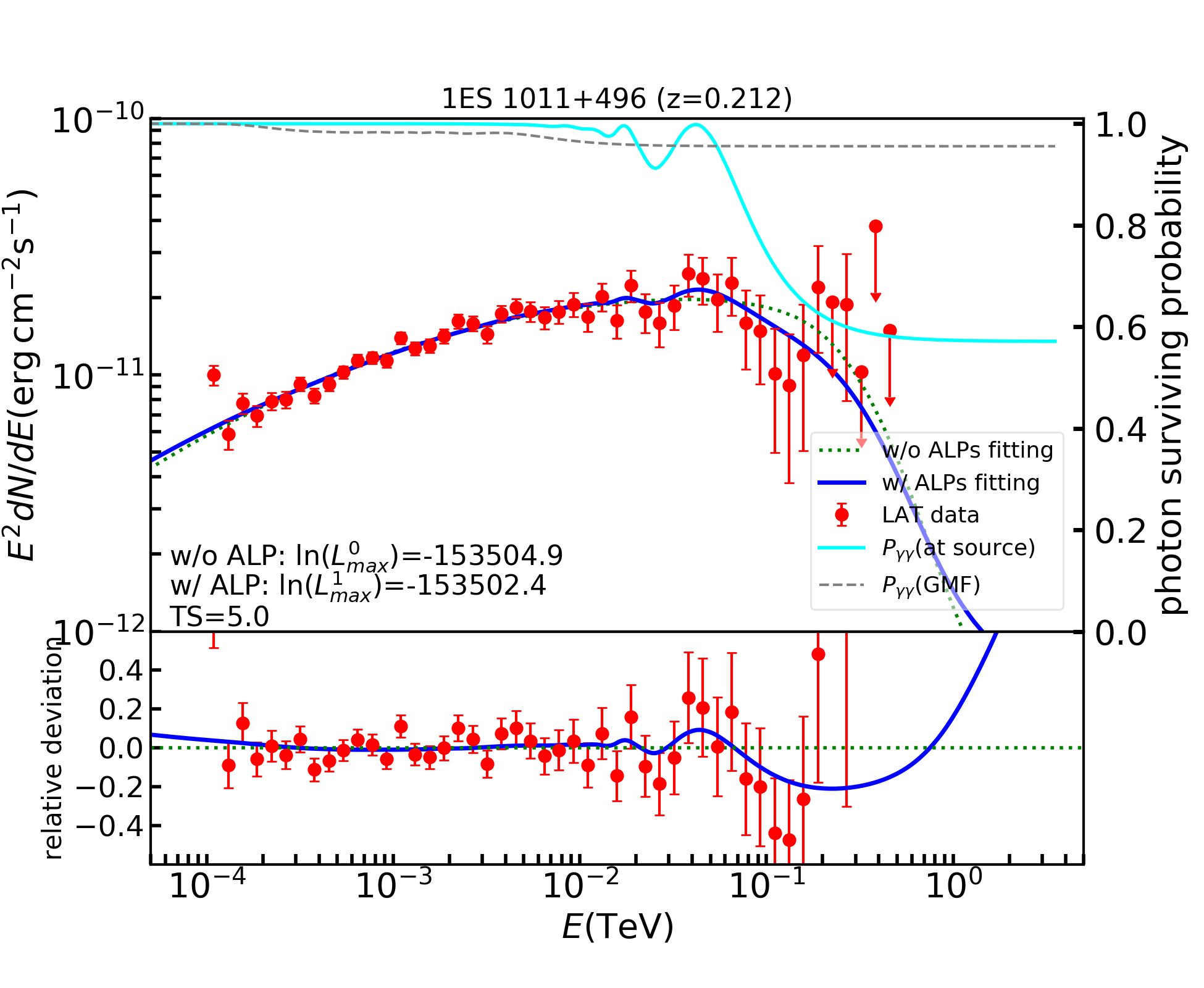}
\end{minipage}%
\begin{minipage}[t]{0.455\linewidth}
\centering
\includegraphics[width=0.9\textwidth]{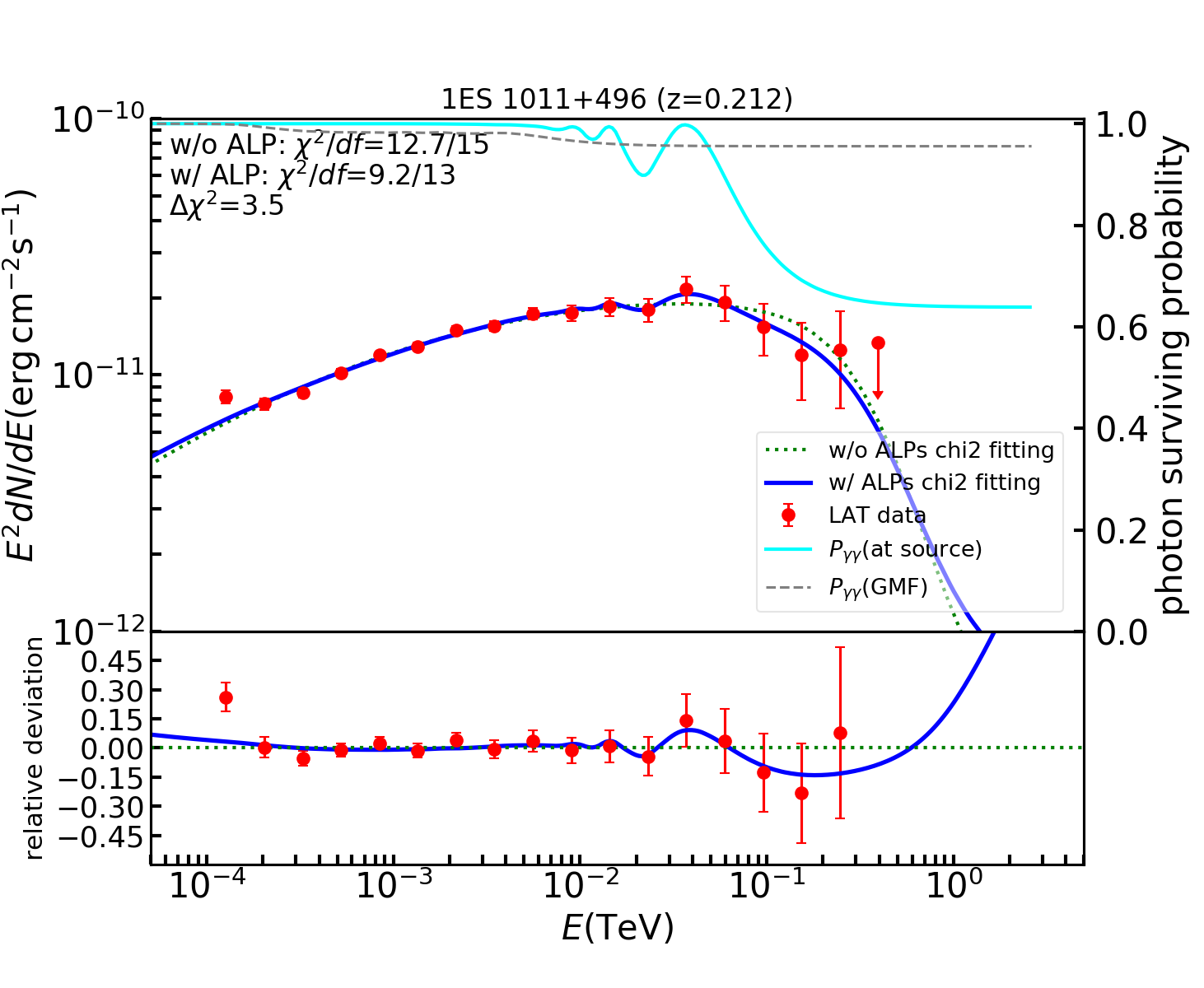}
\end{minipage}
\caption{1ES~1011+496, same as Fig.~\ref{fig:mkn421sed}.}\label{fig:es1011sed}
\end{figure}

\begin{figure}[ht!]
\begin{minipage}[t]{0.455\linewidth}
\centering
\includegraphics[width=0.9\textwidth]{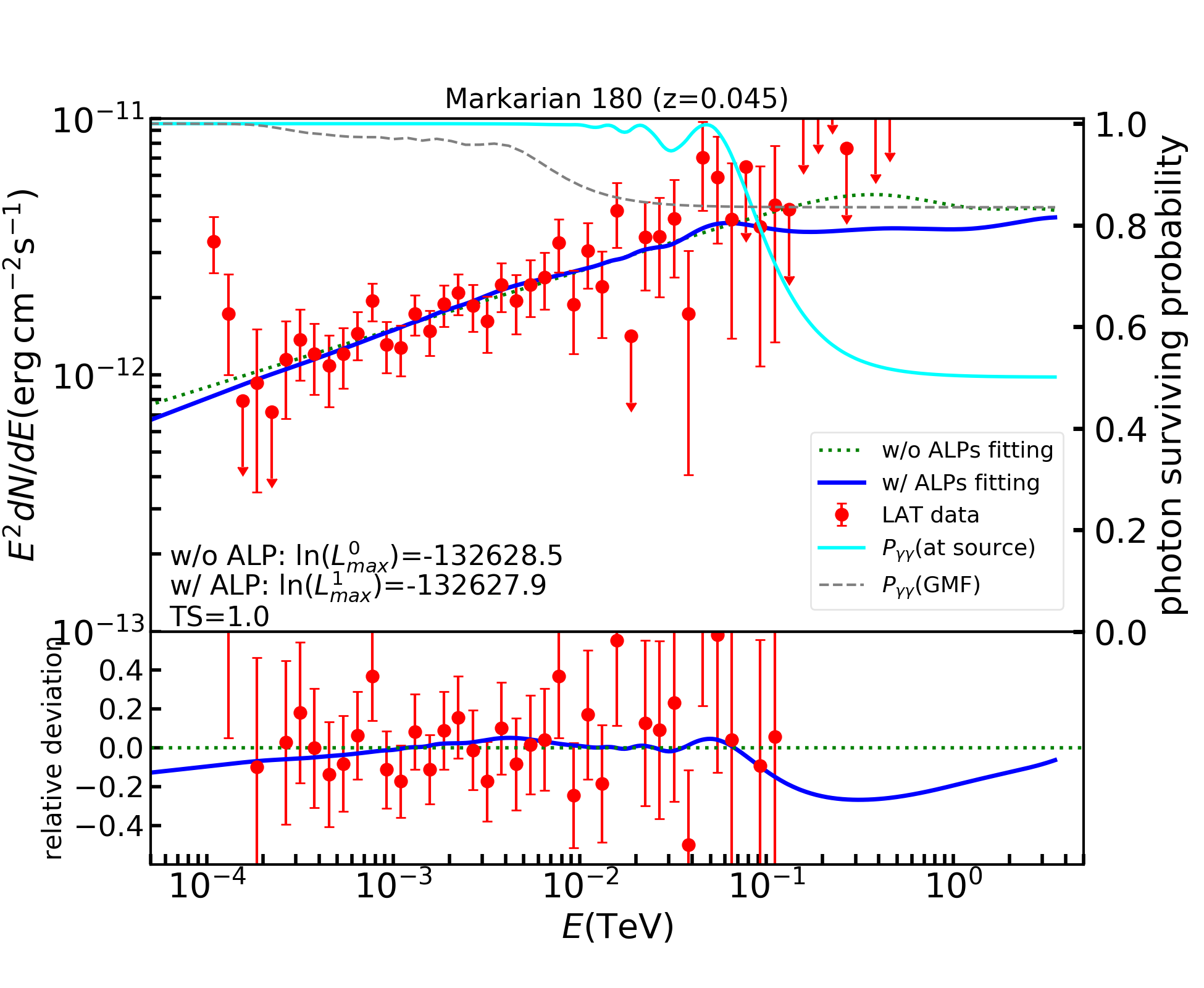}
\end{minipage}%
\begin{minipage}[t]{0.455\linewidth}
\centering
\includegraphics[width=0.9\textwidth]{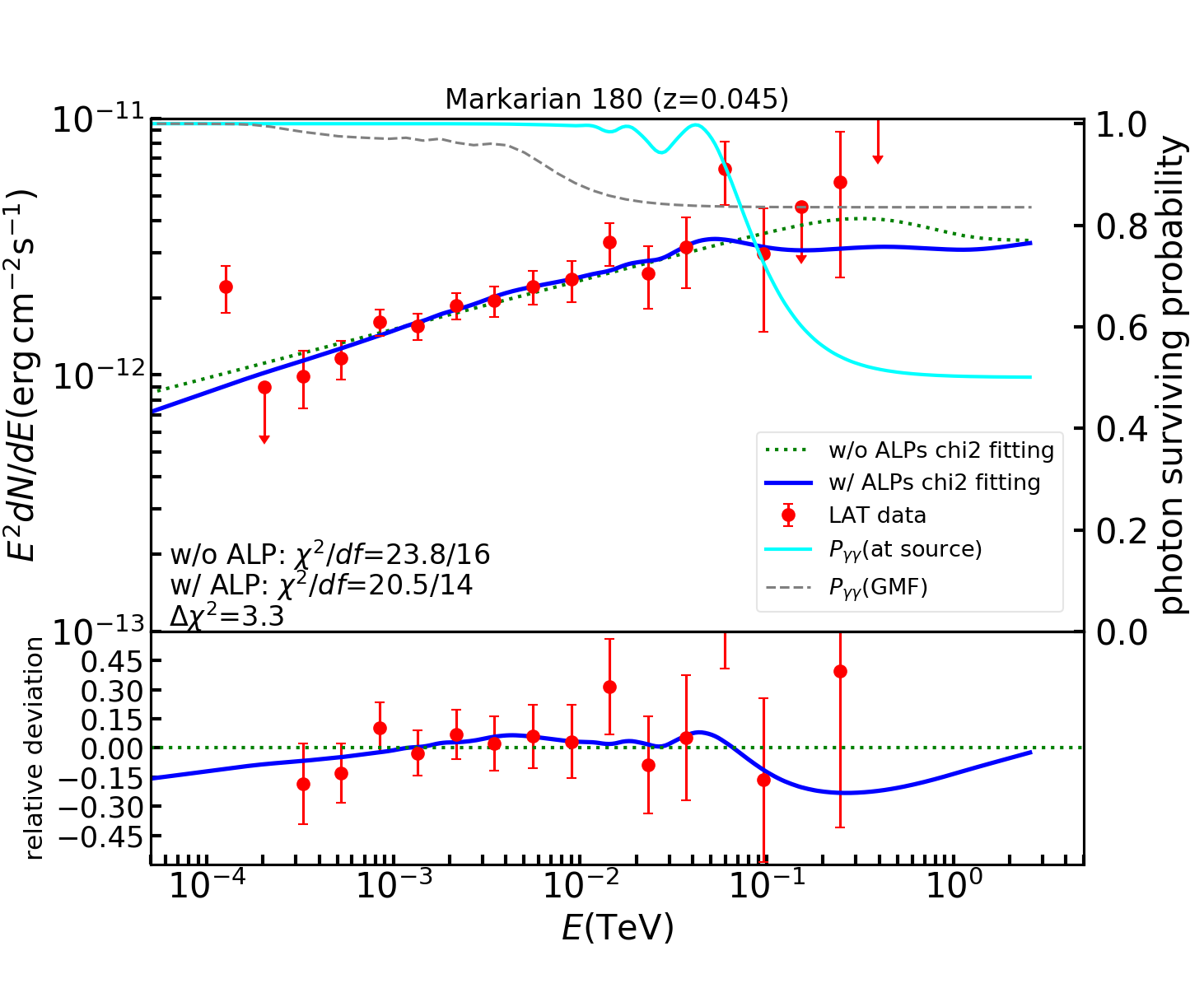}
\end{minipage}
\caption{Markarian~180, same as Fig.~\ref{fig:mkn421sed}.}\label{fig:mkn180sed}
\end{figure}

\begin{figure}[ht!]
\begin{minipage}[t]{0.455\linewidth}
\centering
\includegraphics[width=0.9\textwidth]{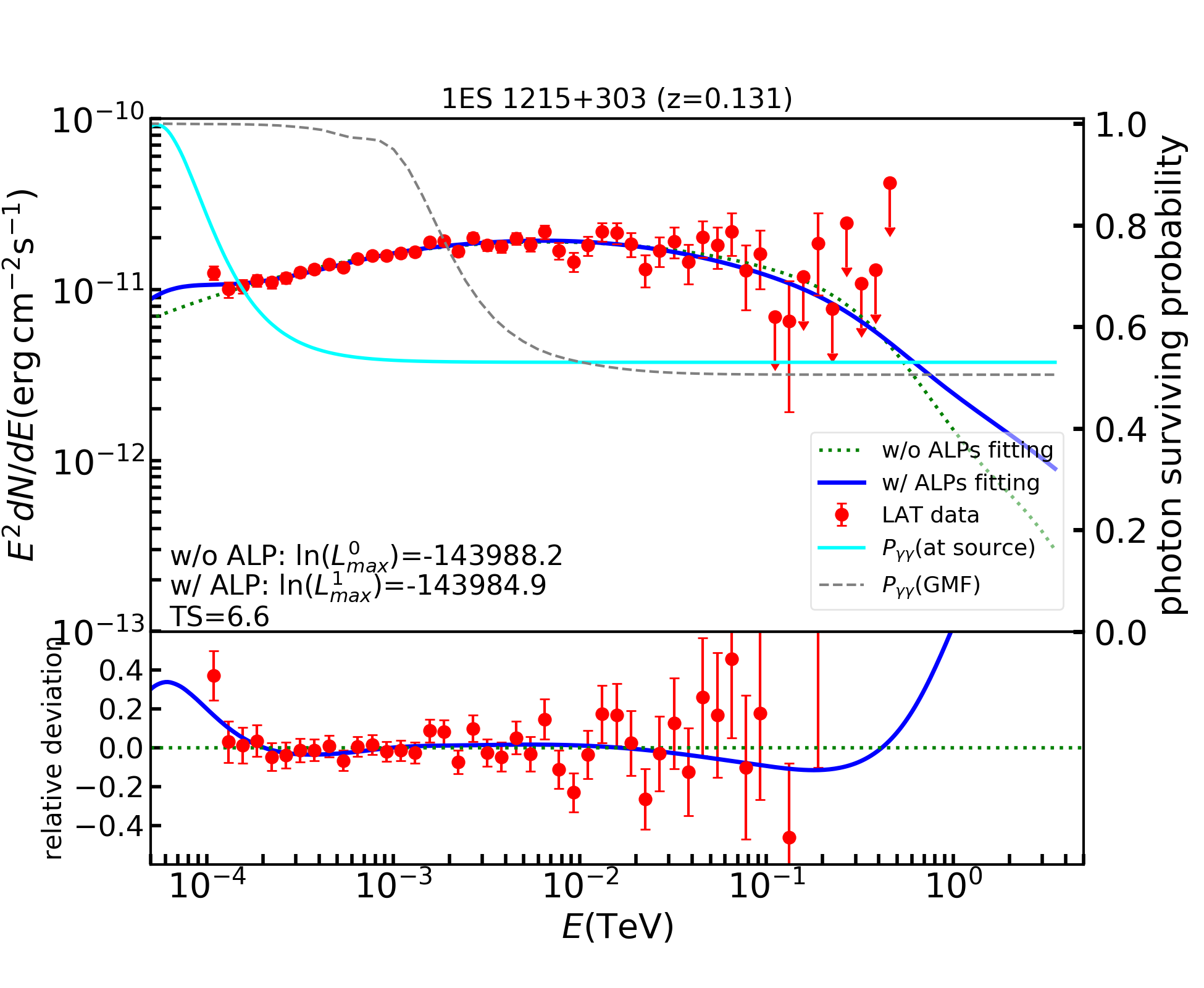}
\end{minipage}%
\begin{minipage}[t]{0.455\linewidth}
\centering
\includegraphics[width=0.9\textwidth]{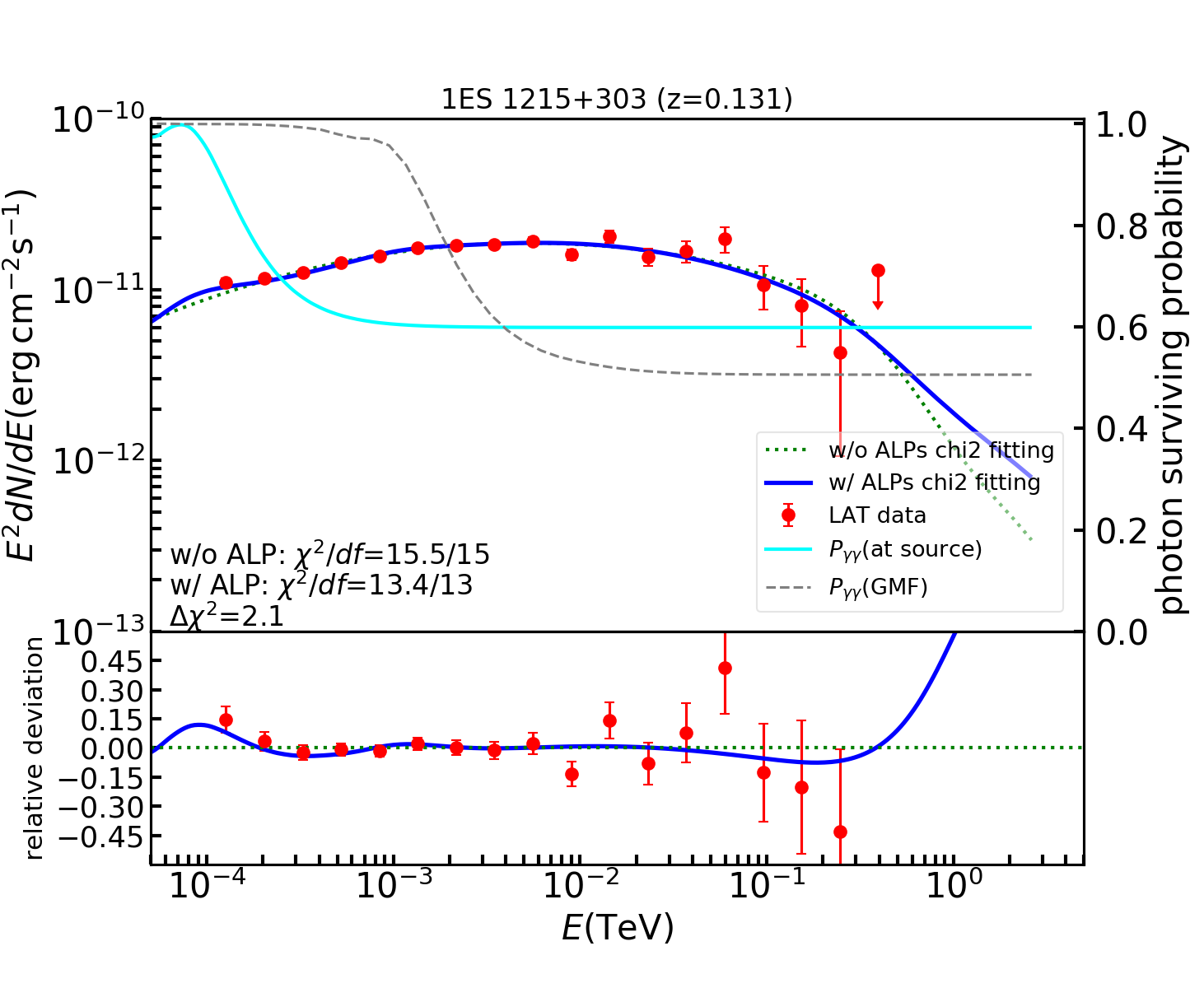}
\end{minipage}
\caption{1ES~1215+303, same as Fig.~\ref{fig:mkn421sed}.}\label{fig:es1215sed}
\end{figure}

\begin{figure}[ht!]
\begin{minipage}[t]{0.455\linewidth}
\centering
\includegraphics[width=0.9\textwidth]{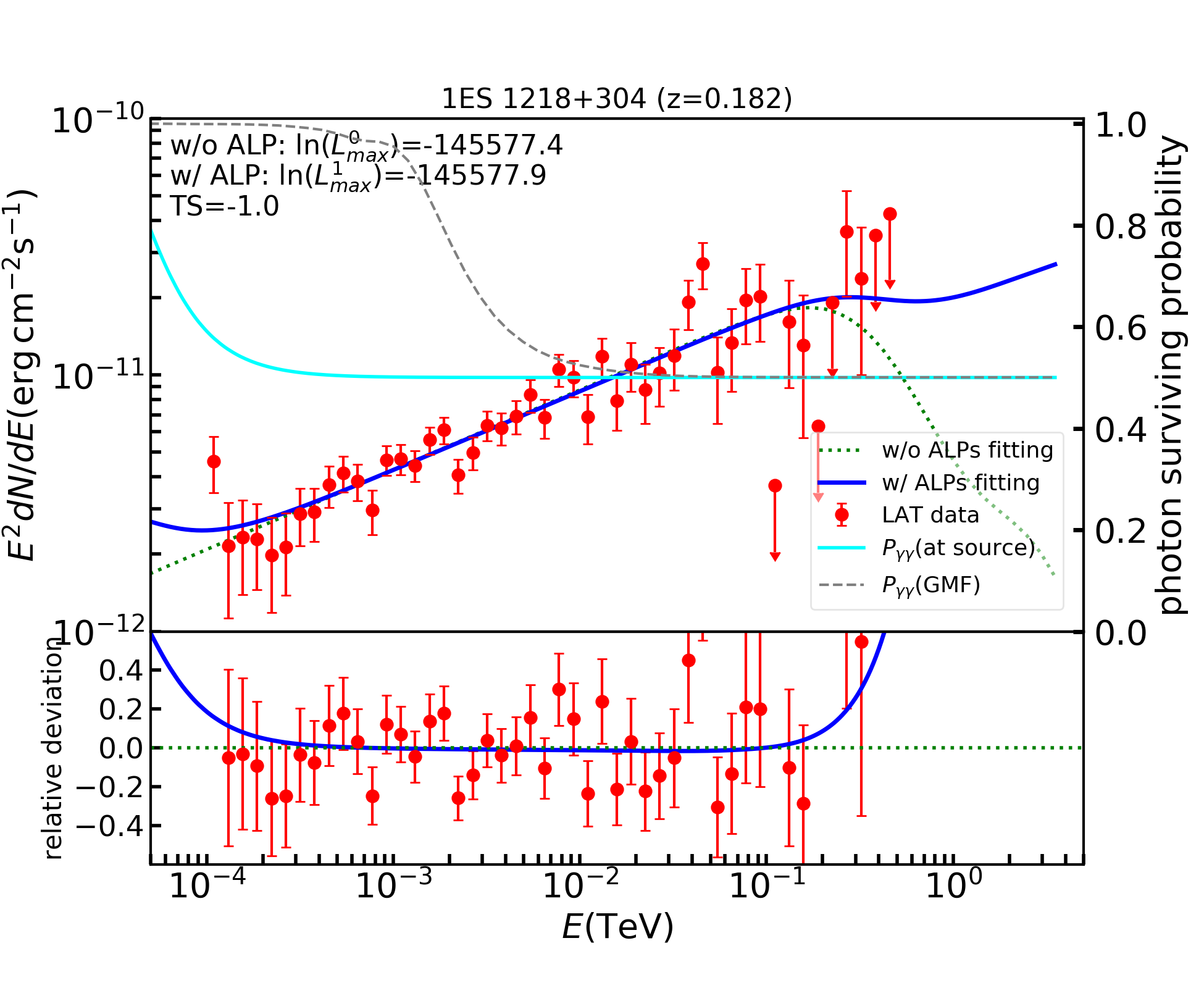}
\end{minipage}%
\begin{minipage}[t]{0.455\linewidth}
\centering
\includegraphics[width=0.9\textwidth]{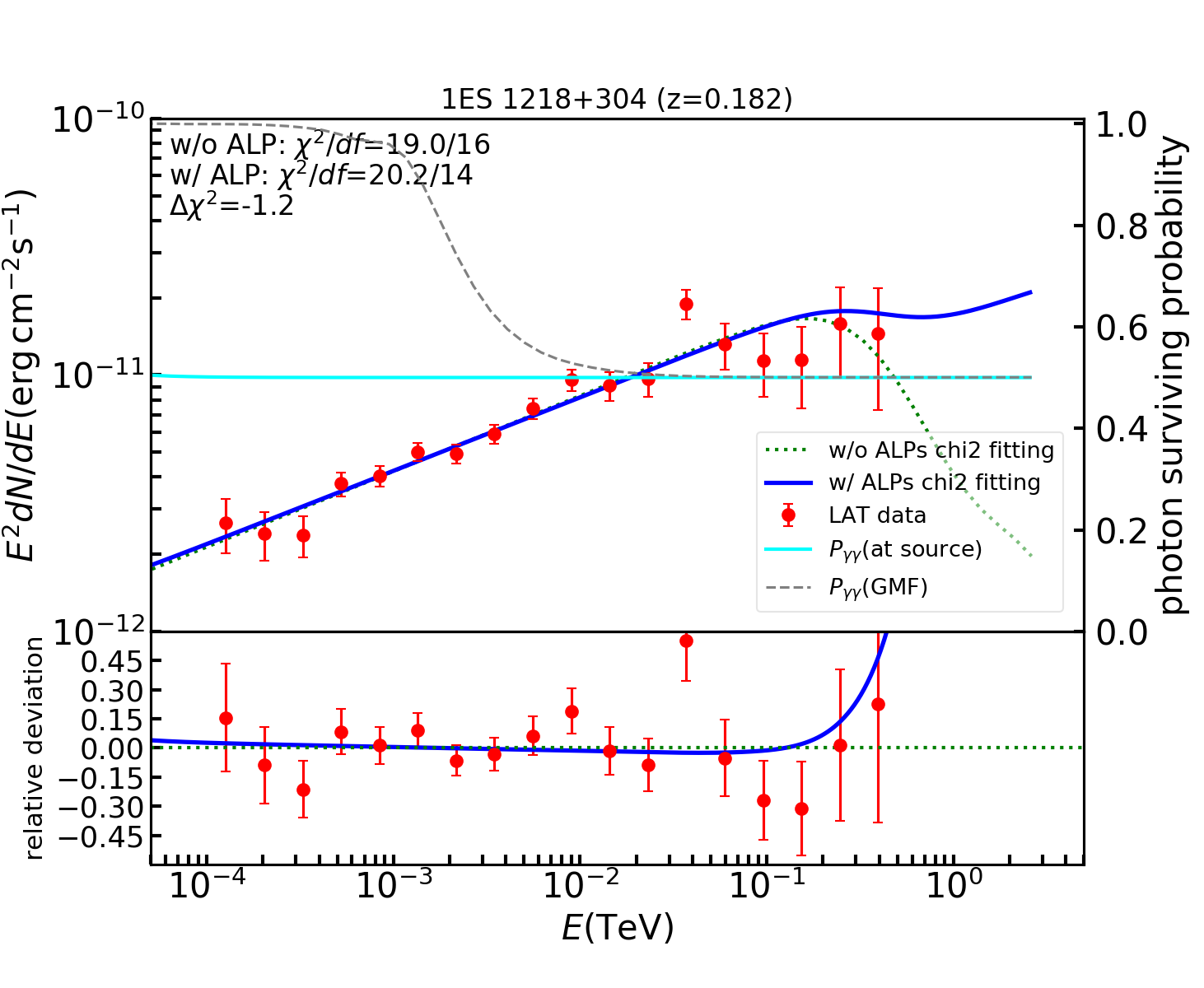}
\end{minipage}
\caption{1ES~1218+304, same as Fig.~\ref{fig:mkn421sed}.}\label{fig:es1218sed}
\end{figure}

\begin{figure}[ht!]
\begin{minipage}[t]{0.455\linewidth}
\centering
\includegraphics[width=0.9\textwidth]{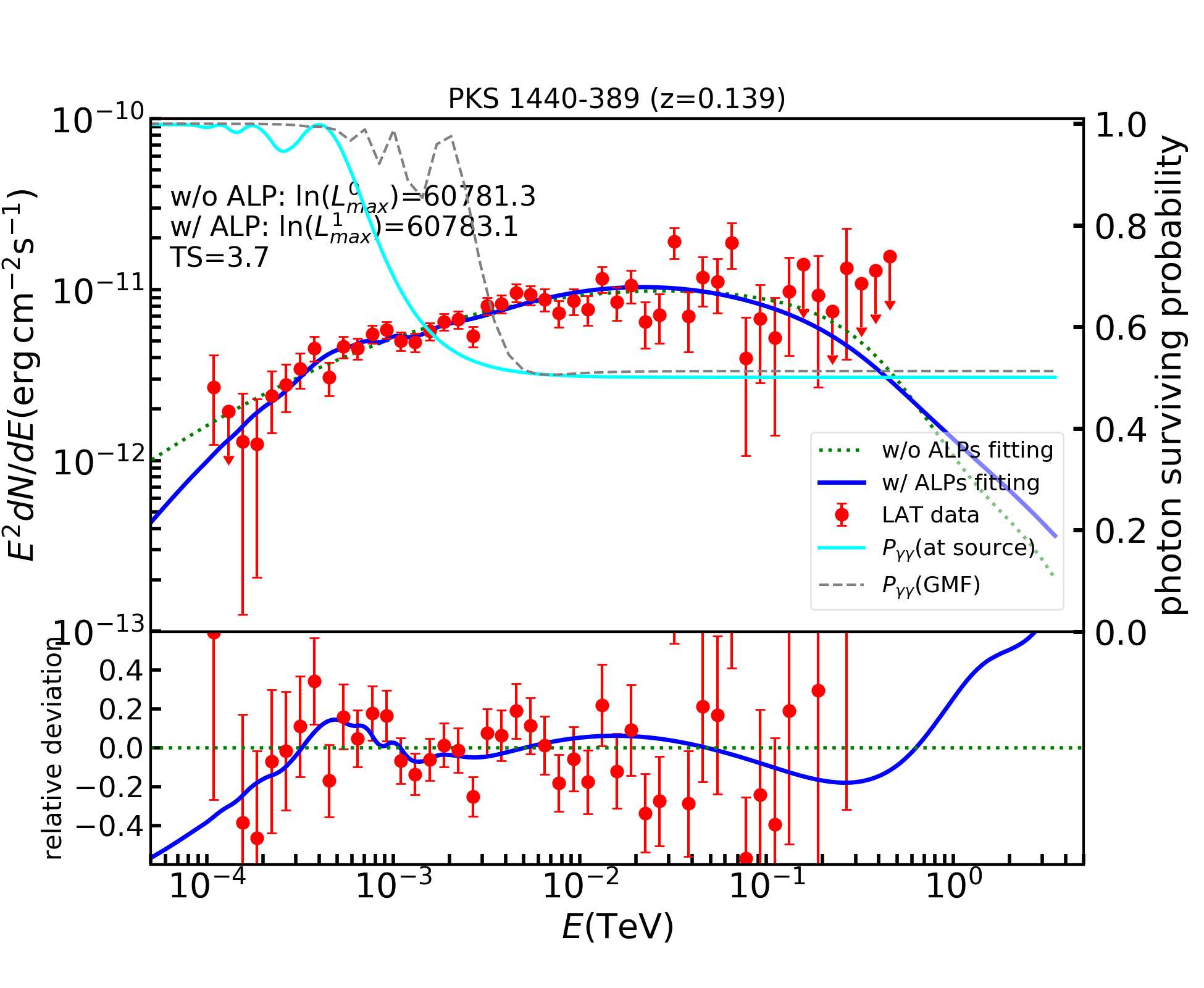}
\end{minipage}%
\begin{minipage}[t]{0.455\linewidth}
\centering
\includegraphics[width=0.9\textwidth]{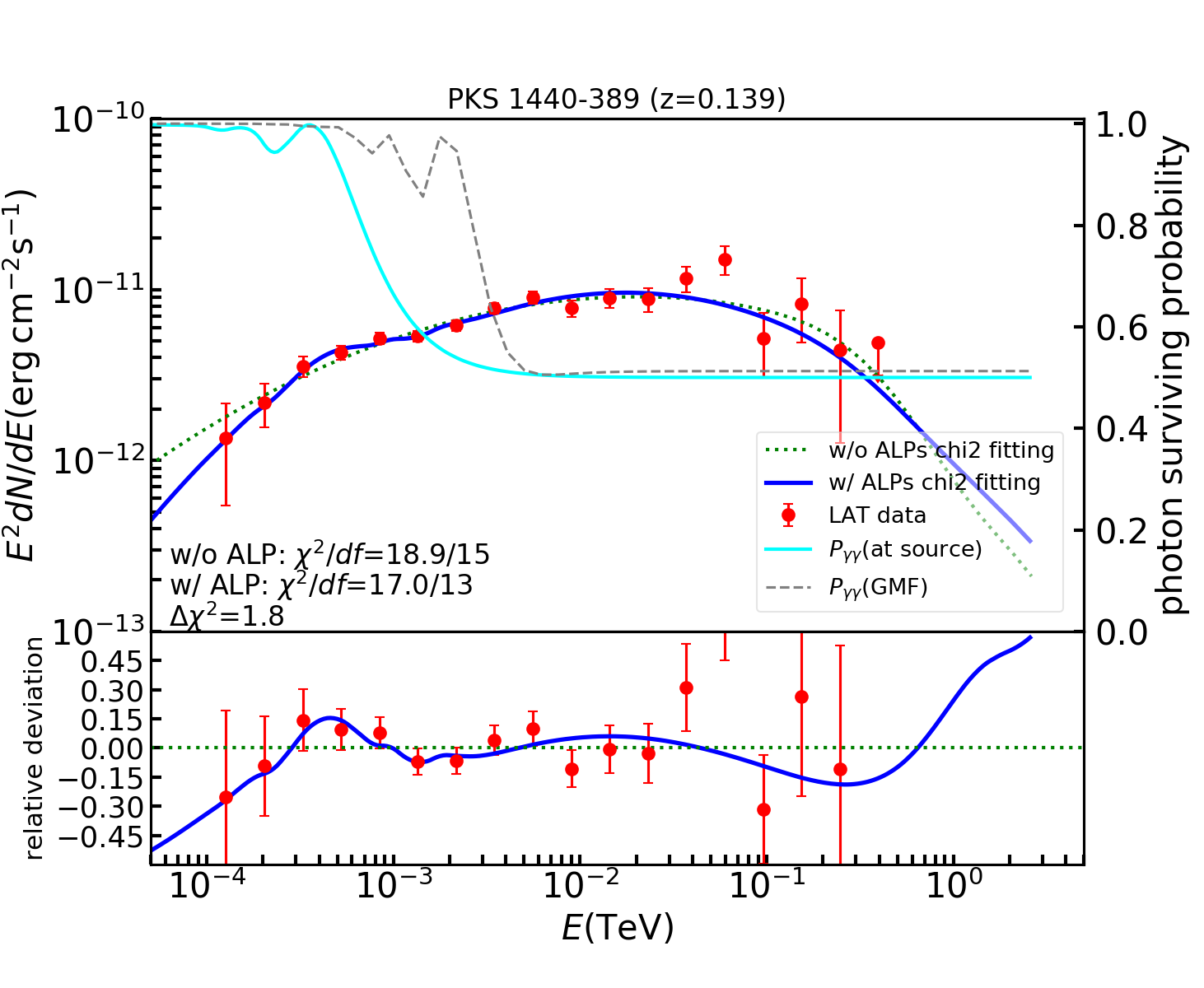}
\end{minipage}
\caption{PKS~1440-389, same as Fig.~\ref{fig:mkn421sed}.}\label{fig:pks1440sed}
\end{figure}

\begin{figure}[ht!]
\begin{minipage}[t]{0.455\linewidth}
\centering
\includegraphics[width=0.9\textwidth]{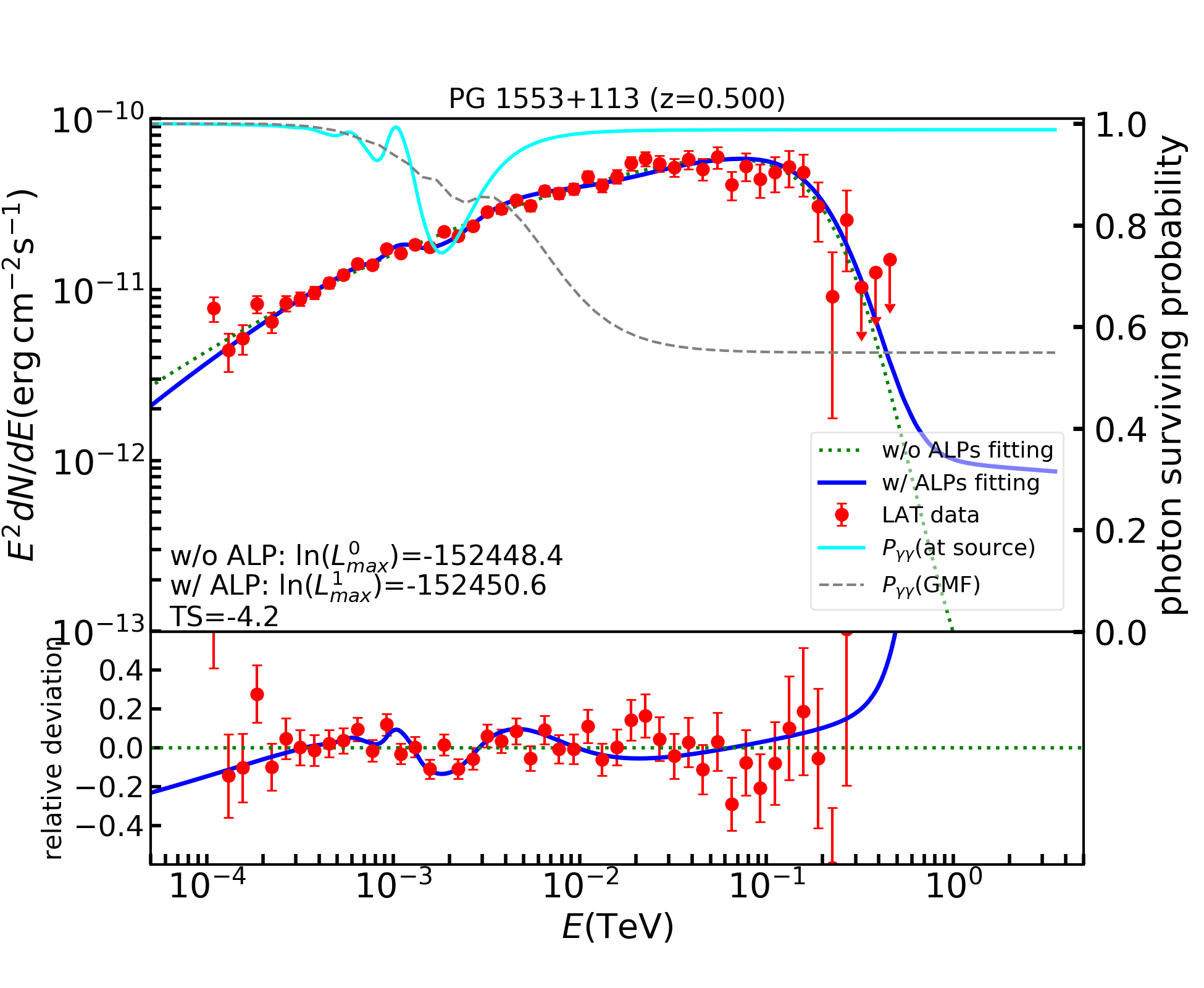}
\end{minipage}%
\begin{minipage}[t]{0.455\linewidth}
\centering
\includegraphics[width=0.9\textwidth]{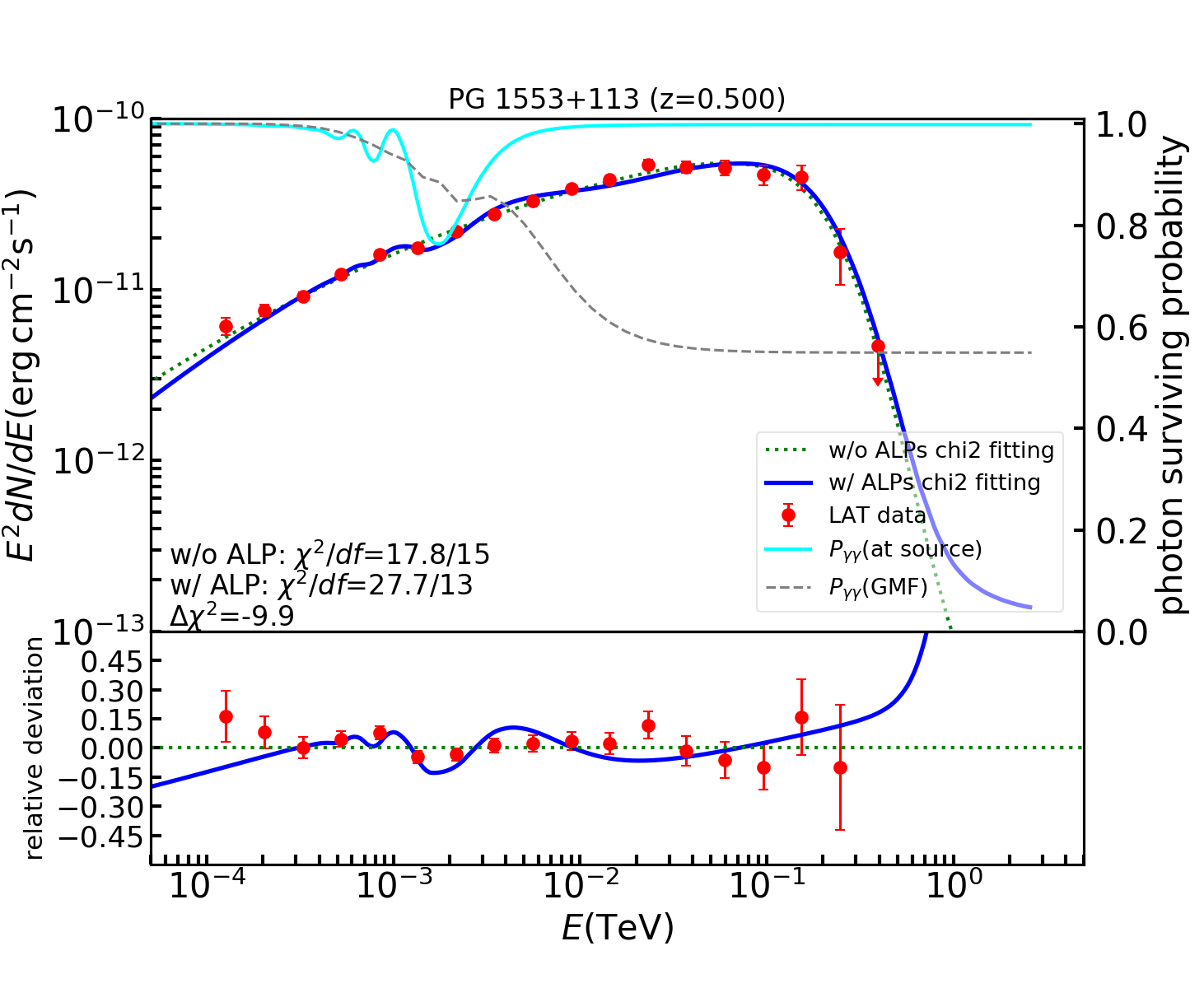}
\end{minipage}
\caption{PG~1553+113, same as Fig.~\ref{fig:mkn421sed}.}\label{fig:pg1553sed}
\end{figure}

\begin{figure}[ht!]
\begin{minipage}[t]{0.455\linewidth}
\centering
\includegraphics[width=0.9\textwidth]{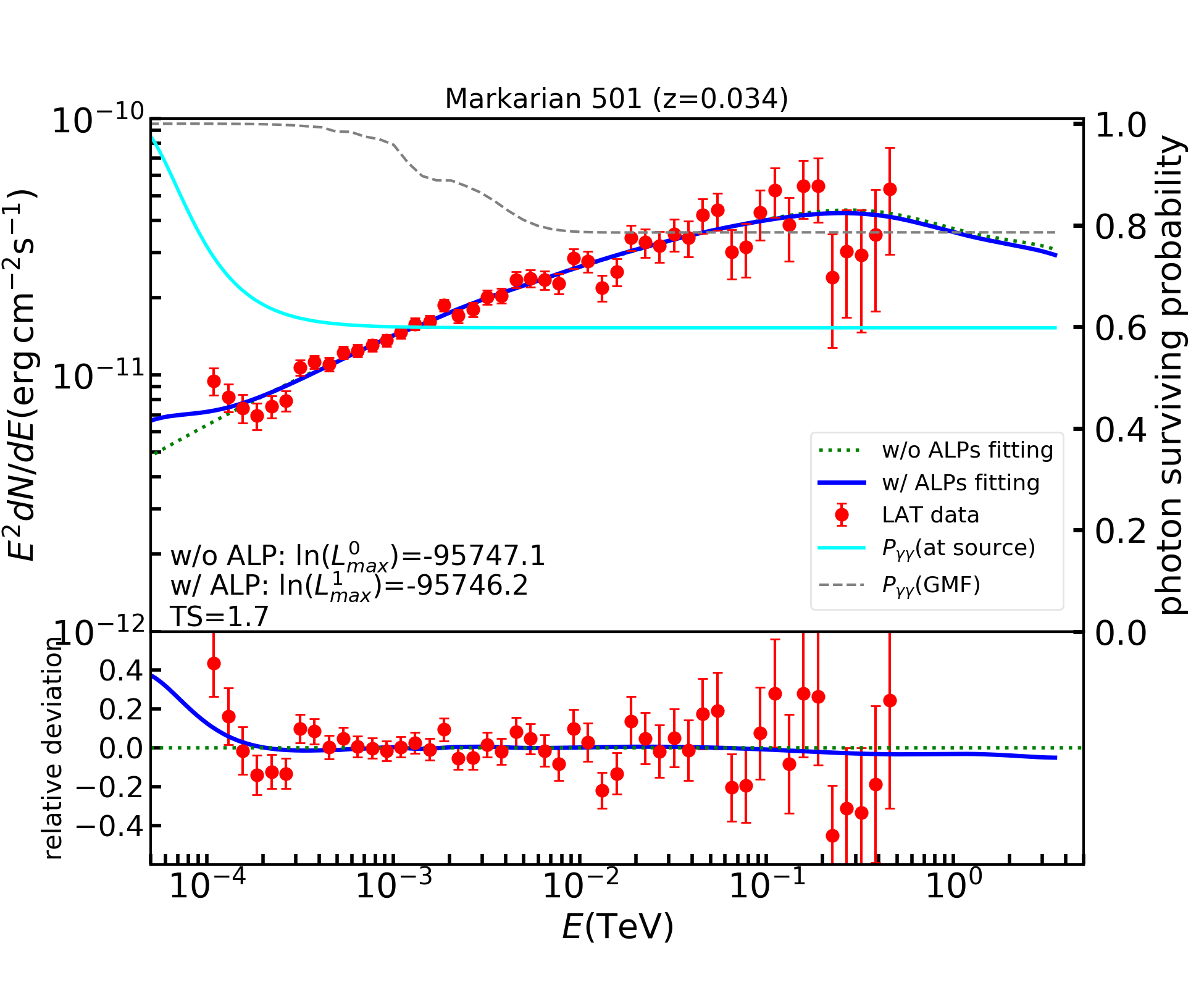}
\end{minipage}%
\begin{minipage}[t]{0.455\linewidth}
\centering
\includegraphics[width=0.9\textwidth]{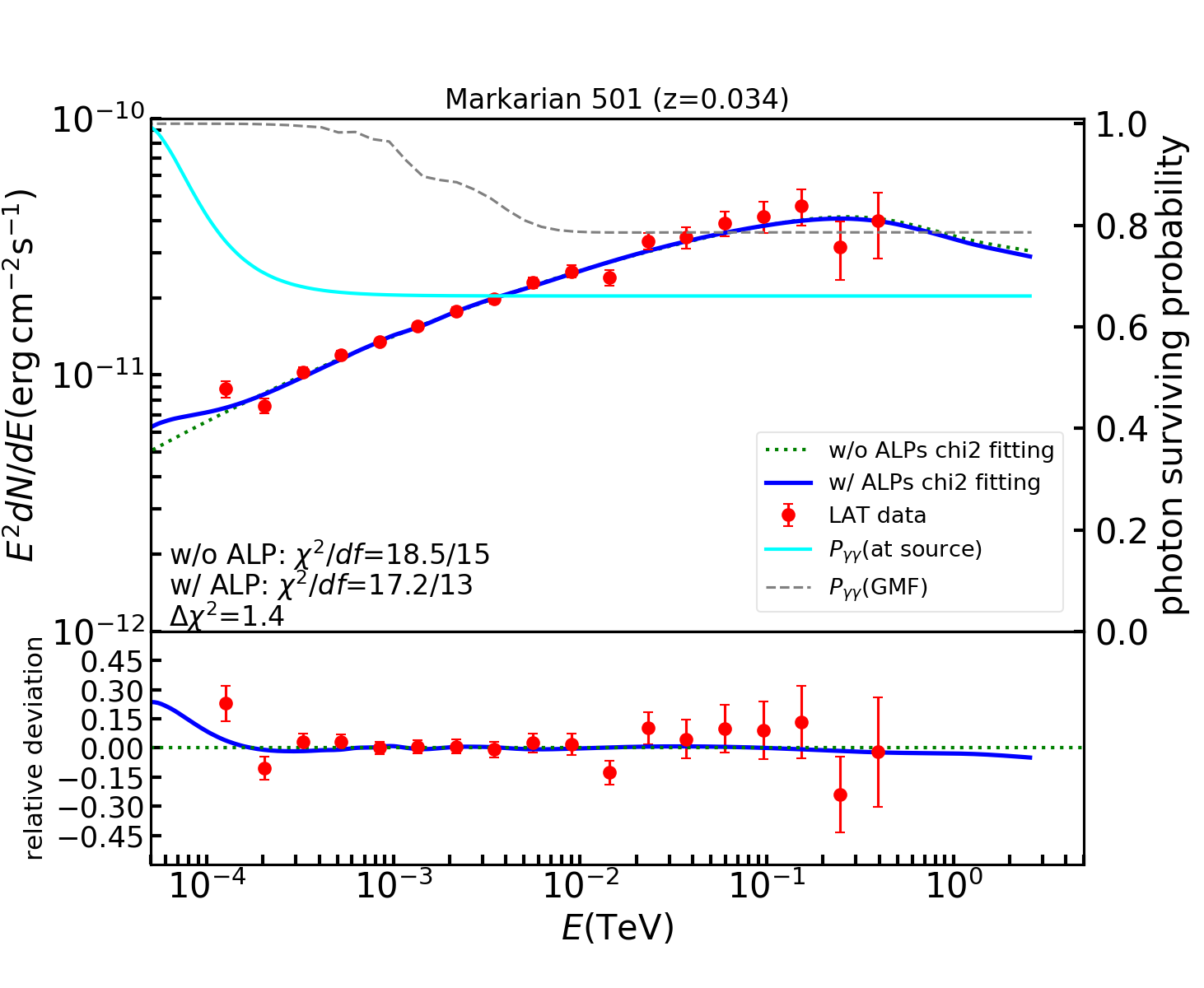}
\end{minipage}
\caption{Markarian~501, same as Fig.~\ref{fig:mkn421sed}.}\label{fig:mkn501sed}
\end{figure}

\begin{figure}[ht!]
\begin{minipage}[t]{0.455\linewidth}
\centering
\includegraphics[width=0.9\textwidth]{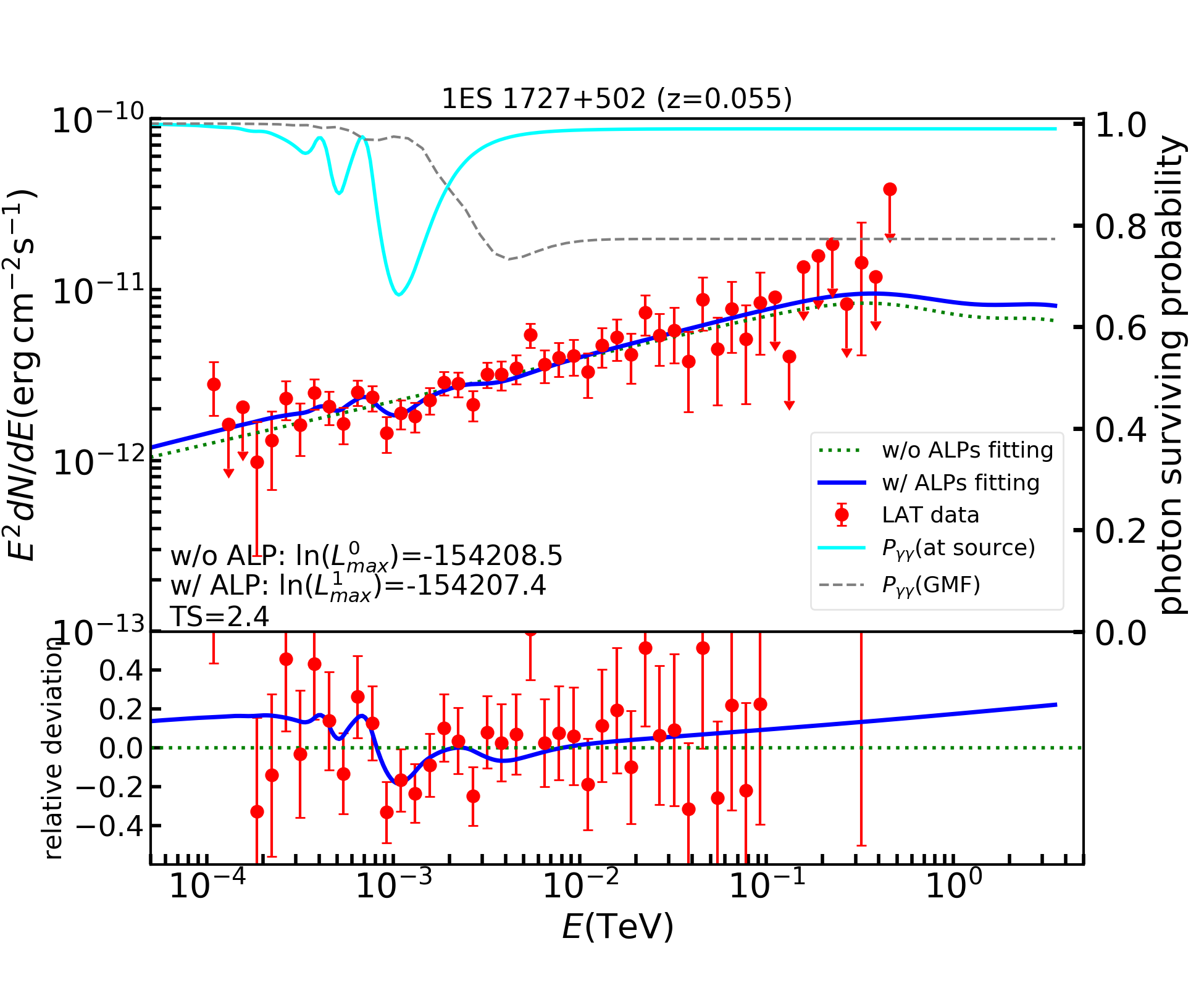}
\end{minipage}%
\begin{minipage}[t]{0.455\linewidth}
\centering
\includegraphics[width=0.9\textwidth]{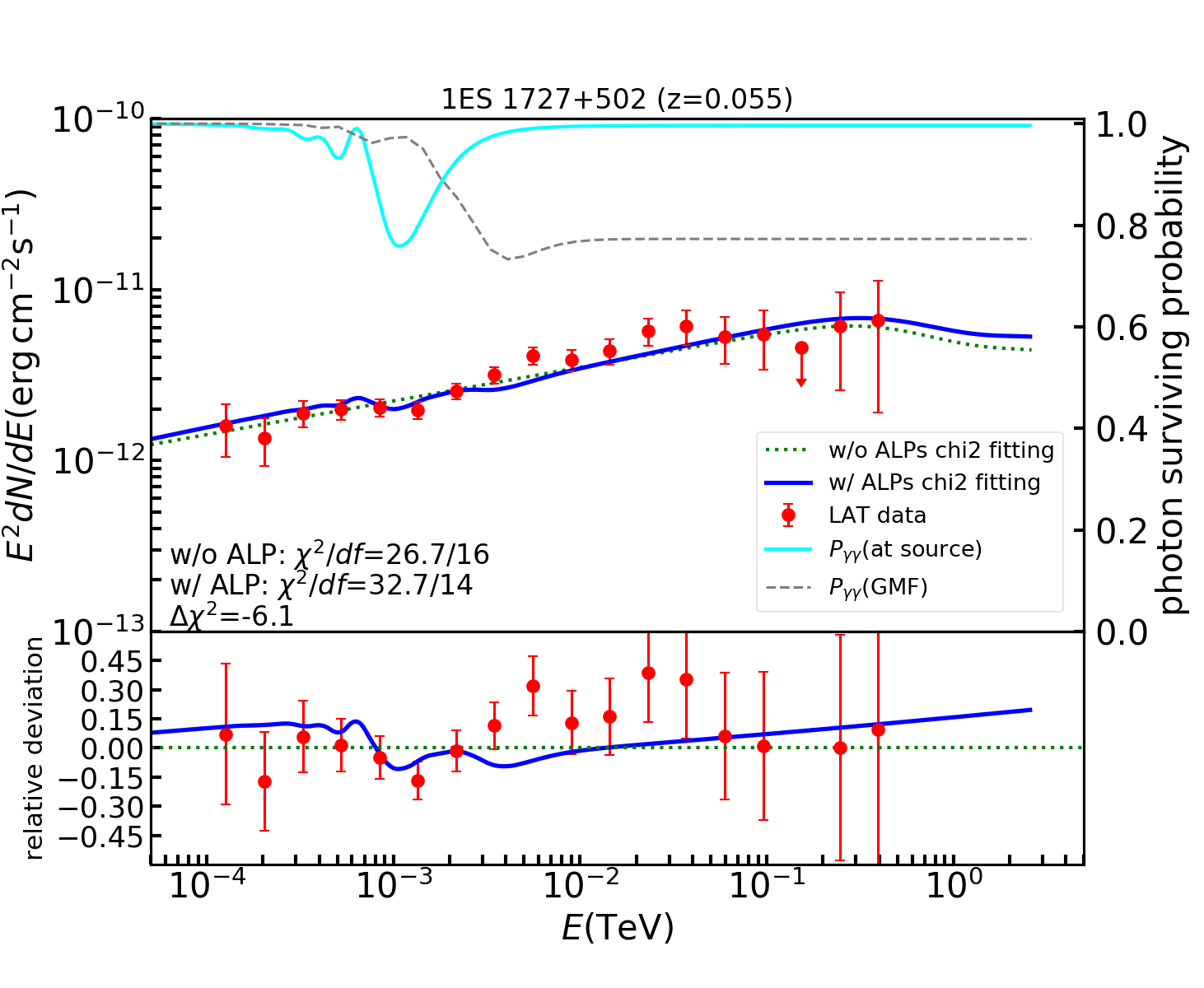}
\end{minipage}
\caption{1ES~1727+502, same as Fig.~\ref{fig:mkn421sed}.}\label{fig:es1727sed}
\end{figure}

\begin{figure}[ht!]
\begin{minipage}[t]{0.455\linewidth}
\centering
\includegraphics[width=0.9\textwidth]{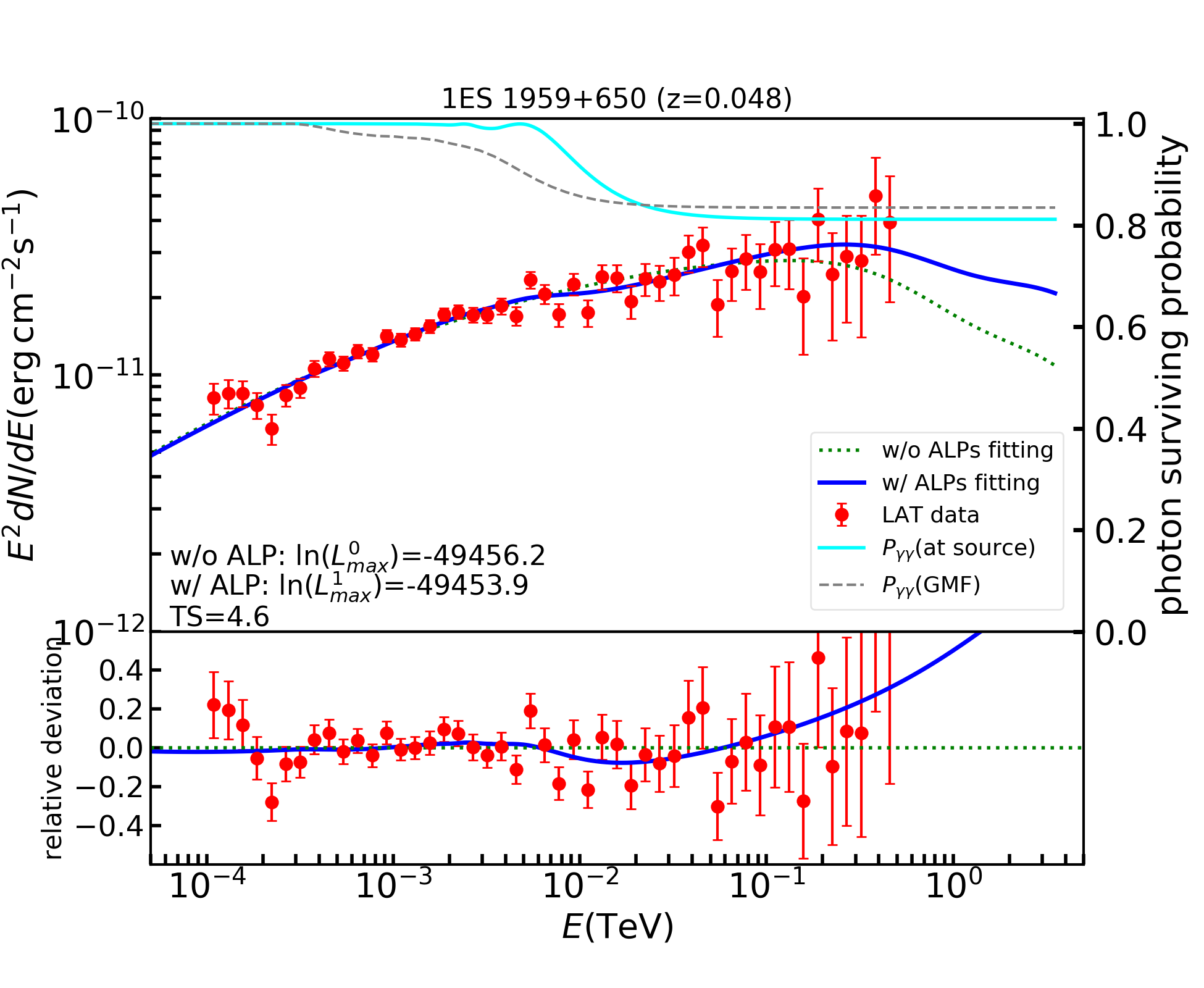}
\end{minipage}%
\begin{minipage}[t]{0.455\linewidth}
\centering
\includegraphics[width=0.9\textwidth]{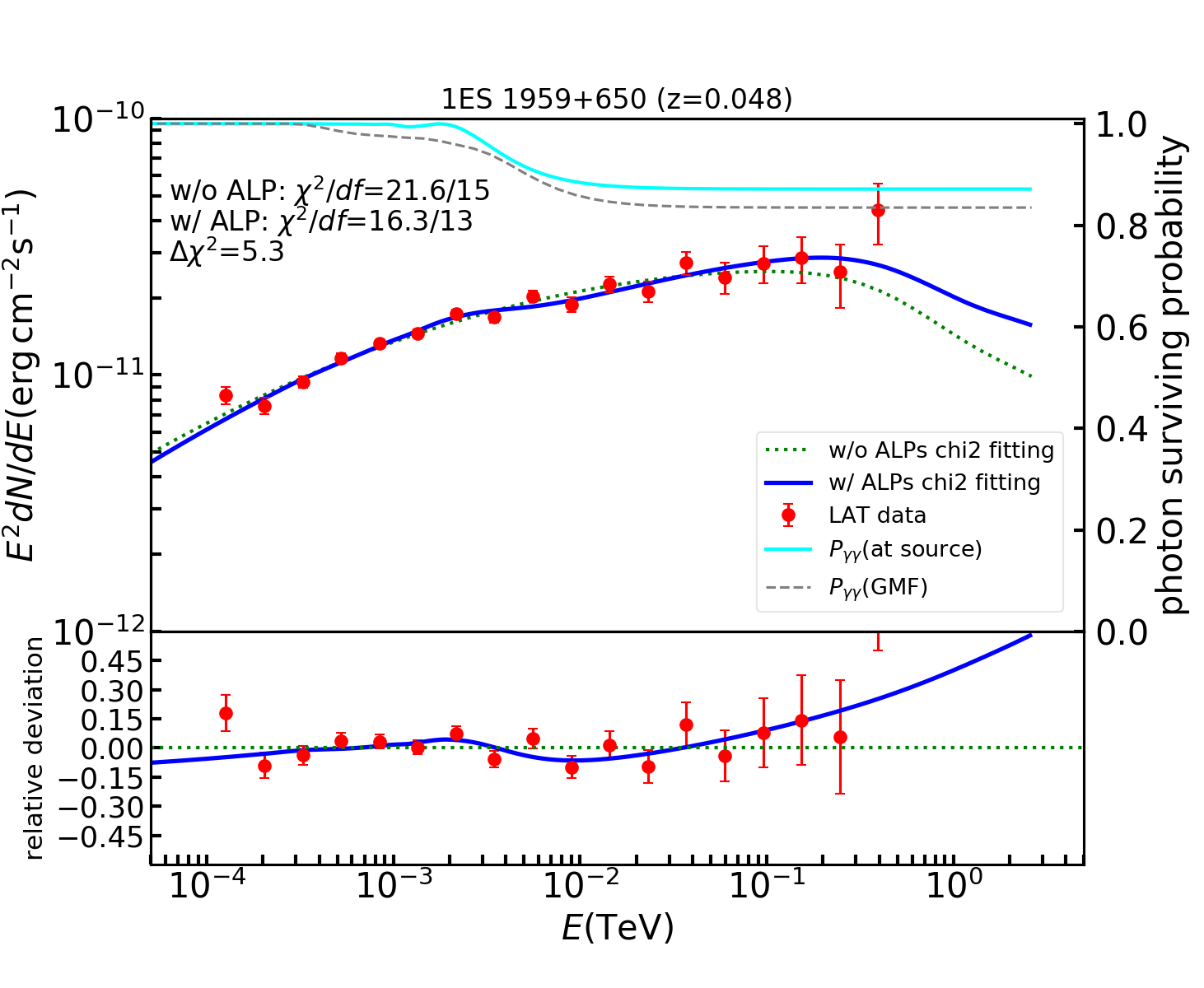}
\end{minipage}
\caption{1ES~1959+650, same as Fig.~\ref{fig:mkn421sed}.}\label{fig:es1959sed}
\end{figure}

\begin{figure}[ht!]
\begin{minipage}[t]{0.455\linewidth}
\centering
\includegraphics[width=0.9\textwidth]{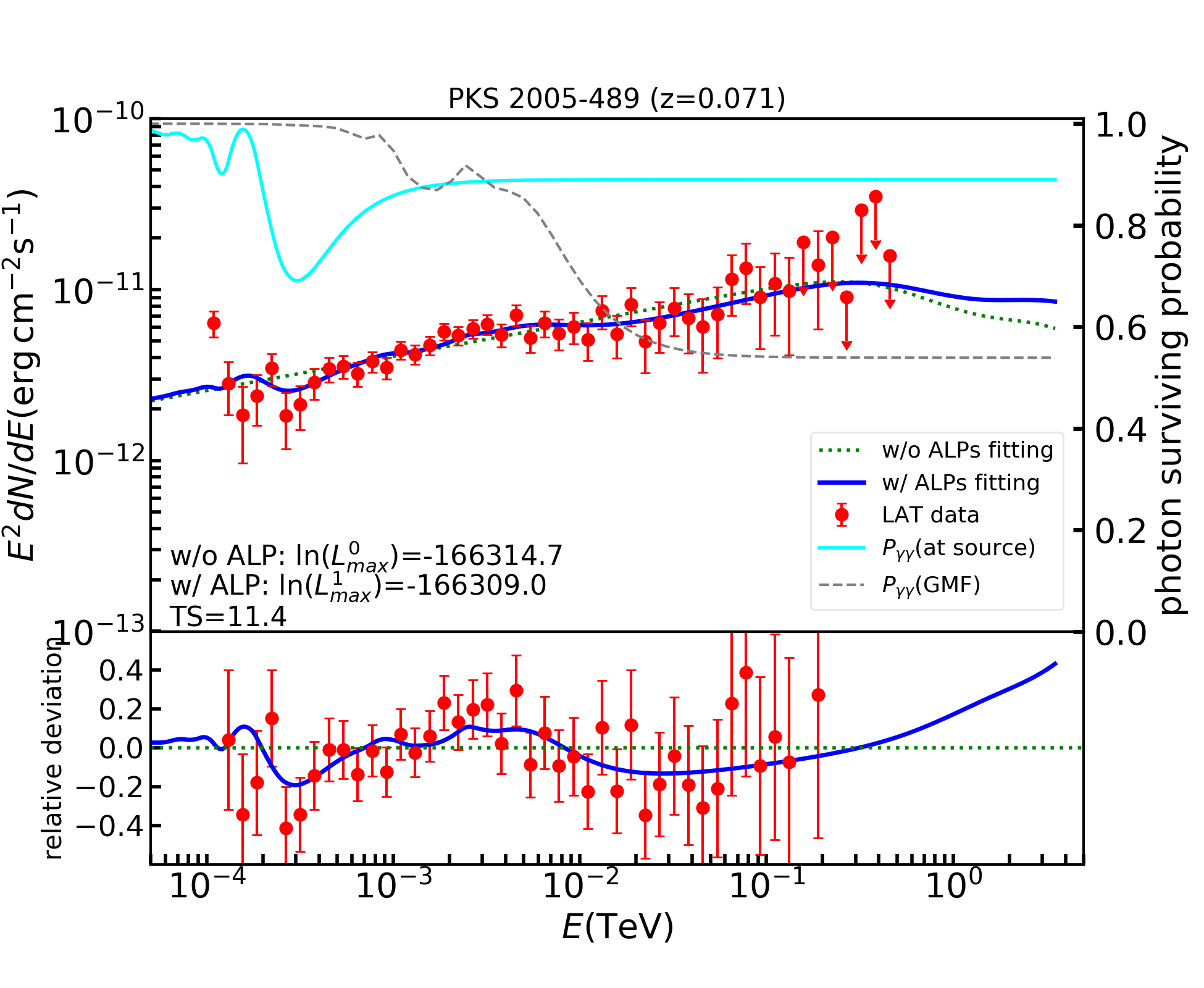}
\end{minipage}%
\begin{minipage}[t]{0.455\linewidth}
\centering
\includegraphics[width=0.9\textwidth]{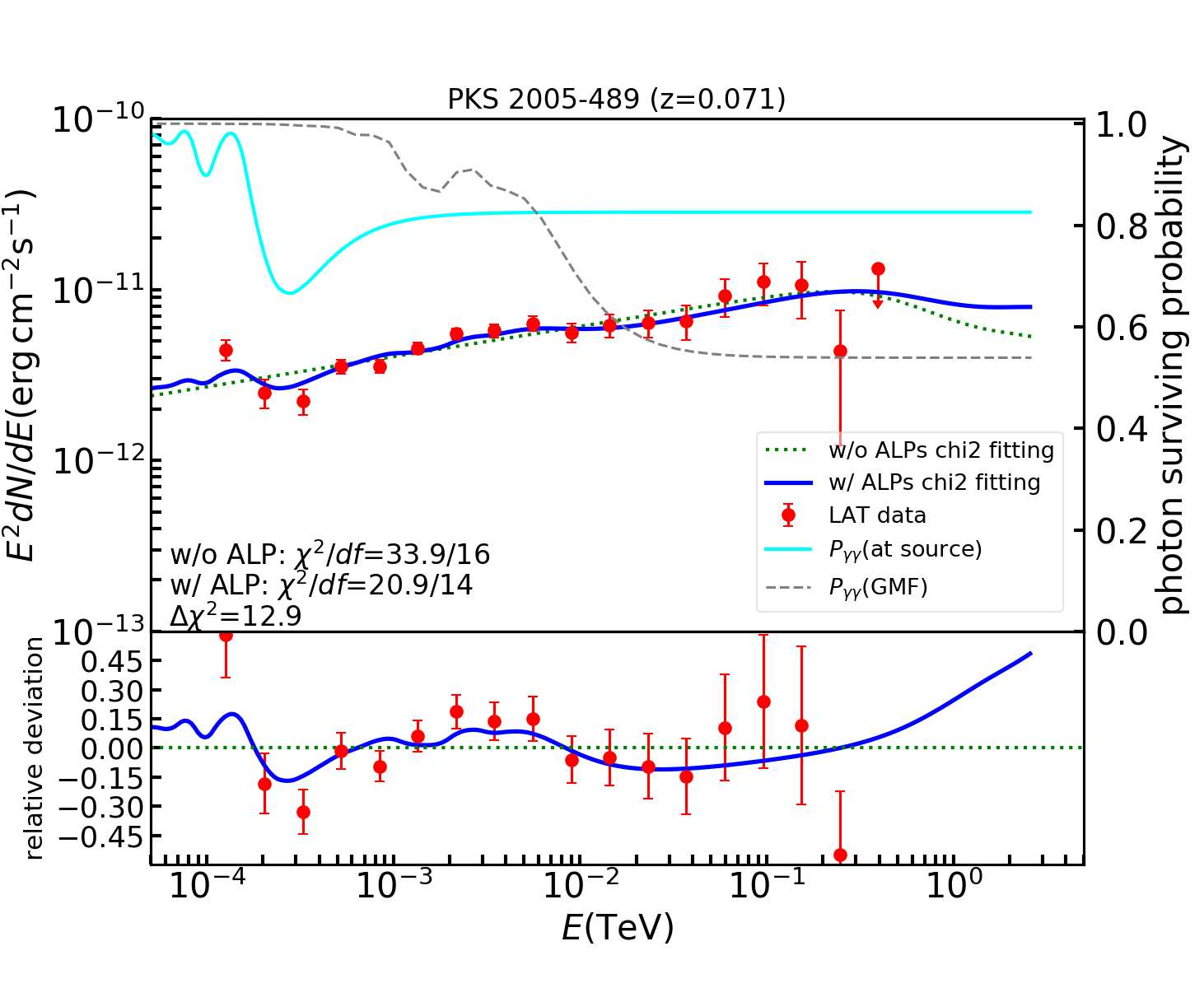}
\end{minipage}
\caption{PKS~2005-489, same as Fig.~\ref{fig:mkn421sed}.}\label{fig:pks2005sed}
\end{figure}

\begin{figure}[ht!]
\begin{minipage}[t]{0.455\linewidth}
\centering
\includegraphics[width=0.9\textwidth]{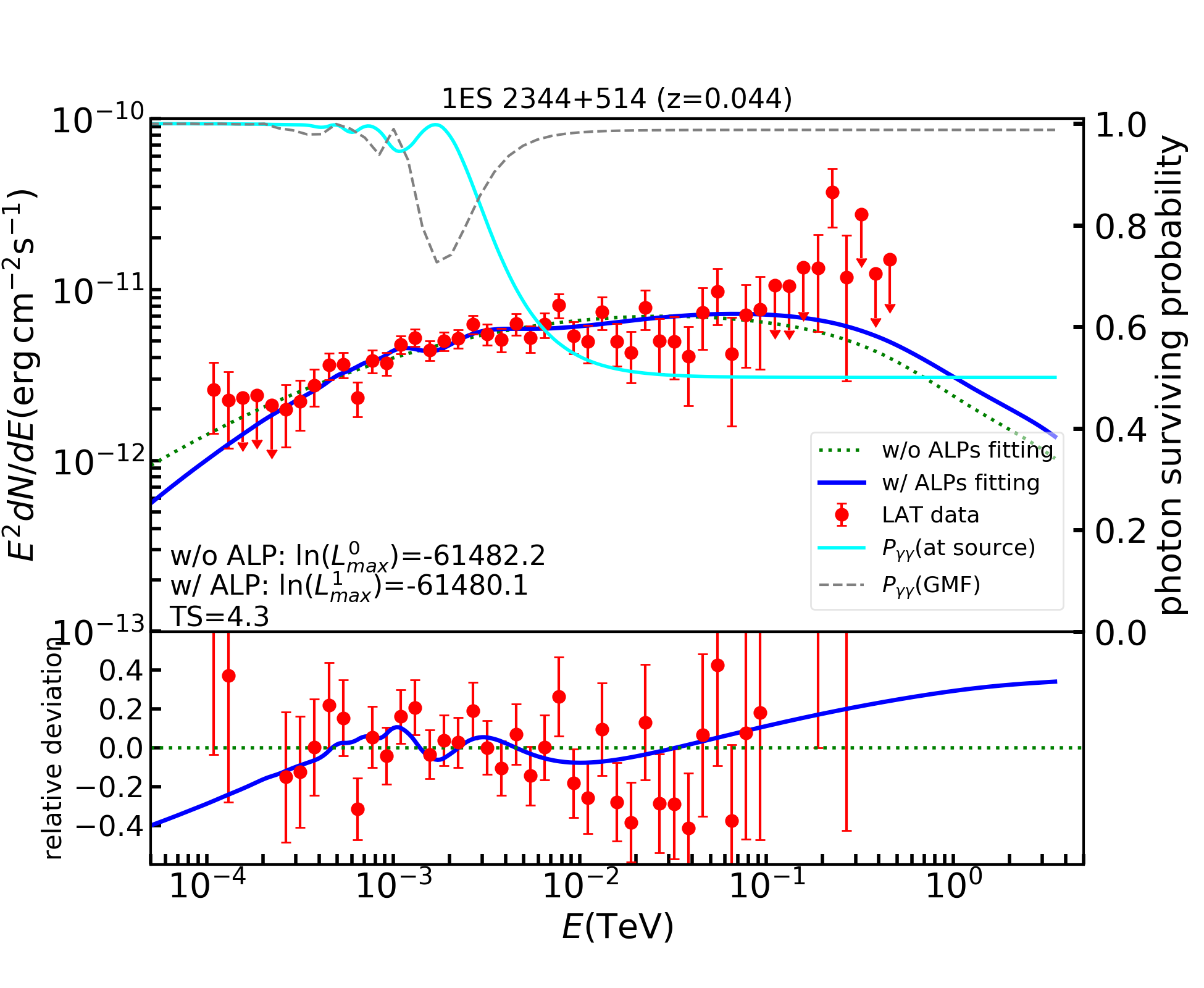}
\end{minipage}%
\begin{minipage}[t]{0.455\linewidth}
\centering
\includegraphics[width=0.9\textwidth]{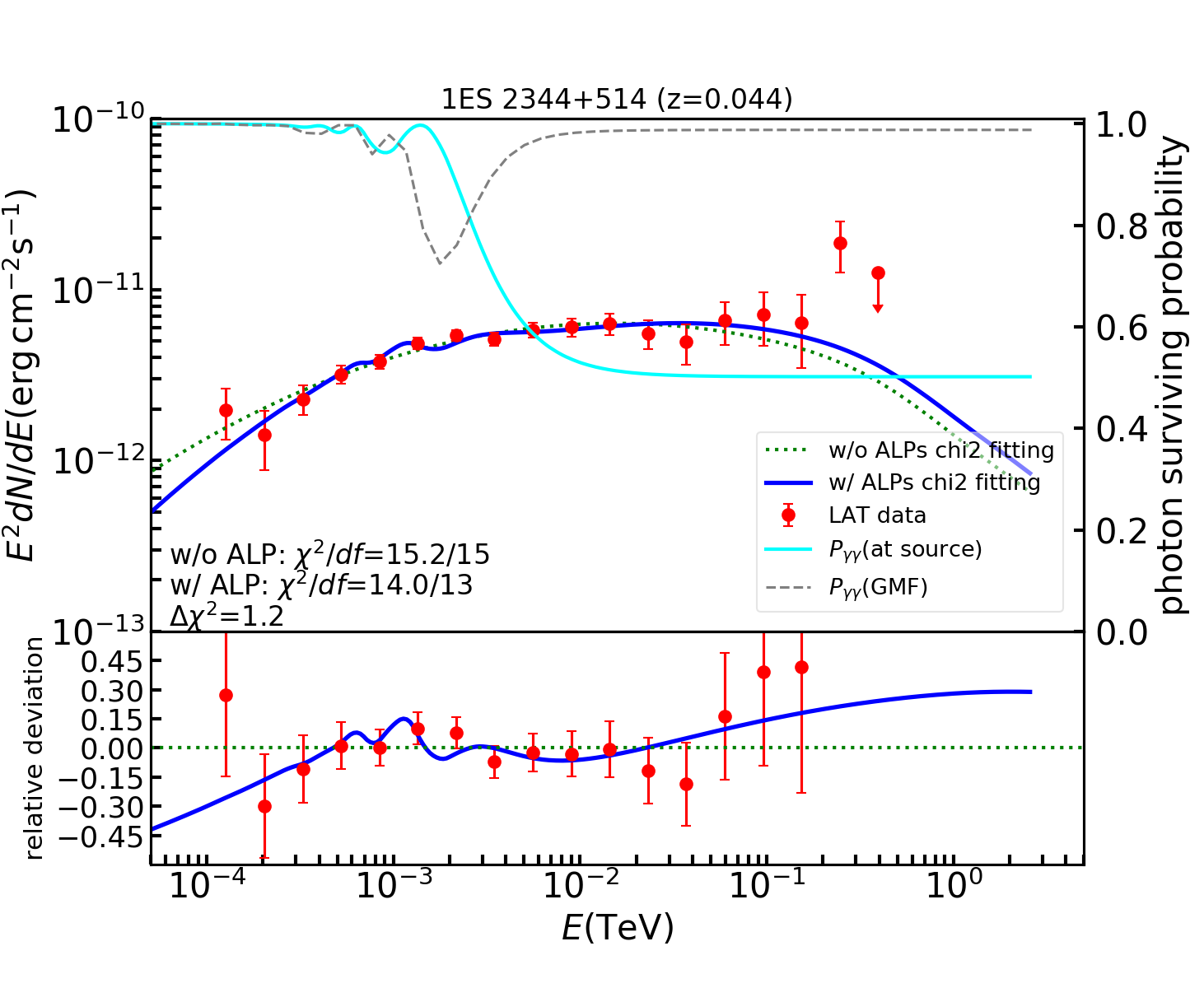}
\end{minipage}
\caption{1ES~2344+514, same as Fig.~\ref{fig:mkn421sed}.}\label{fig:es2344sed}
\end{figure}

In the following figures (Figs.~\ref{fig:es0033} to \ref{fig:es2344}) we present the $(B,s)$ grid maps with likelihood and $\chi^2$ fitting for the sources in our sample collection. The corresponding best-fitting spectral energy 
distributions are provided in Figs.~\ref{fig:es0033sed} to \ref{fig:es2344sed}. 
\begin{figure}[ht!]
\begin{minipage}[t]{0.48\linewidth}
\centering
\includegraphics[width=0.9\textwidth]{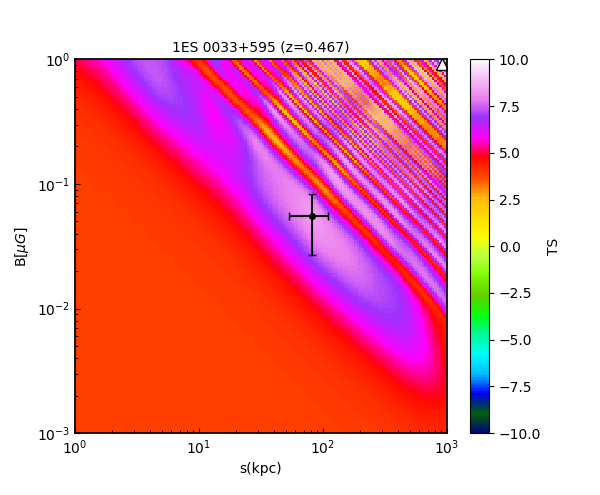}
\end{minipage}%
\begin{minipage}[t]{0.48\linewidth}
\centering
\includegraphics[width=0.9\textwidth]{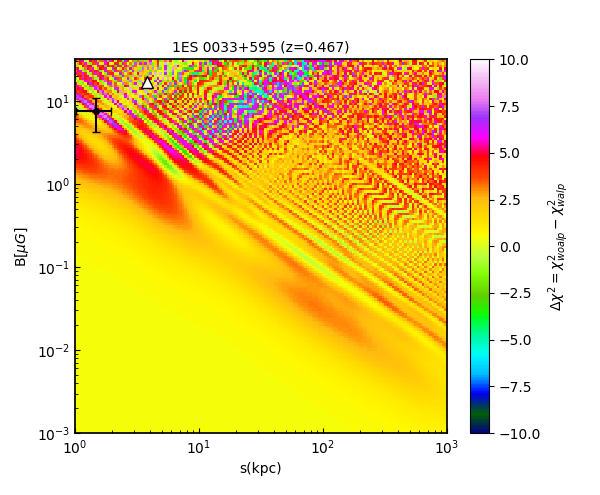}
\end{minipage}
\caption{1ES~0033+595, same as Fig.~\ref{fig:mkn421scan}.}\label{fig:es0033}
\end{figure}
\begin{figure}[ht!]
\begin{minipage}[t]{0.48\linewidth}
\centering
\includegraphics[width=0.9\textwidth]{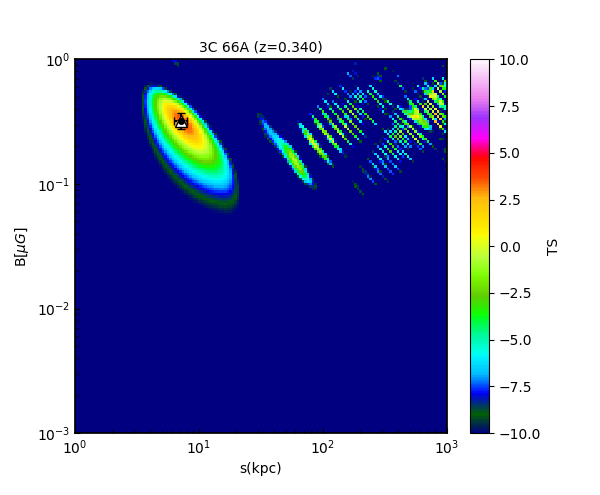}
\end{minipage}%
\begin{minipage}[t]{0.48\linewidth}
\centering
\includegraphics[width=0.9\textwidth]{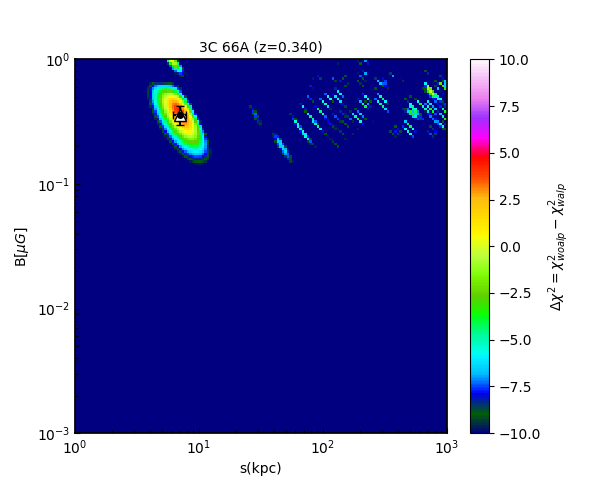}
\end{minipage}
\caption{3C~66A, same as Fig.~\ref{fig:mkn421scan}.}\label{fig:3c66a}
\end{figure}
\begin{figure}[ht!]
\begin{minipage}[t]{0.48\linewidth}
\centering
\includegraphics[width=0.9\textwidth]{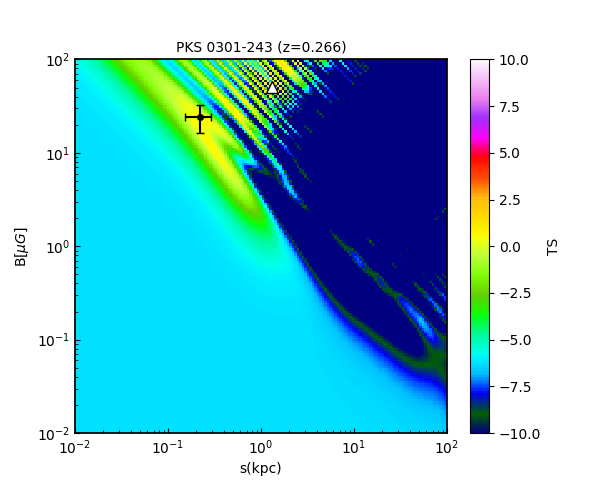}
\end{minipage}%
\begin{minipage}[t]{0.48\linewidth}
\centering
\includegraphics[width=0.9\textwidth]{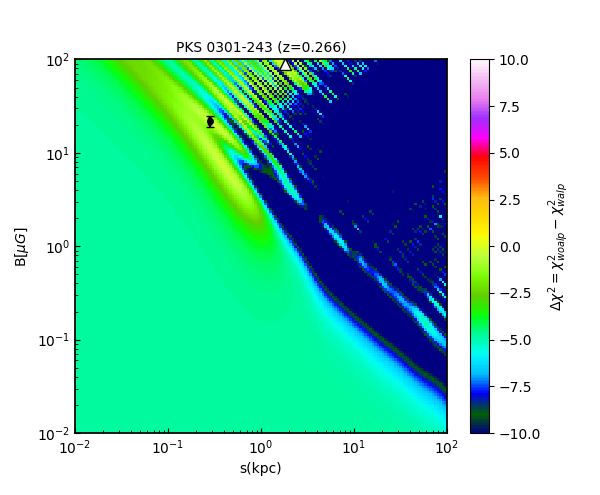}
\end{minipage}
\caption{PKS~0301-243, same as Fig.~\ref{fig:mkn421scan}.}\label{fig:pks0301}
\end{figure}
\begin{figure}[ht!]
\begin{minipage}[t]{0.48\linewidth}
\centering
\includegraphics[width=0.9\textwidth]{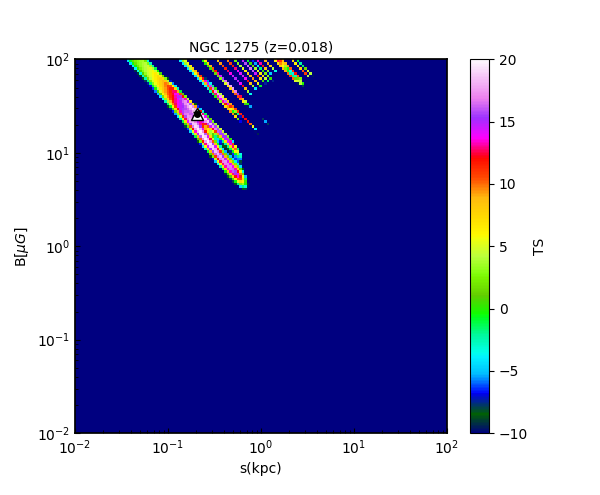}
\end{minipage}%
\begin{minipage}[t]{0.48\linewidth}
\centering
\includegraphics[width=0.9\textwidth]{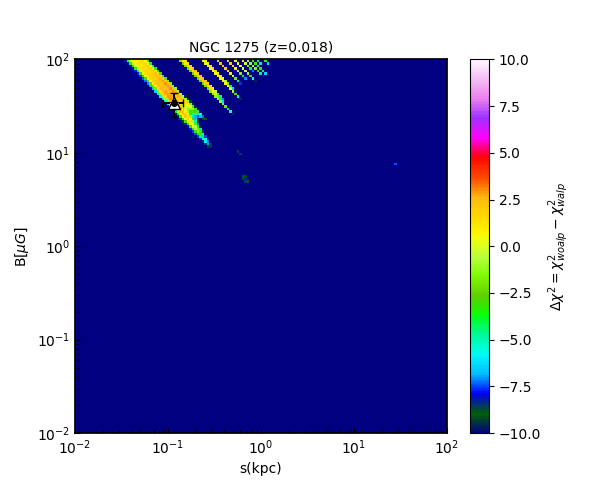}
\end{minipage}
\caption{NGC~1275, same as Fig.~\ref{fig:mkn421scan}.}\label{fig:ngc1275}
\end{figure}
\begin{figure}[ht!]
\begin{minipage}[t]{0.48\linewidth}
\centering
\includegraphics[width=0.9\textwidth]{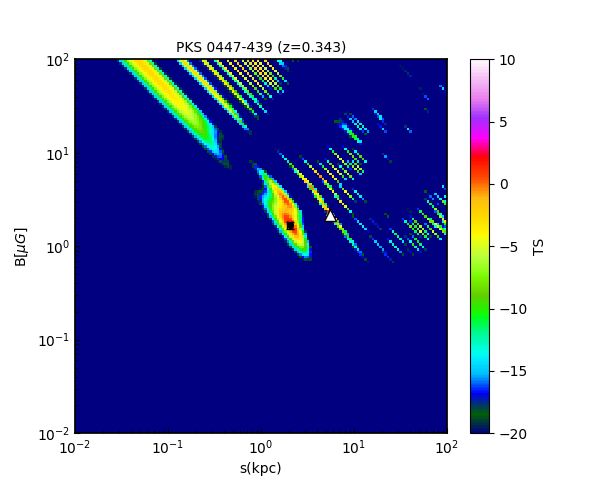}
\end{minipage}%
\begin{minipage}[t]{0.48\linewidth}
\centering
\includegraphics[width=0.9\textwidth]{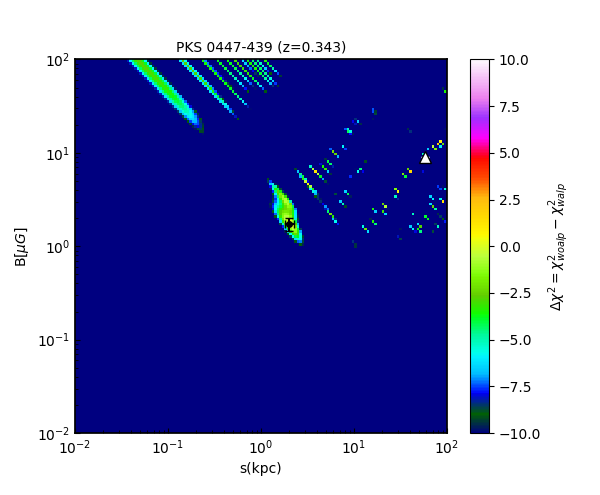}
\end{minipage}
\caption{PKS~0447-439, same as Fig.~\ref{fig:mkn421scan}.}\label{fig:pks0447}
\end{figure}
\begin{figure}[ht!]
\begin{minipage}[t]{0.48\linewidth}
\centering
\includegraphics[width=0.9\textwidth]{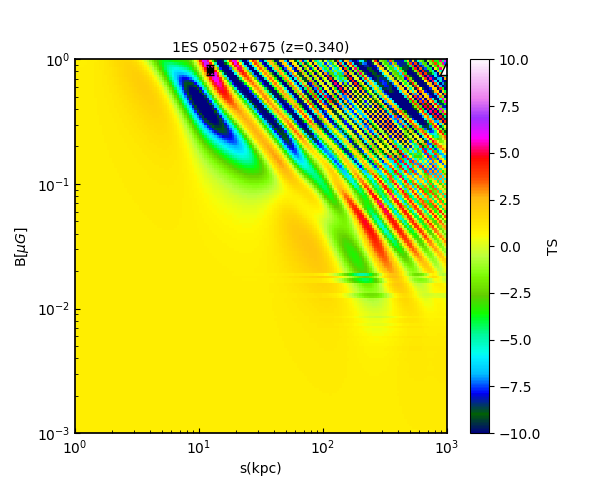}
\end{minipage}%
\begin{minipage}[t]{0.48\linewidth}
\centering
\includegraphics[width=0.9\textwidth]{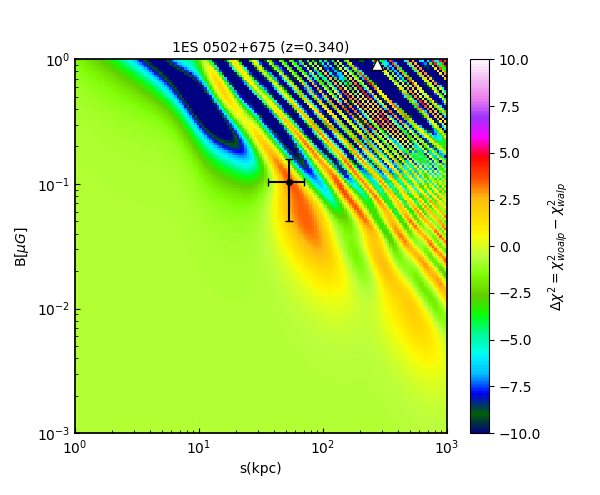}
\end{minipage}
\caption{1E~0502+675, same as Fig.~\ref{fig:mkn421scan}.}\label{fig:es0502}
\end{figure}
\begin{figure}[ht!]
\begin{minipage}[t]{0.48\linewidth}
\centering
\includegraphics[width=0.9\textwidth]{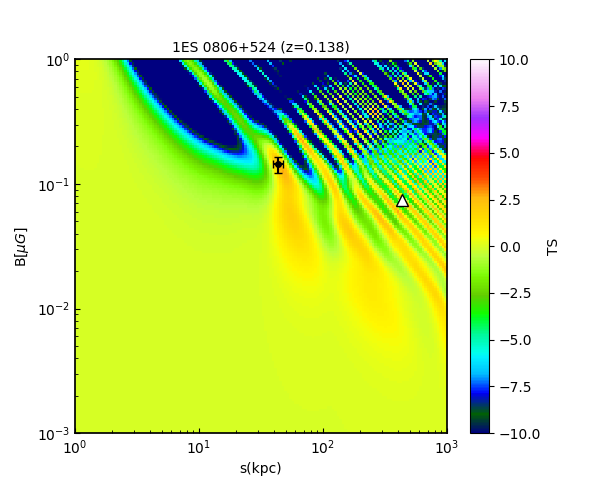}
\end{minipage}%
\begin{minipage}[t]{0.48\linewidth}
\centering
\includegraphics[width=0.9\textwidth]{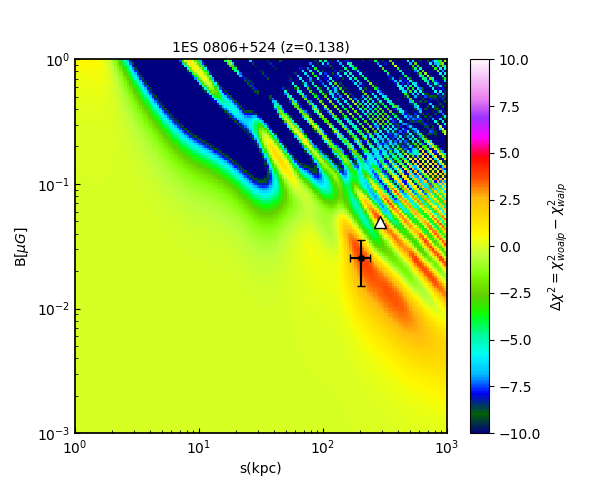}
\end{minipage}
\caption{1ES~0806+524, same as Fig.~\ref{fig:mkn421scan}.}\label{fig:es0806}
\end{figure}
\begin{figure}[ht!]
\begin{minipage}[t]{0.48\linewidth}
\centering
\includegraphics[width=0.9\textwidth]{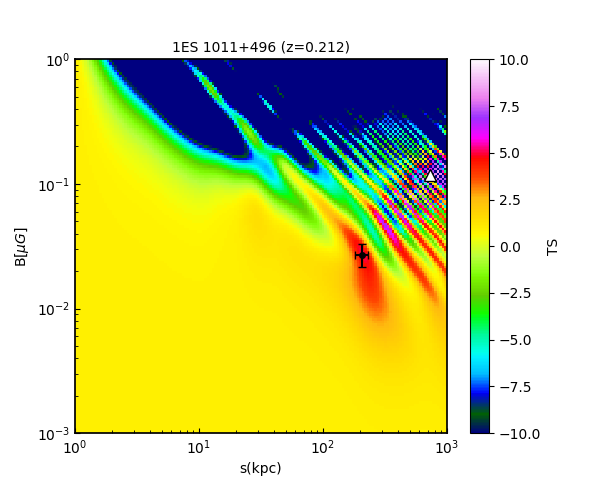}
\end{minipage}%
\begin{minipage}[t]{0.48\linewidth}
\centering
\includegraphics[width=0.9\textwidth]{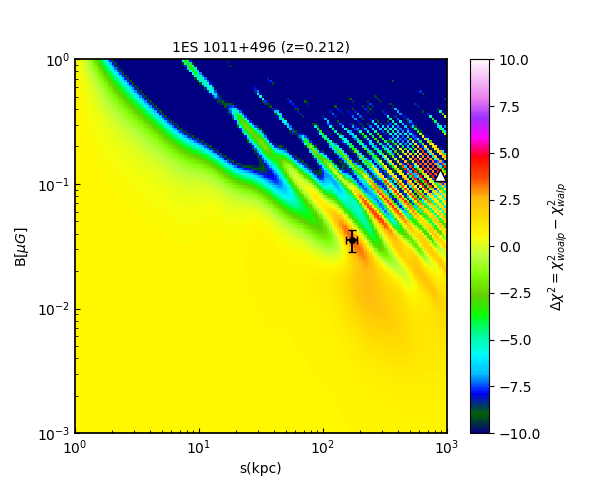}
\end{minipage}
\caption{1ES~1011+496, same as Fig.~\ref{fig:mkn421scan}.}\label{fig:es1011}
\end{figure}
\begin{figure}[ht!]
\begin{minipage}[t]{0.48\linewidth}
\centering
\includegraphics[width=0.9\textwidth]{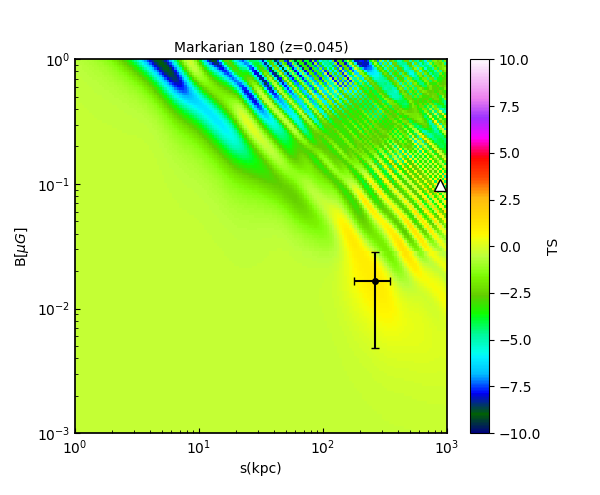}
\end{minipage}%
\begin{minipage}[t]{0.48\linewidth}
\centering
\includegraphics[width=0.9\textwidth]{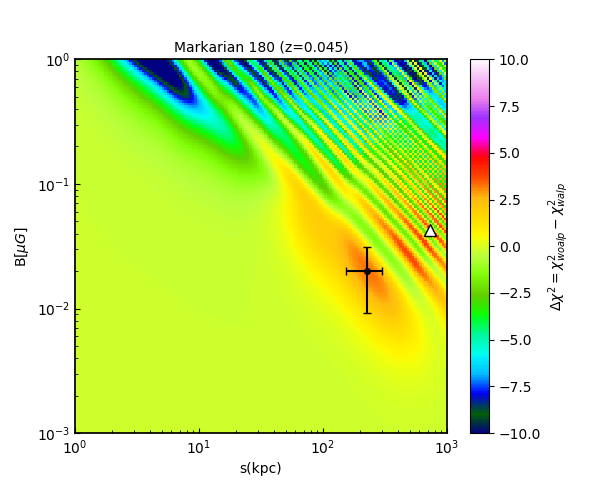}
\end{minipage}
\caption{Markarian~180, same as Fig.~\ref{fig:mkn421scan}.}\label{fig:mkn180}
\end{figure}
\begin{figure}[ht!]
\begin{minipage}[t]{0.48\linewidth}
\centering
\includegraphics[width=0.9\textwidth]{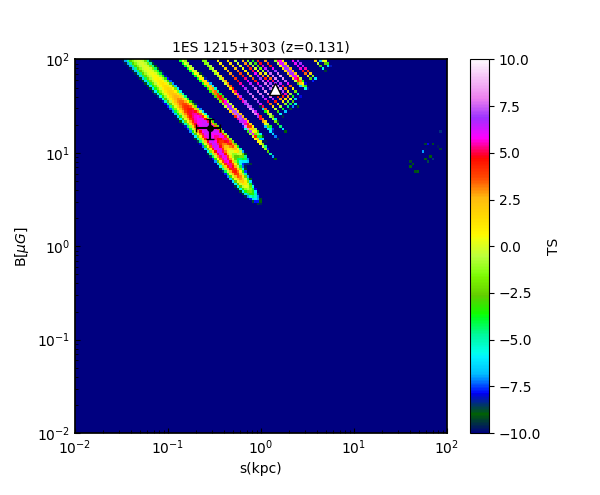}
\end{minipage}%
\begin{minipage}[t]{0.48\linewidth}
\centering
\includegraphics[width=0.9\textwidth]{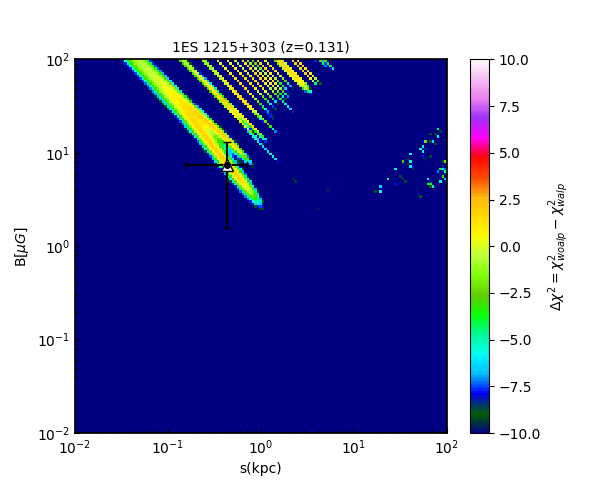}
\end{minipage}
\caption{1ES~1215+303, same as Fig.~\ref{fig:mkn421scan}.}\label{fig:es1215}
\end{figure}
\begin{figure}[ht!]
\begin{minipage}[t]{0.48\linewidth}
\centering
\includegraphics[width=0.9\textwidth]{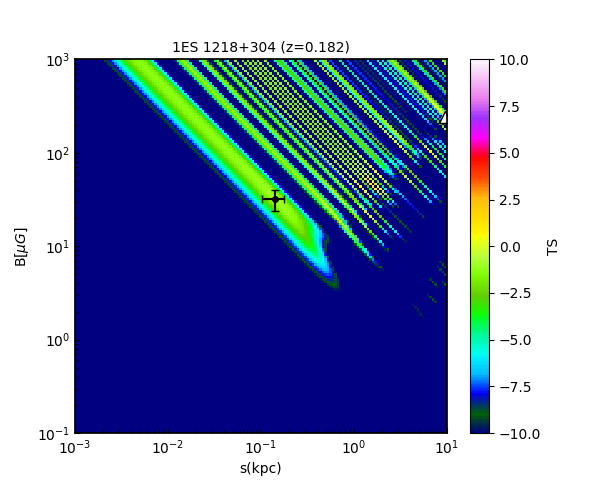}
\end{minipage}%
\begin{minipage}[t]{0.48\linewidth}
\centering
\includegraphics[width=0.9\textwidth]{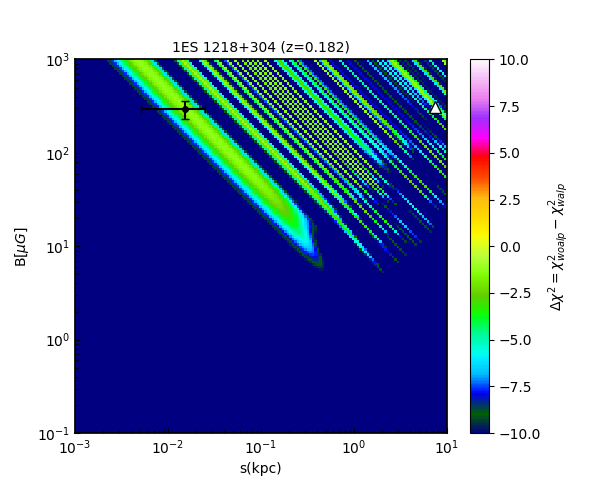}
\end{minipage}
\caption{1ES~1218+304, same as Fig.~\ref{fig:mkn421scan}.}\label{fig:es1218}
\end{figure}
\begin{figure}[ht!]
\begin{minipage}[t]{0.48\linewidth}
\centering
\includegraphics[width=0.9\textwidth]{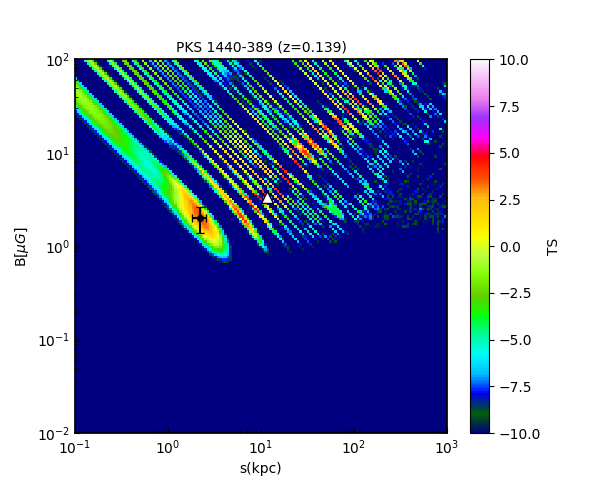}
\end{minipage}%
\begin{minipage}[t]{0.48\linewidth}
\centering
\includegraphics[width=0.9\textwidth]{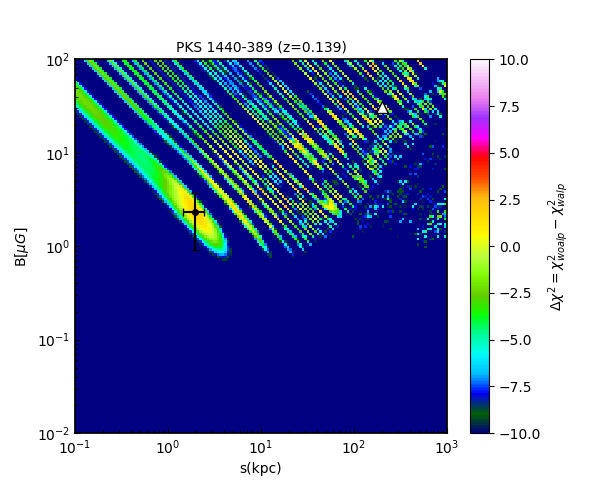}
\end{minipage}
\caption{PKS~1440-389, same as Fig.~\ref{fig:mkn421scan}.}\label{fig:pks1440}
\end{figure}
\begin{figure}[ht!]
\begin{minipage}[t]{0.48\linewidth}
\centering
\includegraphics[width=0.9\textwidth]{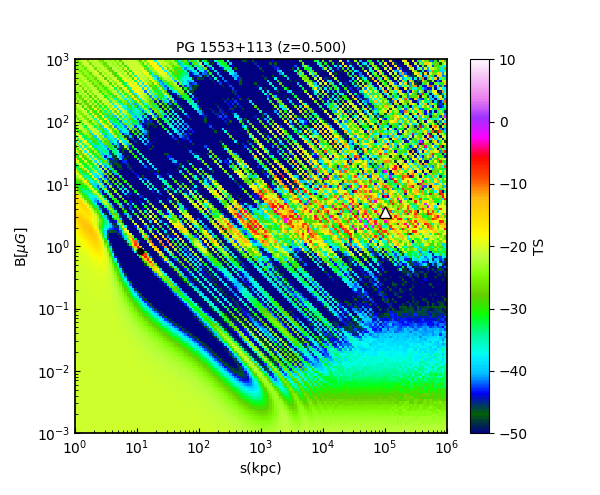}
\end{minipage}%
\begin{minipage}[t]{0.48\linewidth}
\centering
\includegraphics[width=0.9\textwidth]{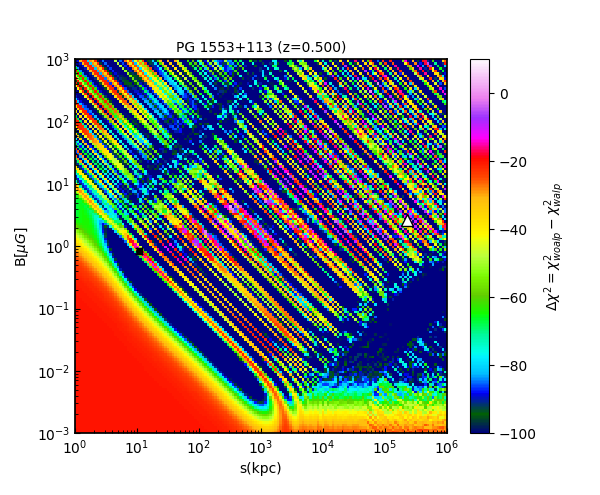}
\end{minipage}
\caption{PG~1553+113, same as Fig.~\ref{fig:mkn421scan}.}\label{fig:pg1553}
\end{figure}
\begin{figure}[ht!]
\begin{minipage}[t]{0.48\linewidth}
\centering
\includegraphics[width=0.9\textwidth]{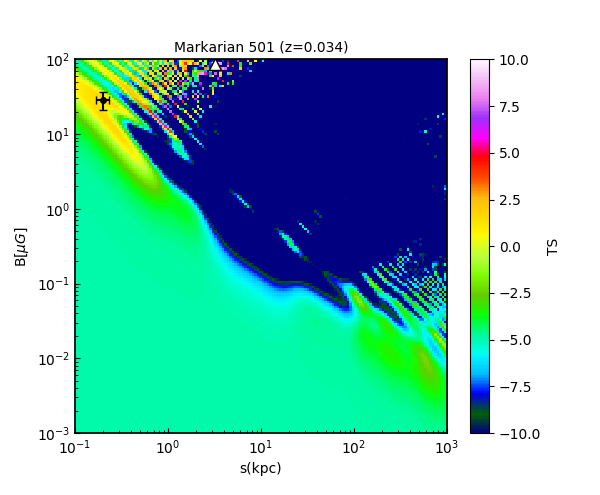}
\end{minipage}%
\begin{minipage}[t]{0.48\linewidth}
\centering
\includegraphics[width=0.9\textwidth]{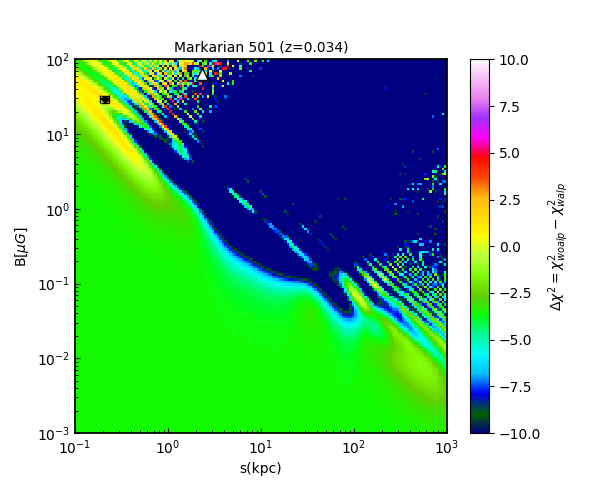}
\end{minipage}
\caption{Markarian~501, same as Fig.~\ref{fig:mkn421scan}.}\label{fig:mkn501}
\end{figure}
\begin{figure}[ht!]
\begin{minipage}[t]{0.48\linewidth}
\centering
\includegraphics[width=0.9\textwidth]{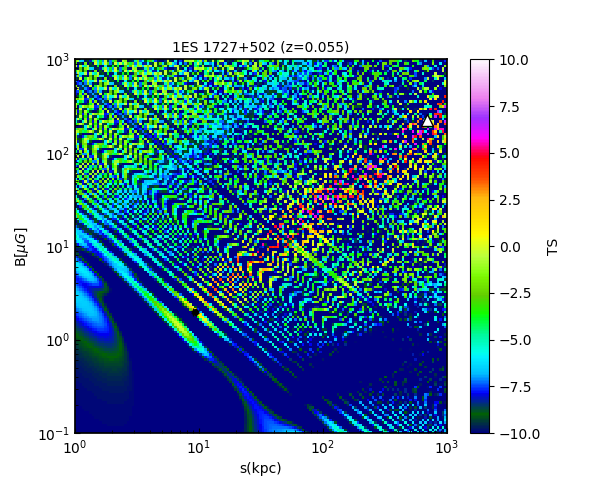}
\end{minipage}%
\begin{minipage}[t]{0.48\linewidth}
\centering
\includegraphics[width=0.9\textwidth]{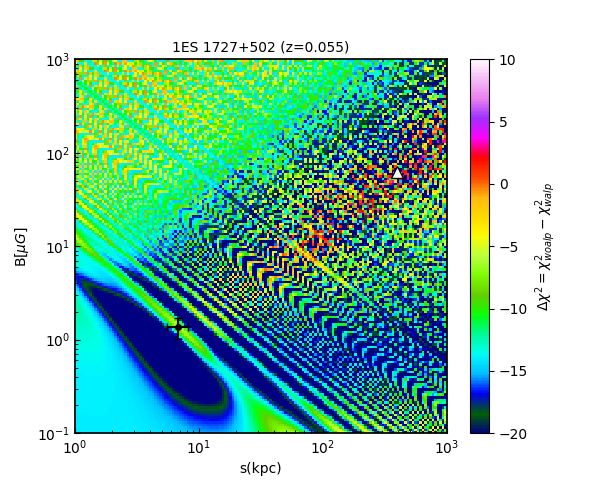}
\end{minipage}
\caption{1ES~1727+502, same as Fig.~\ref{fig:mkn421scan}.}\label{fig:es1727}
\end{figure}
\begin{figure}[ht!]
\begin{minipage}[t]{0.48\linewidth}
\centering
\includegraphics[width=0.9\textwidth]{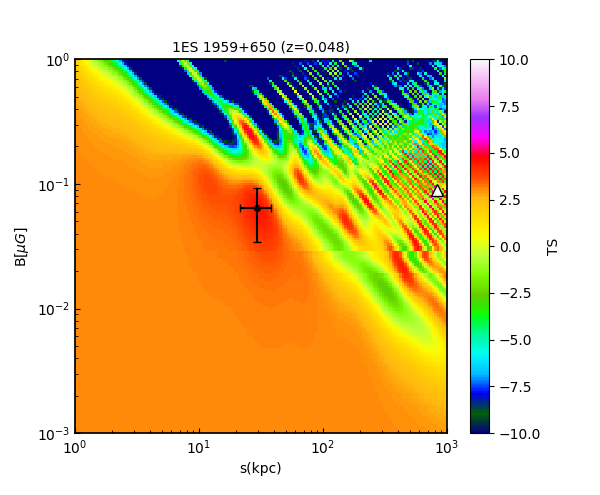}
\end{minipage}%
\begin{minipage}[t]{0.48\linewidth}
\centering
\includegraphics[width=0.9\textwidth]{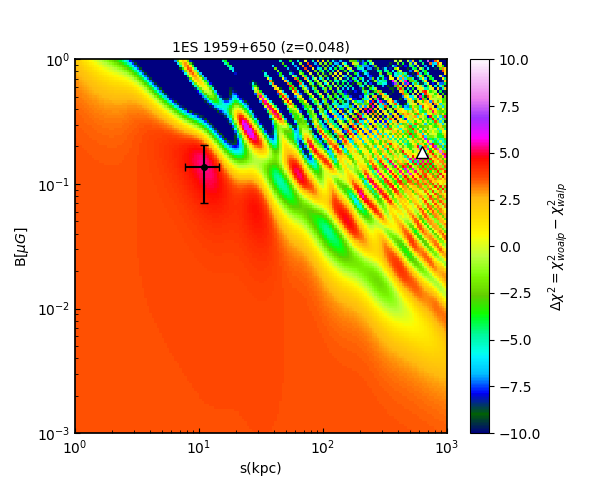}
\end{minipage}
\caption{1ES~1959+650, same as Fig.~\ref{fig:mkn421scan}.}\label{fig:es1959}
\end{figure}
\begin{figure}[ht!]
\begin{minipage}[t]{0.48\linewidth}
\centering
\includegraphics[width=0.9\textwidth]{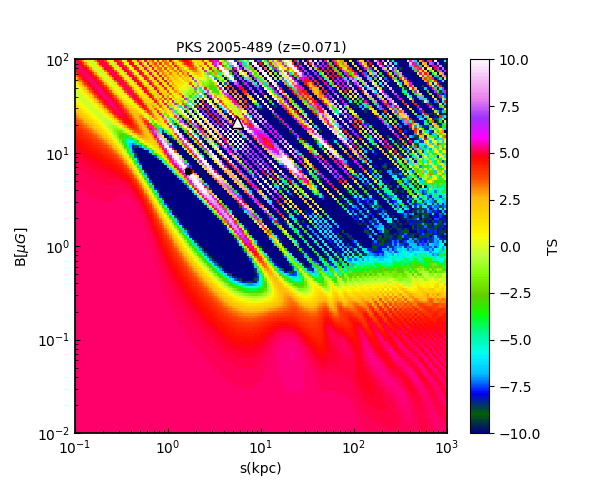}
\end{minipage}%
\begin{minipage}[t]{0.48\linewidth}
\centering
\includegraphics[width=0.9\textwidth]{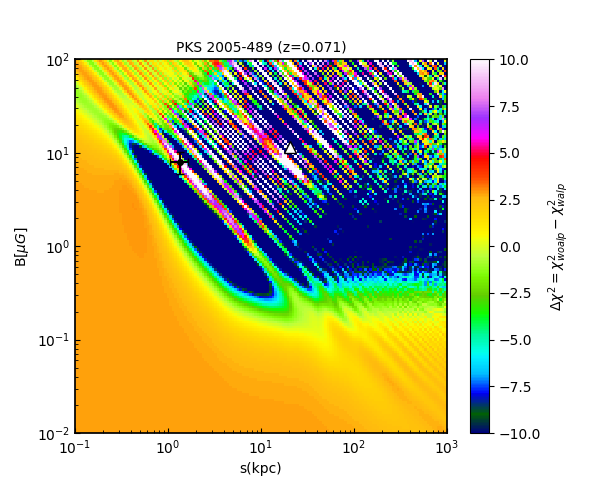}
\end{minipage}
\caption{PKS~2005-304, same as Fig.~\ref{fig:mkn421scan}.}\label{fig:pks2005}
\end{figure}
\begin{figure}[ht!]
\begin{minipage}[t]{0.48\linewidth}
\centering
\includegraphics[width=0.9\textwidth]{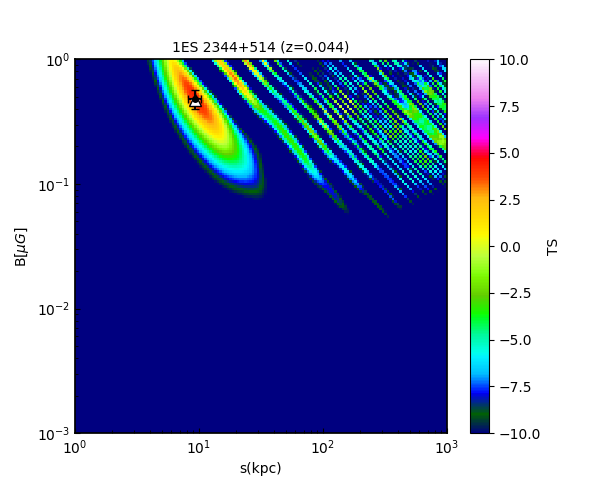}
\end{minipage}%
\begin{minipage}[t]{0.48\linewidth}
\centering
\includegraphics[width=0.9\textwidth]{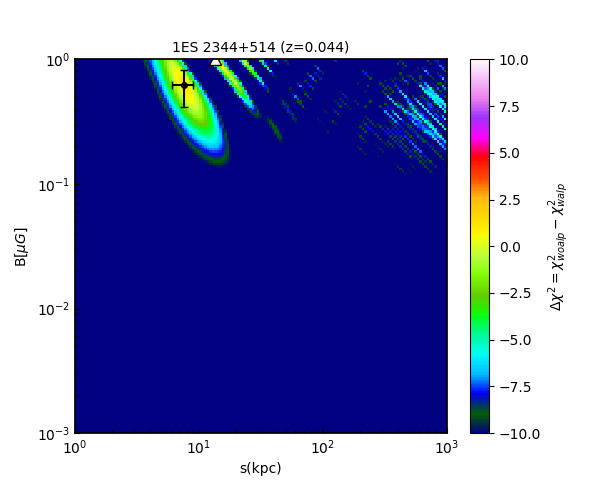}
\end{minipage}
\caption{1ES~2344+514, same as Fig.~\ref{fig:mkn421scan}.}\label{fig:es2344}
\end{figure}

\clearpage
\bibliographystyle{unsrt}
\bibliography{ref.bib}

\begin{thebibliography}{10}

\bibitem{Peccei:1977ur}
R.~D. Peccei and Helen~R. Quinn.
\newblock {Constraints Imposed by CP Conservation in the Presence of
  Instantons}.
\newblock {\em Phys. Rev. D}, 16:1791--1797, 1977.

\bibitem{Peccei:1977hh}
R.~D. Peccei and Helen~R. Quinn.
\newblock {CP Conservation in the Presence of Instantons}.
\newblock {\em Phys. Rev. Lett.}, 38:1440--1443, 1977.

\bibitem{Svrcek:2006yi}
Peter Svrcek and Edward Witten.
\newblock {Axions In String Theory}.
\newblock {\em JHEP}, 06:051, 2006.

\bibitem{Conlon:2006tq}
Joseph~P. Conlon.
\newblock {The QCD axion and moduli stabilisation}.
\newblock {\em JHEP}, 05:078, 2006.

\bibitem{Cicoli:2012sz}
Michele Cicoli, Mark Goodsell, and Andreas Ringwald.
\newblock {The type IIB string axiverse and its low-energy phenomenology}.
\newblock {\em JHEP}, 10:146, 2012.

\bibitem{Preskill:1982cy}
John Preskill, Mark~B. Wise, and Frank Wilczek.
\newblock {Cosmology of the Invisible Axion}.
\newblock {\em Phys. Lett. B}, 120:127--132, 1983.

\bibitem{Abbott:1982af}
L.~F. Abbott and P.~Sikivie.
\newblock {A Cosmological Bound on the Invisible Axion}.
\newblock {\em Phys. Lett. B}, 120:133--136, 1983.

\bibitem{Dine:1982ah}
Michael Dine and Willy Fischler.
\newblock {The Not So Harmless Axion}.
\newblock {\em Phys. Lett. B}, 120:137--141, 1983.

\bibitem{Jaeckel:2013uva}
Joerg Jaeckel.
\newblock {A Family of WISPy Dark Matter Candidates}.
\newblock {\em Phys. Lett. B}, 732:1--7, 2014.

\bibitem{DeAngelis:2007wiw}
Alessandro De~Angelis, Oriana Mansutti, and Marco Roncadelli.
\newblock {Axion-Like Particles, Cosmic Magnetic Fields and Gamma-Ray
  Astrophysics}.
\newblock {\em Phys. Lett. B}, 659:847--855, 2008.

\bibitem{Hooper:2007bq}
Dan Hooper and Pasquale~D. Serpico.
\newblock {Detecting Axion-Like Particles With Gamma Ray Telescopes}.
\newblock {\em Phys. Rev. Lett.}, 99:231102, 2007.

\bibitem{Simet:2007sa}
Melanie Simet, Dan Hooper, and Pasquale~D. Serpico.
\newblock {The Milky Way as a Kiloparsec-Scale Axionscope}.
\newblock {\em Phys. Rev. D}, 77:063001, 2008.

\bibitem{DeAngelis:2011id}
Alessandro De~Angelis, Giorgio Galanti, and Marco Roncadelli.
\newblock {Relevance of axion-like particles for very-high-energy
  astrophysics}.
\newblock {\em Phys. Rev. D}, 84:105030, 2011.
\newblock [Erratum: Phys.Rev.D 87, 109903 (2013)].

\bibitem{Horns:2012kw}
Dieter Horns, Luca Maccione, Manuel Meyer, Alessandro Mirizzi, Daniele
  Montanino, and Marco Roncadelli.
\newblock {Hardening of TeV gamma spectrum of AGNs in galaxy clusters by
  conversions of photons into axion-like particles}.
\newblock {\em Phys. Rev. D}, 86:075024, 2012.

\bibitem{Rubtsov:2014uga}
G.~I. Rubtsov and S.~V. Troitsky.
\newblock {Breaks in gamma-ray spectra of distant blazars and transparency of
  the Universe}.
\newblock {\em JETP Lett.}, 100(6):355--359, 2014.

\bibitem{Libanov:2019fzq}
Maxim Libanov and Sergey Troitsky.
\newblock {On the impact of magnetic-field models in galaxy clusters on
  constraints on axion-like particles from the lack of irregularities in
  high-energy spectra of astrophysical sources}.
\newblock {\em Phys. Lett. B}, 802:135252, 2020.

\bibitem{Majumdar:2018sbv}
Jhilik Majumdar, Francesca Calore, and Dieter Horns.
\newblock {Search for gamma-ray spectral modulations in Galactic pulsars}.
\newblock {\em JCAP}, 04:048, 2018.

\bibitem{Meyer:2013pny}
Manuel Meyer, Dieter Horns, and Martin Raue.
\newblock {First lower limits on the photon-axion-like particle coupling from
  very high energy gamma-ray observations}.
\newblock {\em Phys. Rev. D}, 87(3):035027, 2013.

\bibitem{Kohri:2017ljt}
Kazunori Kohri and Hideo Kodama.
\newblock {Axion-Like Particles and Recent Observations of the Cosmic Infrared
  Background Radiation}.
\newblock {\em Phys. Rev. D}, 96(5):051701, 2017.

\bibitem{Bahre:2013ywa}
Robin B\"ahre et~al.
\newblock {Any light particle search II \textemdash{}Technical Design Report}.
\newblock {\em JINST}, 8:T09001, 2013.

\bibitem{CAST:2017uph}
V.~Anastassopoulos et~al.
\newblock {New CAST Limit on the Axion-Photon Interaction}.
\newblock {\em Nature Phys.}, 13:584--590, 2017.

\bibitem{Pallathadka:2020vwu}
Gautham~Adamane Pallathadka et~al.
\newblock {Reconciling hints on axion-like-particles from high-energy gamma
  rays with stellar bounds}.
\newblock {\em JCAP}, 11:036, 2021.

\bibitem{Raffelt:1987im}
Georg Raffelt and Leo Stodolsky.
\newblock {Mixing of the Photon with Low Mass Particles}.
\newblock {\em Phys. Rev. D}, 37:1237, 1988.

\bibitem{Mirizzi:2005ng}
Alessandro Mirizzi, Georg~G. Raffelt, and Pasquale~D. Serpico.
\newblock {Photon-axion conversion as a mechanism for supernova dimming: Limits
  from CMB spectral distortion}.
\newblock {\em Phys. Rev. D}, 72:023501, 2005.

\bibitem{Mirizzi:2006zy}
Alessandro Mirizzi, Georg~G. Raffelt, and Pasquale~D. Serpico.
\newblock {Photon-axion conversion in intergalactic magnetic fields and
  cosmological consequences}.
\newblock {\em Lect. Notes Phys.}, 741:115--134, 2008.

\bibitem{Galanti:2018upl}
Giorgio Galanti, Fabrizio Tavecchio, Marco Roncadelli, and Carmelo Evoli.
\newblock {Blazar VHE spectral alterations induced by photon\textendash{}ALP
  oscillations}.
\newblock {\em Mon. Not. Roy. Astron. Soc.}, 487(1):123--132, 2019.

\bibitem{Dobrynina:2014qba}
Alexandra Dobrynina, Alexander Kartavtsev, and Georg Raffelt.
\newblock {Photon-photon dispersion of TeV gamma rays and its role for
  photon-ALP conversion}.
\newblock {\em Phys. Rev. D}, 91:083003, 2015.
\newblock [Erratum: Phys.Rev.D 95, 109905 (2017)].

\bibitem{Csaki:2003ef}
Csaba Csaki, Nemanja Kaloper, Marco Peloso, and John Terning.
\newblock {Super GZK photons from photon axion mixing}.
\newblock {\em JCAP}, 05:005, 2003.

\bibitem{Pshirkov:2015tua}
M.~S. Pshirkov, P.~G. Tinyakov, and F.~R. Urban.
\newblock {New limits on extragalactic magnetic fields from rotation measures}.
\newblock {\em Phys. Rev. Lett.}, 116(19):191302, 2016.

\bibitem{Durrer:2013pga}
Ruth Durrer and Andrii Neronov.
\newblock {Cosmological Magnetic Fields: Their Generation, Evolution and
  Observation}.
\newblock {\em Astron. Astrophys. Rev.}, 21:62, 2013.

\bibitem{Jansson:2012pc}
Ronnie Jansson and Glennys~R. Farrar.
\newblock {A New Model of the Galactic Magnetic Field}.
\newblock {\em Astrophys. J.}, 757:14, 2012.

\bibitem{Fermi-LAT:2019pir}
M.~Ajello et~al.
\newblock {The Fourth Catalog of Active Galactic Nuclei Detected by the Fermi
  Large Area Telescope}.
\newblock {\em Astrophys. J.}, 892:105, 2020.

\bibitem{Sanders:2005jx}
Jeremy~S. Sanders, A.~C. Fabian, and R.~J.~H. Dunn.
\newblock {Non-thermal x-rays, a high abundance ridge and fossil bubbles in the
  core of the Perseus cluster of galaxies}.
\newblock {\em Mon. Not. Roy. Astron. Soc.}, 360:133--140, 2005.

\bibitem{Taylor:2006ta}
G.~B. Taylor, N.~E. Gugliucci, A.~C. Fabian, J.~S. Sanders, Gianfranco Gentile,
  and S.~W. Allen.
\newblock {Magnetic fields in the center of the perseus cluster}.
\newblock {\em Mon. Not. Roy. Astron. Soc.}, 368:1500--1506, 2006.

\bibitem{Fermi-LAT:2016nkz}
M.~Ajello et~al.
\newblock {Search for Spectral Irregularities due to
  Photon\textendash{}Axionlike-Particle Oscillations with the Fermi Large Area
  Telescope}.
\newblock {\em Phys. Rev. Lett.}, 116(16):161101, 2016.

\bibitem{Wood:2017yyb}
Matthew Wood, Regina Caputo, Eric Charles, Mattia Di~Mauro, Jeffrey Magill, and
  Jeremy Perkins.
\newblock {Fermipy: An open-source Python package for analysis of Fermi-LAT
  Data}.
\newblock {\em PoS}, ICRC2017:824, 2018.

\bibitem{Fermi-LAT:2019yla}
S.~Abdollahi et~al.
\newblock {$Fermi$ Large Area Telescope Fourth Source Catalog}.
\newblock {\em Astrophys. J. Suppl.}, 247(1):33, 2020.

\bibitem{Dominguez:2010bv}
A.~Dominguez et~al.
\newblock {Extragalactic Background Light Inferred from AEGIS Galaxy SED-type
  Fractions}.
\newblock {\em Mon. Not. Roy. Astron. Soc.}, 410:2556, 2011.

\bibitem{Wilks:1938dza}
S.~S. Wilks.
\newblock {The Large-Sample Distribution of the Likelihood Ratio for Testing
  Composite Hypotheses}.
\newblock {\em Annals Math. Statist.}, 9(1):60--62, 1938.

\bibitem{Chadwick:1999ile}
P.~M. Chadwick, K.~Lyons, T.~J.~L. McComb, S.~McQueen, K.~J. Orford, J.~L.
  Osborne, S.~M. Rayner, S.~E. Shaw, K.~E. Turver, and G.~J. Wieczorek.
\newblock {Pks 2155-304 - a source of vhe gamma-rays}.
\newblock {\em Astropart. Phys.}, 11(1-2):145--148, 1999.

\bibitem{HESS:2010tfm}
A.~Abramowski et~al.
\newblock {VHE gamma-ray emission of PKS 2155-304: spectral and temporal
  variability}.
\newblock {\em Astron. Astrophys.}, 520:A83, 2010.

\bibitem{HESS:2016btr}
H.~Abdalla et~al.
\newblock {Gamma-ray blazar spectra with H.E.S.S. II mono analysis: The case of
  PKS 2155\ensuremath{-}304 and PG 1553+113}.
\newblock {\em Astron. Astrophys.}, 600:A89, 2017.

\bibitem{Punch:1992xw}
M.~Punch et~al.
\newblock {Detection of TeV photons from the active galaxy Markarian 421}.
\newblock {\em Nature}, 358:477--478, 1992.

\bibitem{LAT:2011mmt}
A.~A. Abdo et~al.
\newblock {Fermi large area telescope observations of Markarian 421: The
  missing piece of its spectral energy distribution}.
\newblock {\em Astrophys. J.}, 736:131, 2011.

\bibitem{Meyer:2014gta}
Manuel Meyer and J.~Conrad.
\newblock {Sensitivity of the Cherenkov Telescope Array to the detection of
  axion-like particles at high gamma-ray opacities}.
\newblock {\em JCAP}, 12:016, 2014.

\bibitem{Tavecchio:2014yoa}
Fabrizio Tavecchio, Marco Roncadelli, and Giorgio Galanti.
\newblock {Photons to axion-like particles conversion in Active Galactic
  Nuclei}.
\newblock {\em Phys. Lett. B}, 744:375--379, 2015.

\bibitem{Feain:2009rf}
Ilana Feain, Ronald Ekers, Tara Murphy, Bryan Gaensler, J-P Macquart, Raymond
  Norris, Tim Cornwell, Melanie Johnston-Hollitt, Juergen Ott, and Enno
  Middelberg.
\newblock {Faraday Rotation Structure on Kiloparsec Scales in the Giant Radio
  Lobes of Centaurus A}.
\newblock {\em Astrophys. J.}, 707:114--125, 2009.

\bibitem{Kim:1990}
K.~T. {Kim}, P.~P. {Kronberg}, P.~E. {Dewdney}, and T.~L. {Landecker}.
\newblock {The Halo and Magnetic Field of the Coma Cluster of Galaxies}.
\newblock {\em Astrophys. J.}, 355:29, May 1990.

\bibitem{Carilli:2001hj}
C.~L. Carilli and G.~B. Taylor.
\newblock {Cluster magnetic fields}.
\newblock {\em Ann. Rev. Astron. Astrophys.}, 40:319--348, 2002.

\bibitem{Govoni:2004as}
Federica Govoni and Luigina Feretti.
\newblock {Magnetic field in clusters of galaxies}.
\newblock {\em Int. J. Mod. Phys. D}, 13:1549--1594, 2004.

\bibitem{Subramanian:2005hf}
Kandaswamy Subramanian, Anvar Shukurov, and Nils Erland~L. Haugen.
\newblock {Evolving turbulence and magnetic fields in galaxy clusters}.
\newblock {\em Mon. Not. Roy. Astron. Soc.}, 366:1437--1454, 2006.

\bibitem{Akahori:2010ym}
Takuya Akahori and Dongsu Ryu.
\newblock {Faraday Rotation Measure due to the Intergalactic Magnetic Field}.
\newblock {\em Astrophys. J.}, 723:476--481, 2010.

\bibitem{2023A&A...670L..23H}
V.~{Heesen}, S.~P. {O'Sullivan}, M.~{Br{\"u}ggen}, A.~{Basu}, R.~{Beck},
  A.~{Seta}, E.~{Carretti}, M.~G.~H. {Krause}, M.~{Haverkorn},
  S.~{Hutschenreuter}, A.~{Bracco}, M.~{Stein}, D.~J. {Bomans}, R.~J.
  {Dettmar}, K.~T. {Chy{\.z}y}, G.~H. {Heald}, R.~{Paladino}, and
  C.~{Horellou}.
\newblock {Detection of magnetic fields in the circumgalactic medium of nearby
  galaxies using Faraday rotation}.
\newblock {\em Astron. Astrophys.}, 670:L23, February 2023.

\bibitem{Vernstrom:2017jvh}
Tessa Vernstrom, Bryan~M. Gaensler, Shea Brown, Emil Lenc, and Ray~P. Norris.
\newblock {Low Frequency Radio Constraints on the Synchrotron Cosmic Web}.
\newblock {\em Mon. Not. Roy. Astron. Soc.}, 467(4):4914--4936, 2017.

\bibitem{Brown:2017dwx}
S.~Brown, T.~Vernstrom, E.~Carretti, K.~Dolag, B.~M. Gaensler,
  L.~Staveley-Smith, G.~Bernardi, M.~Haverkorn, M.~Kesteven, and S.~Poppi.
\newblock {Limiting Magnetic Fields in the Cosmic Web with Diffuse Radio
  Emission}.
\newblock {\em Mon. Not. Roy. Astron. Soc.}, 468(4):4246--4253, 2017.

\bibitem{Vacca:2018rta}
Valentina Vacca et~al.
\newblock {Observations of a nearby filament of galaxy clusters with the
  Sardinia Radio Telescope}.
\newblock {\em Mon. Not. Roy. Astron. Soc.}, 479(1):776--806, 2018.

\bibitem{OSullivan:2018shr}
S.~P. O'Sullivan et~al.
\newblock {The intergalactic magnetic field probed by a giant radio galaxy}.
\newblock {\em Astron. Astrophys.}, 622:A16, 2019.

\bibitem{Locatelli:2021byc}
Nicola Locatelli, Franco Vazza, Annalisa Bonafede, Serena Banfi, Gianni
  Bernardi, Claudio Gheller, Andrea Botteon, and Timothy Shimwell.
\newblock {New constraints on the magnetic field in cosmic web filaments}.
\newblock {\em Astron. Astrophys.}, 652:A80, 2021.

\bibitem{Vernstrom:2021hru}
Tessa Vernstrom, George Heald, Franco Vazza, T.~J. Galvin, Jennifer West,
  Nicola Locatelli, Nicolao Fornengo, and Elena Pinetti.
\newblock {Discovery of magnetic fields along stacked cosmic filaments as
  revealed by radio and X-ray emission}.
\newblock {\em Mon. Not. Roy. Astron. Soc.}, 505(3):4178--4196, 2021.

\bibitem{Carretti:2022tbj}
Ettore Carretti et~al.
\newblock {Magnetic field strength in cosmic web filaments}.
\newblock {\em Mon. Not. Roy. Astron. Soc.}, 512(1):945--959, 2022.

\bibitem{2013ApJ...762...33Y}
F.~{Yusef-Zadeh}, J.~W. {Hewitt}, M.~{Wardle}, V.~{Tatischeff}, D.~A.
  {Roberts}, W.~{Cotton}, H.~{Uchiyama}, M.~{Nobukawa}, T.~G. {Tsuru},
  C.~{Heinke}, and M.~{Royster}.
\newblock {Interacting Cosmic Rays with Molecular Clouds: A Bremsstrahlung
  Origin of Diffuse High-energy Emission from the Inner
  2{\textdegree}{\texttimes}1{\textdegree} of the Galactic Center}.
\newblock {\em Astrophys. J.}, 762(1):33, January 2013.

\bibitem{Marinacci:2017wew}
Federico Marinacci et~al.
\newblock {First results from the IllustrisTNG simulations: radio haloes and
  magnetic fields}.
\newblock {\em Mon. Not. Roy. Astron. Soc.}, 480(4):5113--5139, 2018.

\bibitem{Springel:2017tpz}
Volker Springel et~al.
\newblock {First results from the IllustrisTNG simulations: matter and galaxy
  clustering}.
\newblock {\em Mon. Not. Roy. Astron. Soc.}, 475(1):676--698, 2018.

\bibitem{2021JCAP...11..036G}
A.~P. {Gautham}, Francesca {Calore}, Pierluca {Carenza}, Maurizio {Giannotti},
  Dieter {Horns}, Julian {Kuhlmann}, Jhilik {Majumdar}, Alessandro {Mirizzi},
  Andreas {Ringwald}, Anton {Sokolov}, Franziska {Stief}, and Qixin {Yu}.
\newblock {Reconciling hints on axion-like-particles from high-energy gamma
  rays with stellar bounds}.
\newblock {\em JCAP}, 2021(11):036, November 2021.

\bibitem{2022PhRvD.105j3034D}
Christopher {Dessert}, David {Dunsky}, and Benjamin~R. {Safdi}.
\newblock {Upper limit on the axion-photon coupling from magnetic white dwarf
  polarization}.
\newblock {\em Phys. Rev. D}, 105(10):103034, May 2022.

\end{thebibliography}

\end{document}